\documentclass[12pt]{article}
\usepackage{comment}
\usepackage{amsmath}
\usepackage{amssymb}
\usepackage{graphicx}
\usepackage{here}
\usepackage{subcaption}
\usepackage{url}
\usepackage[sort&compress, numbers, merge]{natbib}

\usepackage{physics}
\usepackage[compat=1.1.0]{tikz-feynhand}
\usepackage{slashed}
\usepackage{mathtools}

\numberwithin{equation}{section}

\usepackage[colorlinks=true, linkcolor=blue, citecolor=blue,
urlcolor=black]{hyperref}

\begin{document}
\def\ps{\mathbf{p}}
\def\PS{\mathbf{P}}
\baselineskip 0.6cm
\def\simgt{\mathrel{\lower2.5pt\vbox{\lineskip=0pt\baselineskip=0pt
           \hbox{$>$}\hbox{$\sim$}}}}
\def\simlt{\mathrel{\lower2.5pt\vbox{\lineskip=0pt\baselineskip=0pt
           \hbox{$<$}\hbox{$\sim$}}}}
\def\simprop{\mathrel{\lower3.0pt\vbox{\lineskip=1.0pt\baselineskip=0pt
             \hbox{$\propto$}\hbox{$\sim$}}}}
\def\tr{\mathop{\rm tr}}
\def\SU{\mathop{\rm SU}}

\def\maru{\mathrel{\lower1.8pt\hbox{\scalebox{2.}{$\circ$}}}}   
\def\shikakumaru{\mathrel{\blacksquare\hspace{-2.5pt} \lower1.8pt\hbox{\scalebox{2.}{$\circ$}}}}

\begin{titlepage}

\begin{flushright}
IPMU22-0054
\end{flushright}

\vskip 1.1cm

\begin{center}

{\Large \bf
Precise Estimate of Charged Wino Decay Rate
}

\vskip 1.2cm
Masahiro Ibe$^{a,b}$,
Masataka Mishima$^{a}$,
Yuhei Nakayama$^{a}$ and
Satoshi Shirai$^{b}$
\vskip 0.5cm

{\it

$^a$ {ICRR, The University of Tokyo, Kashiwa, Chiba 277-8582, Japan}

$^b$ {Kavli Institute for the Physics and Mathematics of the Universe
 (WPI), \\The University of Tokyo Institutes for Advanced Study, \\ The
 University of Tokyo, Kashiwa 277-8583, Japan}
}

\vskip 1.0cm

\abstract{
The Wino is an $\mathrm{SU}(2)_{L}$ triplet Majorana fermion and a well-motivated dark matter candidate.
The mass difference between the charged and the neutral Winos is small thanks to the $\mathrm{SU}(2)_{L}$ symmetry.
The small mass difference makes
the charged Wino meta-stable, which provides disappearing charged track signatures at collider experiments.
The constraint on the Wino dark matter at the LHC strongly depends on the Wino lifetime.
We compute the next-to-leading order (NLO) correction of the charged Wino decay and make the most precise estimate of the decay rate.
We find that the NLO decay rate
is determined by the mass difference
and scarcely depend on the Wino mass
itself in the heavy Wino limit.
As a result, we find the NLO correction gives a minor impact on the lifetime of 2--4\% increase.}
\end{center}
\end{titlepage}

\section{Introduction} 
With the establishment of dark matter, we know that the Standard Model alone cannot describe the laws of nature.
The discovery of new physics involving dark matter is one of the most urgent goals for particle physics.
Various types of dark matter have been proposed, and a weakly interacting massive particle (WIMP) is an important dark matter candidate.
Among the various WIMP candidates, the particle called the Wino has attracted particular attention.

The Wino is a Majorana fermion and $\mathrm{SU}(2)_{L}$ triplet
with hypercharge zero.
The Wino has been originally introduced as a superpartner of the weak boson in a supersymmetric (SUSY) model.
In the anomaly mediation model~\cite{Randall:1998uk, Giudice:1998xp}, the Wino is most likely the lightest SUSY particle and dark matter. 
The minimal anomaly mediation realizes ``mini-split SUSY," which has become an increasingly important model after the discovery of the Higgs boson \cite{Hall:2011jd, Hall:2012zp, Nomura:2014asa, Ibe:2011aa, Ibe:2012hu, Arvanitaki:2012ps, ArkaniHamed:2012gw}.
The Wino is also a well-motivated dark matter candidate from a bottom-up perspective, i.e., ``minimal dark matter" model~\cite{Cirelli:2005uq, *Cirelli:2007xd, *Cirelli:2009uv}.

The Wino dark matter $\chi^0$
of a mass $m_0$ 
has an electromagnetically charged partner $\chi^\pm$ of 
a mass $m_\chi$.
The electroweak symmetry breaking makes the charged Wino slightly heavier than the neutral Wino.
The mass difference, $\mathit{\Delta} m = m_\chi - m_0$, is estimated up to the two-loop level~\cite{Yamada:2009ve,Ibe:2012sx,McKay:2017xlc}, which is roughly 160\,MeV. 
Thanks to the small mass difference, the charged Wino is metastable and its decay length, $c\tau_\chi$, 
is about 5\,cm.
This causes the disappearing charged track signals in collider experiments, which play a pivotal role in the exploration of the Wino.

As the Wino has electroweak interactions, it provides either direct or indirect~\cite{Harigaya:2015yaa, Matsumoto:2017vfu, Matsumoto:2018ioi,Katayose:2020one,DiLuzio:2018jwd} signatures at collider experiments.
At the LHC, conventional direct searches based on missing energies and indirect searches via precision measurements on the Drell–Yan processes  provide weak limits of around 100\,GeV on the Wino mass, which is comparable to the LEP constraints~\cite{LEP}.
On the other hand, the disappearing charged track signal is the key to the search for the Wino at the LHC~\cite{Ibe:2006de, Buckley:2009kv, Asai:2007sw, Asai:2008sk, Asai:2008im, Mahbubani:2017gjh, Fukuda:2017jmk}.
The recent ATLAS result excludes the pure Wino mass up to 660\,GeV\,\cite{ATLAS:2022rme}.

The lifetime of the charged Wino is of importance for the disappearing charged track search.
The current reconstruction of the disappearing charged track at ATLAS is based on the innermost pixel detectors.
It requires that the charged Wino decays at a distance of at least 
12 cm from the beam line, which is larger than the prediction of 
$c\tau_\chi$.
The signal acceptance depends exponentially on $c\tau_\chi$.
Therefore, precise estimate of the charged Wino lifetime is essential for the probe of the Wino dark matter at collider experiments.

The charged Wino mainly decays into the charged pion, $\chi^\pm \to \chi^0+ \pi^\pm$.
The most important parameter for estimate of the lifetime 
is the mass difference, $\mathit{\Delta}m$, 
since the decay rate
is approximately proportional to $(\mathit{\Delta}m)^3$ at tree-level.
The remaining theory error from the three-loop contributions 
to the mass difference is about 
$\delta\mathit{\Delta}m=\pm 0.3$\,MeV~\cite{Ibe:2012sx,McKay:2017xlc}, 
which results in about $1$\,\% uncertainty of the decay rate.

While the mass difference has been calculated to the two-loop level,
the tree-level amplitude has been used to calculate the Wino decay rate. 
In this work, we study the next-to-leading order (NLO)
correction to the charged Wino decay rate into the pion.

At the tree-level, the decay amplitude is proportional to $\mathit{\Delta} m$.
At the one-loop, however, it is not obvious whether NLO corrections induce only amplitudes proportional to $\mathit{\Delta} m$.
For the NLO estimate, we need to 
consider the extended Chiral perturbation theory (ChPT) including QED and the Wino.
There, 
the amplitudes not proportional to
$\mathit{\Delta}m$  
are generated 
in contrast to the tree-level amplitude.
If those contributions remained in the decay rate, the one-loop contributions would dominate over the tree-level decay rate and the prediction would change drastically.
In addition, there could be the large amplitude enhanced by $\log (m_\chi/m_\pi)$,
which also would change the decay rate significantly.
Therefore, it is a non-trivial question whether the one-loop correction is indeed sub-leading compared to the 
tree-level decay rate.

As we will show, such Wino mass dependences cancel in the decay rate once we match the low energy effective theory, i.e.,
the extended ChPT,
to the electroweak theory.
The resultant NLO decay rate
is determined by the mass difference
and scarcely depends 
on the Wino mass itself in the heavy Wino limit.
As a result, we found that the NLO correction gives a minor impact on the lifetime of 2--4\% increase.

This result shows that a decoupling theorem similar to the Appelquist-Carazzone theorem~\cite{Appelquist:1974tg} holds for the Wino decay at the one-loop level.
That is, the radiative corrections depend on the Wino mass only through $\order{\alpha m_{\chi}^{s}}$  with $s \le 0$ and there is no logarithmically enhanced dependencies on $m_\chi$, and the decay rate becomes constant in the limit of $m_\chi \to \infty$.
Here, $\alpha$ is the QED fine-structure constant.
This result is non-trivial 
since the Appelquist--Carazzone decoupling theorem is not applicable to the decay of the Wino, where the external lines of the diagrams include heavy particles.

This paper is organized as follows.
In Sec.\,\ref{sec:Tree-level},
we summarise the tree-level 
Wino decay.
In Sec.\,\ref{sec:procedure}, we explain the procedure of the NLO analysis and
summarize the result.
In Sec.\,\ref{sec:EWtoFF}, 
we match the electroweak theory 
and the Four-Fermi theory 
at the one-loop level for the first step.
In Sec.\,\ref{sec:FFtoChPT}, 
we match the Four-Fermi theory 
to the ChPT at the one-loop level.
In Sec.\,\ref{sec:QED}, we evaluate QED correction to the Wino decay rate
in the effective field theory
where the counterterms are provided in Sec\,\ref{sec:FFtoChPT}.
Our result is given in Sec.\,\ref{sec:Results}.
Finally, Sec.\,\ref{sec:conclusions} is devoted to conclusions and discussions.

\section{Tree-Level Wino Interactions and Decays}
In this section, we review the tree-level analysis of the charged Wino decay.
Throughout this paper, $m_\chi$ and $m_0$ denote the pole masses of the 
charged and the neutral Winos, respectively.
The mass difference between them
is denoted by,
$\mathit{\Delta}m$.
In the pure-Wino case,
the mass difference is a function of the Wino mass $m_\chi$.
The NLO correction to the decay rate, 
however, can be calculated for a generic $\mathit{\Delta}m$.
In the following, we take $m_{\chi}$ and $\mathit{\Delta} m$ as free parameters,
 where we assume $m_\pi < \mathit{\Delta} m< m_K$ with $m_{\pi,K}$ being the charged $\pi/K$-meson masses,
 although we consider the pure-Wino scenario.

\label{sec:Tree-level}
\subsection{Tree-Level Lagrangian}
In this subsection, we summarize the Wino interactions.
The tree-level Lagrangian of the Wino is given by,
\begin{align}
    \mathcal{L}_{\mathrm{tree}}= \chi^{i\dagger} i \bar{\sigma}^\mu \left(\delta^{ij} \partial_{\mu} +g\epsilon^{ikj}W^k_\mu \right)\chi^j - \frac{1}{2}m
    \chi^i\chi^i 
    - \frac{1}{2}m\chi^{i\dagger}\chi^{i\dagger}\ ,
\end{align}
where $g$ is the SU(2)$_L$ gauge coupling constant, $W^i_\mu$  ($i=1,2,3$) are the $\mathrm{SU}(2)_L$ gauge bosons, and the two-component Weyl fermions, $\chi^i$, are the Winos.
We follow the conventions of the spinor indices in Ref.\,\cite{Dreiner:2008tw}.
The tree-level Wino mass parameter, $m$, is taken to be real and positive without loss of generality.

With the electroweak symmetry breaking, 
we rewrite the Lagrangian as,
\begin{align}
\mathcal{L}_{\mathrm{Wino,tree}}=& \chi^{3\dagger} i \bar{\sigma}^\mu  \partial_{\mu} \chi^3
        +\chi^{+\dagger} i \bar{\sigma}^\mu  \partial_{\mu} \chi^{+}
         +\chi^{-\dagger} i \bar{\sigma}^\mu  \partial_{\mu} \chi^{-}
         - \frac{1}{2}m
    \left(\chi^3\chi^3 +h.c.\right)
        - m
        \left(\chi^{+}\chi^{-} + h.c.\right)\ \cr
        &- g \chi^{-\dagger} \bar{\sigma}^\mu W^3_\mu \chi^{-} 
        + g \chi^{+\dagger} 
        \bar{\sigma}^\mu W^3_\mu\chi^{+}
        -g\left(\chi^{+\dagger} \bar{\sigma}^\mu W^+_\mu \chi^3 +h.c.\right)+ 
        g\left( \chi^{-\dagger} \bar{\sigma}^\mu W_\mu^- \chi^3 + h.c.\right)\ , \cr
\end{align}
where we have defined
\begin{align}
    W^\pm &= \frac{1}{\sqrt{2}}(W^1 \mp iW^2) \ ,\\
    \chi^\pm &= \frac{1}{\sqrt{2}}(\chi^1 \mp i\chi^2)\ .
\end{align}
The neutral Wino, $\chi^0=\chi^3$, becomes the 
dark matter.
In terms of the four-component fermions, the above Lagrangian is reduced to
\begin{align}
\mathcal{L}_{\mathrm{Wino,tree}}    =& \frac{1}{2}\bar{\psi}_0 \left(i\slashed{\partial}- m \right) \psi_0 
+ \bar{\psi}_- \left(i\slashed{\partial}- m \right) \psi_-\cr
&-g \bar{\psi}_-\slashed{W}^3\psi_- + g \bar{\psi}_-\slashed{W}^- \psi_0 + g \bar{\psi}_0 \slashed{W}^+ \psi_- \ .
\end{align}
Here we defined, 
\begin{align}
    \psi_0 &= (\chi^0_\alpha, \epsilon^{\dot{\alpha}\dot{\beta}}\chi^{0*}_{\dot{\beta}})^T\ , \\
    \psi_- &= (\chi^-_{\alpha},\epsilon^{\dot{\alpha}\dot{\beta}}\chi^{+\dagger}{}_{\dot{\beta}})^T\ ,
\end{align}
where $\psi_0$ is a Majorana fermion.%
\footnote{By using $C=-i\gamma^2 \gamma^0$, we define $ \psi^c :=-i\gamma^2 \psi^* = C \bar{\psi}^T$. The Majorana fermion satisfies $\psi_0^c = \psi_0$.}
Due to the electroweak symmetry breaking, the charged and the neutral Wino masses are split.
In the following, we take the tree-level mass parameter $m$ to be equal to the physical charged Wino mass, $m_\chi$, by adjusting the mass counterterm.
The neutral Wino mass is, on the other hand, given by
$m_0 = m_{{\chi}}-\mathit{\Delta} m$.

We obtain the Wino coupling to the $Z$-boson and the photon by replacing
\begin{align}
    W^3_\mu = c_W Z_\mu + s_W A_\mu\ .
\end{align}
The weak mixing angle, $s_W=\sin\theta_W$ ($c_W=\sqrt{1-s_W^2}$), 
and the QED coupling constant, $e$, are defined by,
\begin{align}
\label{eq:W}
    s_W^2  &=  1- \frac{M_W^2}{M_Z^2}\ , \\
    e &= g s_W\ .
    \label{eq:g}
\end{align}
In our analysis, we adopt the on-shell 
renormalization 
scheme in the electroweak theory developed in Refs.\,\cite{Bohm:1986rj,Hollik:1988ii}.
In this scheme, the tree-level masses, $M_{Z,W}$, are 
taken to be the physical gauge boson masses by setting the counterterms in the electroweak theory appropriately.
The weak mixing angle, $s_W$, is defined  so that 
the relation in Eq.\,\eqref{eq:W} is valid even in higher order. 
The electromagnetic gauge coupling constant $e$
is determined by the Thomson limit.
The tree-level gauge coupling constant $g$ is defined so that Eq.\,\eqref{eq:g} is valid also in higher order.
The physical quantities used in our analysis 
are summarized in Tab.\,\ref{tab:input}.

\begin{table}[t]
    \caption{Input parameters used in the analysis taken from Ref.\,\cite{Workman:2022ynf}. We use the weak mixing angle defined in the on-shell scheme in Eq.\,\eqref{eq:W}.
    Note that the values of the Fermi constant and the pion decay constant does not affect the final result as we take the ratio between the decay rates of the charged Wino and the charged pion.}
    \centering
    \begin{tabular}{ccc}
    \hline
    \bf{Quantity} & \bf{symbol} & \bf{Value}\\
    \hline
    QED fine-structure constant    & $\alpha$ & 1/137.035 999 084(21)\\
$W^\pm$-boson mass   &  $M_W$ &80.379(12)\,GeV  \\
$Z$-boson mass & $M_Z$ &91.1876(21)\,GeV \\
$e$ mass & $m_e$ & 
0.51099895000(15)\,MeV\\
$\mu$ mass & $m_\mu$ & 105.6583755(23)\,MeV\\
$\pi^\pm$ mass & $m_\pi$ & 139.57039(18)\,MeV\\
Charged pion lifetime &$\tau_\pi$&
$ 2.6033(5)\times 10^{-8}\,\mathrm{sec}$\\
$B(\pi^\pm \to \mu^\pm + \nu(+\gamma))$
& &$ 99.98770(4)$\%
\\
\hline\\
    \end{tabular}
    \label{tab:input}
\end{table}

By integrating out the $W$-boson from
\begin{align}
 {\cal L}_{\mathrm{weak,tree}} = M_W^2 W^+_\mu W^{-\mu}  +\left( g \bar{\psi}_-\slashed{W}^- \psi_0  
 + \frac{g}{\sqrt{2}}\,\bar{\ell}\,
 \slashed{W}^- P_L \nu_\ell  
+ \frac{V_{ud}^*g}{\sqrt{2}}\,\bar{d}\, \slashed{W}^- P_L u + h.c.\right)\ ,
\end{align}
we obtain 
the Four-Fermi interactions relevant
for the Wino decay;
\begin{align}
\label{eq:WinoWeak}
    \mathcal{L}_{\mathrm{tree}}^\mathrm{FF} \supset - 2G^0_F
    \left[(\bar{\nu}_{\ell}
 \gamma^\mu 2P_L \ell) 
 (\bar{\psi}_-\gamma_\mu \psi_0)
  + V_{ud}
(\bar{u}\gamma^\mu 2P_L d  )
(\bar{\psi}_-\gamma_\mu \psi_0) + h.c.\right]\ .
\end{align}
Here, $G^0_F \equiv g^2/(4\sqrt{2}M_W^2)$ and $P_L = (1-\gamma_5)/2$ 
is the left-handed projection operator.
Note that $G_F^0$ defined here is 
not equal to the conventional Fermi constant $G_F = 1.1663788\times 10^{-5}\,\mathrm{GeV}^{-2}$~\cite{Workman:2022ynf}, which is determined 
by the muon lifetime.
For the leptons, $\ell$, we have taken summation over the three generations. 
Since the mass difference $\mathit{\Delta}m$ is smaller than the $K$-meson masses, we leave only the up and down quarks. 
We have taken the mass basis of the quarks
where $V_{ud}$ denotes
the $(1,1)$ element of the CKM matrix.

\begin{figure}[t]
    \centering
    \includegraphics[]{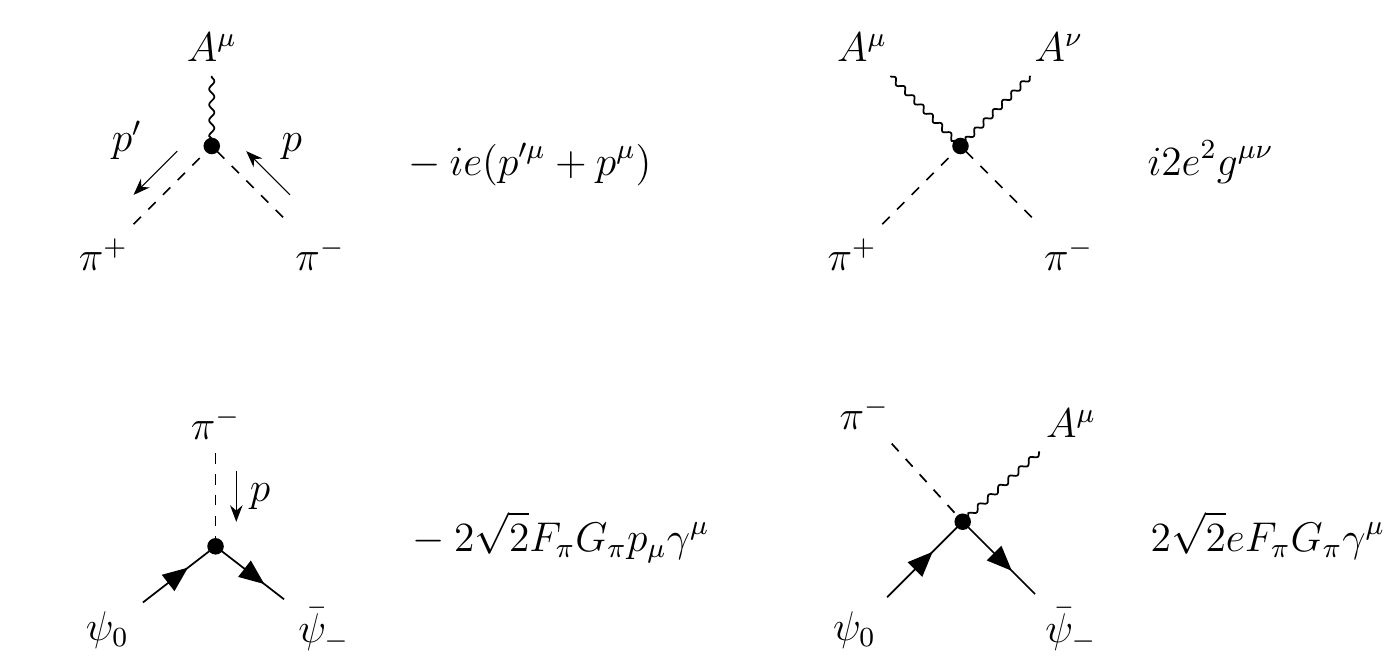}
    \caption{The Feynman rules of the Wino--pion interactions. The arrow on the fermion propagator denotes the flow of the fermion number. We do not show the flow of the QED charge.
    In the figure, we show the names of the fields.}
    \label{fig:Wino-Pion-QED}
\end{figure}

Only 
the charged pion contributes to 
the charged Wino decay,
since we assume that the mass difference
is smaller than the $K$-meson mass.
The coupling to the 
pions 
can be obtained
through 
the axial quark current,
\begin{align}
\label{eq:JA}
    J_{A\mu}^a(x)=  \left(\bar{u}(x),\bar{d}(x)\right)
    \gamma_\mu \gamma_5 \frac{\sigma^a}{2} 
    \left(
    \begin{array}{c}
         u(x)  \\
         d(x)
    \end{array}\right)\ ,
\end{align}
with $\sigma^a$ $(a=1,2,3)$ being the Pauli matrices.
The axial current contains the pions as,
\begin{align}
\label{eq:pion}
    \langle 0 | J_{A\mu}^a(0)|\pi^b(p)\rangle = i F_\pi  p_\mu \delta^{ab}\ ,
\end{align}
where $\pi^a(p)$ denotes the pions with 
momentum $p$, and 
$F_\pi = 92.2\pm 0.2$\,MeV
is the pion decay constant~\cite{Descotes-Genon:2005wrq}. 
The charged pion is defined by,
\begin{align}
    \pi^-(x) = \frac{1}{\sqrt{2}} (\pi^1(x)+i\pi^{2}(x)) \ .
\end{align}
By substituting 
\begin{align}
\label{eq:JmutoPi}
    J_\mu^{\mathrm{(quark)}} = V_{ud}
(\bar{u}\gamma^\mu 2P_L d  )\to  \sqrt{2}F_\pi V_{ud} D_\mu \pi^- \ ,
\end{align}
into Eq.\,\eqref{eq:WinoWeak}
, we obtain the Wino-pion interaction,
\begin{align}
\label{eq:WinoPion}
    {\mathcal{L}}_{\mathrm{Wino-Pion}} =- 2\sqrt{2}F_\pi{G_\pi^0} (D_\mu \pi^-) \times (\bar{\psi}_- \gamma^\mu \psi_0) + h.c.\ ,
\end{align}
where $G_\pi^0 =V_{ud} G_F^0$.
The Feynman rules for the Wino-pion interactions are
given in Fig.\,\ref{fig:Wino-Pion-QED}.

\subsection{Tree-Level Wino Decay}
Since we assume
$\mathit{\Delta}m >m_\pi$, 
the charged Wino can decay into the neutral Wino and the charged pion.
In this subsection, we calculate the charged Wino decay rate
at the leading order.

The tree level amplitude of the charged Wino decay, $\psi_-(p) \to  \psi_0(q)+\pi^-(r)$, is given by,
\begin{align}
\label{eq:iMtree}
    i\mathcal{M}_{{\mathrm{tree}}} &=  2\sqrt{2}F_\pi G^0_\pi \mathit{\Delta} m\,\bar{u}_{0}(q)  u_{-}(p)\ ,
\end{align}
where $p$, $q$, $r$ are the four momenta of the charged Wino, the neutral Wino and the pion, respectively. 
The $u$'s denote the fermion wave functions.
The spin-summed squared matrix element averaged by the charged Wino spin is given by,
\begin{align}
    \overline{|\mathcal{M}_{\mathrm{tree}}|^2}
    &=  8 F_\pi^2 (G_\pi^0)^2
    \mathit{\Delta} m^2
    ((m_{{\chi}} + m_0)^2 - m_\pi^2)\\
    &=  32 F_\pi^2 (G^0_\pi)^2  \mathit{\Delta} m^2m_{{\chi}}^2
    \left(1-\frac{\mathit{\Delta}m}{m_\chi}+\frac{\mathit{\Delta}m^2 - m_\pi^2}{4m_\chi^2}\right)\ .
\end{align}
As a result, 
 the tree-level charged Wino decay rate is given by,
\begin{align}
    \Gamma_{\mathrm{tree}}(\chi^-\to \pi^- + \chi^0)=\frac{4}{\pi} F_\pi^2 (G^0_\pi)^2 {\mathit{\Delta} m}^3\left(1- \frac{m_\pi^2}{\mathit{\Delta} m^2}\right)^{1/2} \left(1-\frac{\mathit{\Delta}m}{m_\chi}+\frac{\mathit{\Delta}m^2 - m_\pi^2}{4m_\chi^2}\right) \ .
\end{align}
By taking the ratio of the above expression and the tree-level pion decay rate,
we obtain
\begin{align}
\label{eq:treeRatio}
    \frac{\Gamma_{\mathrm{tree}}(\chi^-\to \pi^- + \chi^0)}{\Gamma_{\mathrm{tree}}(\pi^- \to \mu^- + \bar{\nu})} 
= 16 \frac{{\mathit{\Delta} m}^3}{m_\pi m_\mu^2}
\left(1- \frac{m_\pi^2}{\mathit{\Delta} m^2}\right)^{1/2}\left(1-\frac{\mathit{\Delta}m}{m_\chi}+\frac{\mathit{\Delta}m^2 - m_\pi^2}{4m_\chi^2}\right)
\left(1- \frac{m_\mu^2}{m_\pi^2}\right)^{-2}\ ,
\end{align}
where $\Gamma_{\mathrm{tree}}(\pi^- \to \mu^- + \bar{\nu}_\mu)$ is given by,
\begin{align}
\label{eq:GammaPI}
    \Gamma_{\mathrm{tree}}(\pi^- \to \mu^- + \bar{\nu}_\mu) = \frac{(G_\pi^0)^2 F_\pi^2 m_\mu^2 m_\pi}{4\pi} \left(1-\frac{m_\mu^2}{m_\pi^2}\right)^2\ .
\end{align}
As a result, from the pion lifetime in Tab.\,\ref{tab:input}, 
the Wino decay length, $c\tau_\chi$, turns out to be
$5\,\mathrm{cm}$
for $\mathit{\Delta} m \sim 160$\,MeV.
In what follows, we denote $\Gamma_\chi =\Gamma_{\mathrm{tree}}(\chi^- \to \chi^0+\pi^-)$ and 
$\Gamma_\pi=\Gamma_{\mathrm{tree}}(\pi^- \to \mu^- + \bar{\nu}_\mu)$, respectively.

The charged Wino also decays into the neutral Wino, a charged lepton $\ell$ and a neutrino $\bar{\nu}_{\ell}$. 
The decay rate is given by
\begin{align}
\label{eq:3body}
    &\Gamma_{\mathrm{tree}}(\chi^- \to \ell^-+\bar{\nu}_\ell + \chi^0 ) = \frac{2 (G_F^0)^2{\mathit{\Delta}m^5}}{15\pi^3}F_{\mathrm{lep}}(m_\ell/\mathit{\Delta}m)  \ , \\
    &F_{\mathrm{lep}}(x) = \frac{1}{2} \left(\sqrt{1-x^2} \left(-8 x^4-9 x^2+2\right)+15 x^4 \tanh
   ^{-1}\left(\sqrt{1-x^2}\right)\right) \ ,
\end{align}
in the large $m_\chi$ limit where $m_\ell$ is the mass of the lepton $\ell$~\cite{Thomas:1998wy}.
For the electron and muon,
the function $F_{\mathrm{lep}}(x)$ are $F_{\mathrm{lep}}(m_e/\mathit{\Delta}m) \simeq 1$
and $F_{\mathrm{lep}}(m_\mu/\mathit{\Delta}m)\simeq 0.1$ for $\mathit{\Delta}m \simeq 160$\,MeV. 
Accordingly, the branching ratios of
the electron mode and the muon mode are about $2$\% and $0.1$\% for $\mathit{\Delta}m \simeq 160$\,MeV, respectively.
We denote $\Gamma_\ell =\Gamma_{\mathrm{tree}}(\chi^- \to \ell^-+\bar{\nu}_\ell + \chi^0)$ in the following.

\section{Procedure and Summary
}
\label{sec:procedure}
\subsection{NLO Matching from EW Theory to ChPT}
In this subsection, we summarize the procedure for the calculation of the NLO corrections to the charged Wino decay rate. 
The electroweak theory with the Wino is renormalizable and has only one new parameter in addition to those in the electroweak theory, i.e., the Wino mass $m_\chi$.
Accordingly, we can predict the charged Wino decay rate for a given Wino mass. 

For the charged Wino decay, however, 
we use the Wino-pion interactions, which are not renormalizable.
Moreover, the radiative corrections to the Wino-pion interactions require counterterms 
which are not obtained by the multiplicative renormalization to the tree-level Lagrangian.
The same problem has also arised in the analysis of the radiative corrections to the charged pion decay. 
The counterterms necessary
for the pion decay
have been introduced in Ref.\,\cite{Knecht:1999ag} by extending the ChPT to include QED and the weak interactions.
Following Ref.\,\cite{Knecht:1999ag}, we will introduce counterterms necessary for
the charged Wino decay (Sec.\,\ref{sec:FFtoChPT}).

Within the extended ChPT including the Wino-pion interactions, the divergent parts of the counterterms are set to cancel the ultraviolet (UV) divergences.
On the other hand, the finite parts of the counterterms are not determined.
To ensure the predictability, we therefore 
need to 
determine the finite parts of the counterterms
in the extended ChPT from the electroweak theory.

\begin{figure}[t]
\centering{{\includegraphics[width=0.7\textwidth]{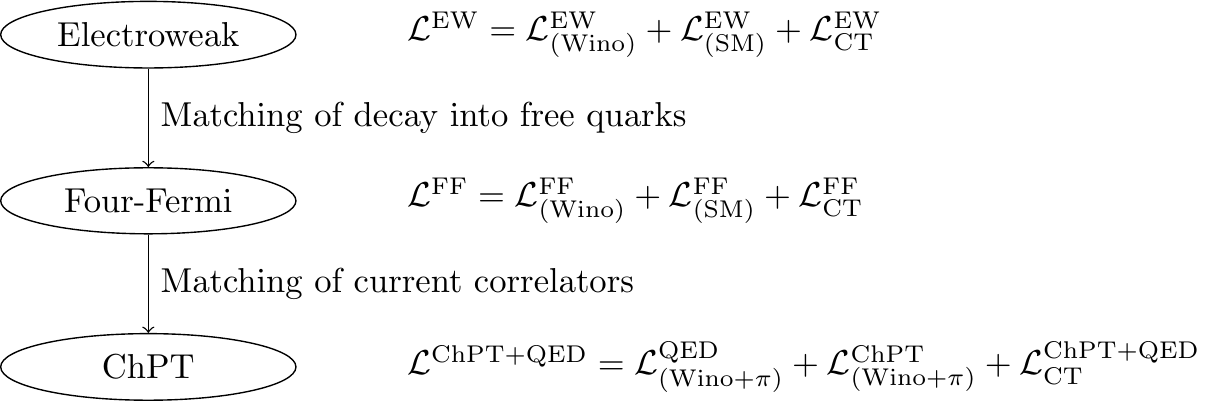}} }
 \caption{
Flowchart of the matching procedure.
 }
\label{fig:flowchart}
\end{figure}

The finite part of the counterterms for the pion decay has been determined by Descotes-Genon and Moussallam
by matching the extended ChPT with 
the electroweak theory \cite{Descotes-Genon:2005wrq}.
In our analysis, we follow the matching procedure taken by 
Descotes-Genon and Moussallam (D\&M).
The matching procedure is as follows (see Fig.\,\ref{fig:flowchart});
\vspace{.5cm}
\begin{enumerate}
    \item We determine the counterterms in the effective Four-Fermi theory of the Wino and quarks, $\mathcal{L}_{\mathrm{CT}}^{\mathrm{FF}}$, by matching the charged Wino decay amplitude into the free quarks in the electroweak theory and in the Four-Fermi theory.
    The pure QCD corrections do not affect the matching condition because the they are common in both theories. Therefore, we do not need to include the pure QCD loop diagrams.
    \item We determine the counterterms in the extended ChPT including the Winos, 
    the weak interaction, and QED,  $\mathcal{L}_{\mathrm{CT}}^{\mathrm{ChPT+QED}}$, from $\mathcal{L}_{\mathrm{CT}}^{\mathrm{FF}}$ 
    by matching the current correlator calculations 
    in the extended ChPT and in the Four-Fermi theory. 
\end{enumerate}
Once we prepare the counterterms, $\mathcal{L}_{\mathrm{CT}}^{\mathrm{ChPT+QED}}$, 
we can
make a prediction of the Wino decay rate at the NLO. 
Finally, by taking the ratio
between the NLO decay rates 
of the Wino and the pion, we obtain precise estimate of the Wino decay rate.

The ChPT and the Four-Fermi theory are matched by comparing 
the identical current correlators calculated in these two different theories.
In the Four-Fermi theory, the currents are given in terms of the quarks, and hence, the evaluation of the current correlators go beyond the perturbative analysis due to the strong dynamics.
To overcome this difficulty and to obtain an analytical result, we will use a phenomenological hadronic model.

In this respect, we use the minimal resonance model (MRM)~\cite{Weinberg:1967kj} (see also Ref.\,\cite{Moussallam:1997xx}).
The MRM comprises the 
$\pi$, $\rho$ and $a_1$ resonances, which satisfies
the Weinberg sum rules and 
the leading QCD asymptotic constraints.
The former feature is important to yield reasonable estimate of the hadronic contributions. 
The latter feature, on the other hand, ensures that the UV dependence
of the hadronic model 
is consistent with that of the 
Four-Fermi theory.

\subsection{Summary of  Result}
\begin{figure}[tb]
\centering
  \includegraphics[width=0.7\linewidth]{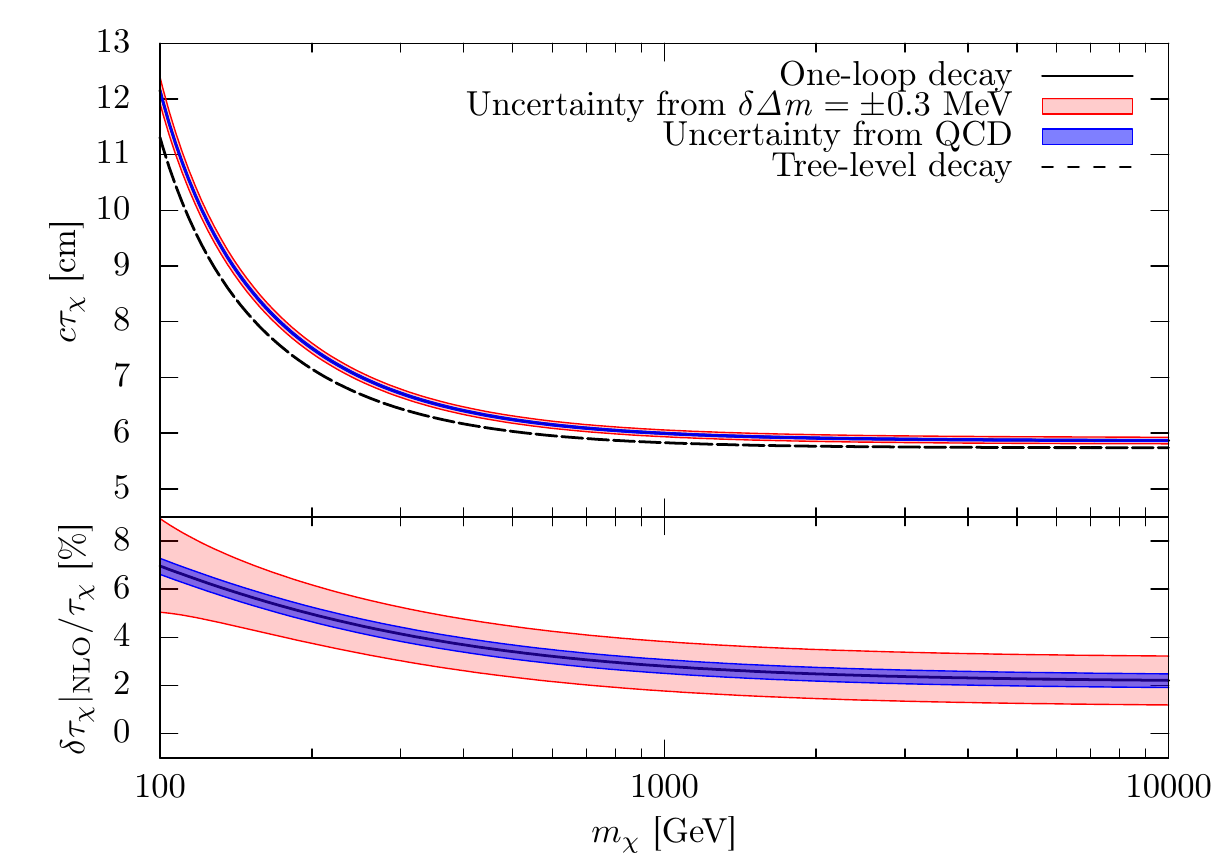}
\caption{
The Wino decay length as a function of the Wino mass $m_\chi$.
We show the central value of our estimation in black solid lines.
We also show the tree-level decay length in a black dashed line.
The blue bands show the uncertainty of our one-loop estimation from the QCD dynamics.
The red bands show the uncertainty from the three-loop correction to the Wino mass difference,
$\delta\mathit{\Delta}m = \pm 0.3\, \mathrm{MeV}$
in Ref.\,\cite{Ibe:2012sx}. Here we define $\delta \tau_\chi|_\mathrm{NLO} \equiv \tau_\chi|_\mathrm{NLO}-\tau_\chi|_\mathrm{LO}$.}
\label{fig:lifetime}
\end{figure}

Applying the above procedure, the final result 
of the NLO result of the Wino decay length is given by,
\begin{align}
c \tau_\chi /  \mathrm{cm} \simeq \frac{-7.19014\times 10^6 + 103791 t+2487.92 t^2+  22.9056   t^3 }{ 1 -34759.7 t +  376.148 t^2+ 3.90903t^3 }  ~~~ \left(t = \frac{m_\chi}{\mathrm {GeV}} \right)
\end{align}
for the pure Wino case.
This fitting is valid for $m_\chi>90$ GeV.
Here, we have used the prediction on $\mathit{\Delta}m$ as a function 
of $m_\chi$ in GeV (see Ref.\,\cite{Ibe:2012sx})
at the two-loop level.\footnote{
Here we adopt a fitting formula for the two-loop mass difference for the pure Wino as:
\begin{align}
    \mathit{\Delta m}/\mathrm{MeV} =  
\frac{ 21.8641 + 8.68343 t + 0.0568066 t^2}{1 + 0.0530366 t + 0.000345101t^2} ~~~ \left(t = \frac{m_\chi}{\mathrm {GeV}} \right),
\end{align}
which provides a stable fitting for $m_{\chi} > 90$ GeV.
}
The analytic and numerical NLO decay rates as a function of $m_{\chi}$
and $\mathit{\Delta}m$ are also given in Eqs.\,\eqref{eq:analyticNLO} and \eqref{eq:numericalNLO}.
The result confirms that the decoupling theorem similar to the Appelquist-Carazzone theorem holds for the Wino decay at the one-loop level, where the radiative corrections include only $\order{\alpha m_{\chi}^{s}}$ effects with $s \le 0$ and no logarithmic enhancement factor of $\order{\log m_\chi}$.

 Fig.\,\ref{fig:lifetime} shows the numerical estimate of the NLO Wino decay length 
(black solid line).
The blue bands show the uncertainty of the NLO decay rate estimation from the QCD dynamics.
The red bands show the uncertainty to that from the higher loop corrections to the Wino mass difference, $\delta{\mathit{\Delta}m}=0.3$\,MeV, in Ref.\,\cite{Ibe:2012sx}.

\section{Matching Between EW and Four-Fermi Theories}
\label{sec:EWtoFF}
Let us begin with the matching between the Four-Fermi theory and the electroweak theory.
As we explained in the previous section,
we match these two theories through the decay amplitudes of the charged Wino into the free quarks,
\begin{align}
    \chi^- \to \chi^0 + d_L + \bar{u}_L\ ,
\end{align}
by assuming that there is no QCD strong dynamics.
Here, we assume that the quark masses are zero.
We adopt the on-shell renormalization scheme in the electroweak theory in Refs.\,\cite{Bohm:1986rj,Hollik:1988ii}, 
while we use the $\overline{\mathrm{MS}}$ scheme in the Four-Fermi theory.

 At the tree-level, the Four-Fermi theory between the Wino and the quarks is given in Eq.\,\eqref{eq:WinoWeak}
\begin{align}
\label{eq:WinoQuark}
    \mathcal{L}^{\mathrm{FF}}_{\mathrm{(Wino)}}
    =&-4 G_\pi^0 (\bar{u}\gamma^\mu P_L d)(\bar{\psi}_-\gamma^\mu \psi_0) +h.c. 
\end{align}
At the tree-level, 
the decay amplitude of the process 
$\chi^-(p) \to \chi^0(q) + d(r_1)+\bar{u}(r_2) $ is given by,
\begin{align}
\label{eq:quarktree}
i\mathcal{M}_{\mathrm{tree}}^{\mathrm{quark}} = -4i G_\pi^0 \bar{u}_{0}(q) \gamma_\mu u_{-}(p) \bar{u}_d(r_1)\gamma^\mu P_L v_u(r_2)\ .
\end{align}
Here $p$, $q$, $r_{1,2}$ are the four momenta of the Winos and the quarks, respectively,
and 
$u_d$ and $v_{\bar{u}}$ are the wave functions
of the quarks.
Since the mass difference $\mathit{\Delta}m$ is tiny, 
the decay amplitude into the free quarks 
in the electroweak theory is given by 
\begin{align}
i\mathcal{M}_{\mathrm{tree}}^{\mathrm{quark}}|_{\mathrm{EW}} =
i\mathcal{M}_{\mathrm{tree}}^{\mathrm{quark}} \times (1+ \order{\mathit{\Delta}m^2/M_W^2} )\ ,
\end{align}
where the 
$\order{\mathit{\Delta}m^2/M_W^2}$ contributions are 
negligibly small.

Following D\&M analysis,
we add counterterms to the Four-Fermi theory,
\begin{align}
\label{eq:FF counterterms}
     \mathcal{L}^{\mathrm{FF}}_{\mathrm{CT}}
\supset -4 G_\pi^0 e^2 & (\bar{u}\gamma^\mu P_L d)(\bar{\psi}_-\gamma^\mu \psi_0 +h.c.)
     \cr
 &\times\left[f_{\chi\chi}Q_\chi^2 + f_{d\bar{u}}(Q_d+Q_{\bar{u}})^2 + f_{\chi d}Q_\chi Q_d - f_{\chi \bar{u}}Q_\chi Q_{\bar{u}}\right]\ ,
\end{align}
where $f$'s are the coefficients of the counterterms.
The coefficients
$f's$ depend on the arbitrary mass scale $\mu$ of the $\overline{\mathrm{MS}}$ scheme.
In addition to $f$'s, we need a counterterm for the charged Wino mass.
With these counterterms, the UV divergence proportional to $\mathcal{M}_{\mathrm{tree}}^{\mathrm{quark}}$ 
can be eliminated.
The QED charges of the charged Wino and the down and the anti-up quarks are denoted by $Q_{\chi,d,\bar{u}}$, respectively.
We eventually take $Q_{\chi}=-1$ and $Q_{d}+Q_{\bar{u}}=-1$ while we leave
$Q_{d}-Q_{\bar{u}}= \bar{Q}$ as a free parameter.
As we will see, $\bar{Q}$ dependence allows us to
impose two independent conditions on $f$'s by matching the amplitude of a single decay process.

In this convention for 
the counterterms in Eq.\,\eqref{eq:FF counterterms}, 
we have eliminated 
the counterterms to the kinetic terms of the Wino and the quarks
by field redefinitions.
The relations between the counterterms
proportional to $f_{\chi\chi}$ and $f_{d\bar{u}}$ and those 
to the kinetic terms are explained in 
 Eq.\,\eqref{eq:quark counterterms} and in  Appendix\,\ref{sec:quark counterterm}.
We also implicitly take the Wino mass counterterm so that $m_\chi$ becomes the physical charged Wino mass.

\subsection{Wino Decay into Free Quarks in Electroweak Theory}
\subsubsection{Two-Point Functions}
Let us begin with the fermion two-point function, $\Sigma_\mathrm{fermion}$, which appears in the one-particle irreducible (1PI) 
effective action through 
\begin{align}
    \Gamma_{\mathrm{fermion}}^{(2)}(p) = (\slashed{p}-m_{\mathrm{fermion}}) - \Sigma_{\mathrm{fermion}}(\slashed{p}) + (\mathrm{CT})\ ,
\end{align}
where $m_{\mathrm{fermion}}$ is the tree-level fermion mass, and (CT) denotes the counterterms.
At the one-loop level, the relevant two-point functions come from the photon and the $W/Z$-boson loop diagrams.
As stated in Sec.~\ref{sec:procedure}, the pure QCD corrections do not affect the matching conditions and hence we omit these corrections.
In the dimensional regularization ($d=4-2\epsilon_{\mathrm{EW}}$),
they are given by,
\begin{align}
& \frac{d}{d\slashed{p}}\Sigma_-^\mathrm{\gamma(EW)}(m_{{\chi}})= - 
    \frac{Q_\chi^2\alpha}{4\pi}
    \left(\frac{1}{\bar{\epsilon}_{\mathrm{EW}}} + \log \frac{\mu_{\mathrm{EW}}^2}{m_{{\chi}}^2} -2 \log \frac{m_{{\chi}}^2}{m_\gamma^2} + 4\right) \ , \label{eq:sigmaEWgamma wino}\\
   & \frac{d}{d\slashed{p}}\Sigma_-^{Z\mathrm{(EW)}}(m_{{\chi}})= - 
    \frac{Q_\chi^2c_W^2\alpha}{4\pi s_W^2}
    \left(\frac{1}{\bar{\epsilon}_{\mathrm{EW}}} + \log \frac{\mu_{\mathrm{EW}}^2}{m_{{\chi}}^2} -2 \log \frac{m_{{\chi}}^2}{M_Z^2} + 4\right) \ , \label{eq:sigmaEWZ wino} \\
  & \frac{d}{d\slashed{p}}\Sigma_-^{W\mathrm{(EW)}}(m_{{\chi}})= - 
    \frac{\alpha}{4\pi s_W^2}
    \left(\frac{1}{\bar{\epsilon}_{\mathrm{EW}}} + \log \frac{\mu_{\mathrm{EW}}^2}{m_{{\chi}}^2} -2 \log \frac{m_{{\chi}}^2}{M_W^2} + 4\right)\ ,
    \\
    & \frac{d}{d\slashed{p}}\Sigma_0^{W\mathrm{(EW)}}(m_{{\chi}})= - 
    \frac{2\alpha}{4\pi s_W^2}
    \left(\frac{1}{\bar{\epsilon}_{\mathrm{EW}}} + \log \frac{\mu_{\mathrm{EW}}^2}{m_{{\chi}}^2} -2 \log \frac{m_{{\chi}}^2}{M_W^2} + 4\right)\ , \\
        & \frac{d}{d\slashed{p}}\Sigma_{u_L,d_L}^\mathrm{\gamma(EW)}(0) = -
     \frac{Q_{\bar{u},d}^2\alpha}{4\pi}
     \left(
     \frac{1}{\bar{\epsilon}_{\mathrm{EW}}} + \log\frac{\mu_{\mathrm{EW}}^2}{m_\gamma^2} - \frac{1}{2}
     \right)\ ,
     \\
    & \frac{d}{d\slashed{p}}\Sigma_{u_L,d_L}^{Z\mathrm{(EW)}}(0)= - 
    \frac{(g_L^{u,d})^2\alpha}{4\pi c_W^2s_W^2}
    \left(\frac{1}{\bar{\epsilon}_{\mathrm{EW}}} + \log \frac{\mu_{\mathrm{EW}}^2}{M_Z^2}-\frac{1}{2}\right)\ ,
     \\
    & \frac{d}{d\slashed{p}}\Sigma_{u_L,d_L}^{W\mathrm{(EW)}}(0)= - 
    \frac{\alpha}{8\pi s_W^2}
    \left(\frac{1}{\bar{\epsilon}_{\mathrm{EW}}} + \log \frac{\mu_{\mathrm{EW}}^2}{M_W^2}-\frac{1}{2}\right)\ ,
\end{align}
where $\alpha=e^2/4\pi$, $\mu_{\mathrm{EW}}$ an arbitrary 
mass parameter, $\bar{\epsilon}_{\mathrm{EW}}^{-1}=\epsilon_{\mathrm{EW}}^{-1}-\gamma_E + \log4\pi$ with $\gamma_E$ being the Euler-Mascheroni constant, 
and $m_\gamma$ a photon mass to regulate the infrared (IR) singularity.
Hereafter, to distinguish the divergences 
in the electroweak theory, the Four-Fermi theory,
and the ChPT, we use the regularization parameters $\bar\epsilon_{\mathrm{EW}}$,
$\bar\epsilon_{\mathrm{FF}}$ and
$\bar\epsilon_{\mathrm{ChPT}}$.
We also distinguish the arbitrary 
mass parameters associated with 
the dimensional regularization in each theory by $\mu_{\mathrm{EW}}$,
$\mu_{\mathrm{FF}}$ and
$\mu_{\mathrm{ChPT}}$
for each theory.
The subscripts, $-,0,u_L,d_L$, denote the charged Wino, the neutral Wino, the left-handed up-quark, the left-handed down-quark, respectively. 
The quark couplings $g_L$ to the $Z$-boson are given by
\begin{align}
    g_L^u = \frac{1}{2}+ s_W^2 Q_{\bar{u}}\ ,
   \quad g_L^d = -\frac{1}{2}- s_W^2 Q_{d}
     \ .
\end{align}
In this work, we perform loop calculation by using the program 
\texttt{Package-X\,v2.1.1}~\cite{Patel:2016fam}.

To keep the SU(2)$_L$ symmetry, we take 
the wave function renormalization factors 
for each SU(2)$_L$ multiplet, i.e., $Z_\chi \equiv 1 + \delta_\chi$ 
for the Wino triplet, and 
  $Z_L \equiv 1 + \delta_L$ for the quark doublet.
Following Ref.\,\cite{Hollik:1988ii},
we take the on-shell renormalization scheme where
\begin{align}
\label{eq:OnShellWF}
    \delta_{\chi} &= \frac{d}{d\slashed{p}}\Sigma_{-}^{(\mathrm{EW})}(m_{{\chi}})\ , \\
    \delta_{L} & = \frac{d}{d\slashed{p}}
\Sigma_{d_L}^{(\mathrm{EW})}(0) \ .
\end{align} 
With these choices,
the residues of 
mass poles of the propagators of
the charged Wino and the down-type quark fields become unity, while those of the neutral Wino and the up-type quark are not unity.
As a result, the two-point functions appear in combination with the wave function renormalization factor as,
\begin{align}
\label{eq:WFEW}
\mathcal{M}^{\mathrm{WF(EW)}} =& \mathcal{M}_{\mathrm{tree}}^{\mathrm{quark}} \times\Bigg[
\frac{1}{2}\frac{d}{d\slashed{p}}\Sigma_{-}^{(\mathrm{EW})}
+\frac{1}{2}\frac{d}{d\slashed{p}}
\Sigma_{0}^{(\mathrm{EW})}
+\frac{1}{2}\frac{d}{d\slashed{p}}\Sigma_{u_L}^{(\mathrm{EW})}
+\frac{1}{2}\frac{d}{d\slashed{p}}
\Sigma_{d_L}^{(\mathrm{EW})} - (\delta_{\chi}+\delta_{L})\Bigg] \\
=&\mathcal{M}_{\mathrm{tree}}^{\mathrm{quark}} \times\Bigg\{
\frac{\alpha}{4\pi }
\left[
 \left(1+ \frac{\bar{Q}}{2}\right)\log\frac{m_\gamma^2}{M_Z^2}
-\frac{1}{s_W^2}\log c_W^2  
\right]\Bigg\}\ ,
\label{eq:WAVE}
\end{align}
where we have used $Q_{d} + Q_{\bar{u}} = -1$ while leaving $\bar{Q}=Q_{d}-Q_{\bar{u}}$ as a free parameter.
Note that the final result does not depend on the choice of $\delta_\chi$ and $\delta_L$ (see Eqs.\,\eqref{eq:EWcounter} and \eqref{eq:MVertexEW}).

\subsubsection{Box Contributions in Electroweak Theory}
\begin{figure}[t]
    \centering
    \includegraphics[]{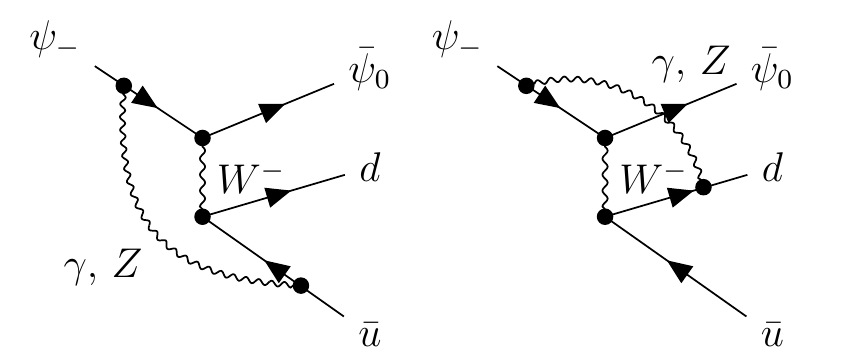}
    \caption{The box contributions to the Wino decay 
    into the free quarks in the electroweak theory.}
    \label{fig:EW_box}
\end{figure}
The box diagrams in Fig.\,\ref{fig:EW_box}
also contribute to the charged Wino decay.%
\footnote{In this paper, the Feynman diagrams are drawn by using \texttt{TikZ-FeynHand}\,
\cite{Ellis:2016jkw,Dohse:2018vqo}.}
The two photon contributions are summarized as,
\begin{align}
\label{eq:MaPhotonEW}
    \mathcal{M}_a^{\gamma(\mathrm{EW})}  
&= \mathcal{M}_{\mathrm{tree}}^{\mathrm{quark}} \times \frac{Q_\chi Q_{\bar{u}}\alpha}{4\pi}
    \left(
    \log\frac{M_W^2}{m_\gamma^2}+
   \frac{\pi M_W}{m_{{\chi}}}\right) -\mathcal{M}_{A} \times \frac{Q_\chi Q_{\bar{u}}\alpha}{6}\frac{M_W}{m_{{\chi}}}+ \order{M_W^2/m_\chi^2}
\ , \\
  \mathcal{M}_b^{\gamma(\mathrm{EW})} 
  \label{eq:MbPhotonEW}
 &= \mathcal{M}_{\mathrm{tree}}^{\mathrm{quark}} \times \frac{Q_\chi Q_{d}\alpha}{4\pi}
    \left(
    \log\frac{M_W^2}{m_\gamma^2}
    +
   \frac{\pi M_W}{m_{{\chi}}}\right) +\mathcal{M}_{A} \times \frac{Q_\chi Q_{d}\alpha}{6}
    \frac{M_W}{m_\chi} + \order{M_W^2/m_\chi^2}\ .
\end{align}
Here, the amplitude $\mathcal{M}_A$ is given by,
\begin{align}
    &\mathcal{M}_A = -4G_\pi^0\big[\bar{u}_d(r_1)\gamma^\mu P_L v_u(r_2)\big] \big[\bar{u}_{0}(q)\gamma^\mu \gamma_5 u_{-}(p)\big] \ ,
\end{align}
which is not proportional to the tree-level amplitude, $\mathcal{M}_{\mathrm{tree}}^{\mathrm{quark}}$.
When we perform
loop momentum integration,
we neglected 
the external quark momenta $r_{1,2}$.
We have numerically checked that the small quark momenta  scarcely affect the matching conditions.
We have also used the Gordon equations in the limit of $q = p$ and $m_0 = m_{{\chi}}$ (see Appendix\,\ref{sec:GordonEquations}),
\begin{align}
    p_\mu\bar{u}_{0}(q)P_{L,R} u_{-}(p)|_{p=q,m_0=m_\chi}  
    =\frac{1}{2}m_\chi
    \bar{u}_{0}(q)\gamma_\mu u_{-}(p)|_{p=q,m_0=m_\chi} \ .
\end{align}

Similarly, 
the $Z$-boson contributions become 
\begin{align}
  \label{eq:MaZEW}
    \mathcal{M}_a^{Z(\mathrm{EW})}  =& \mathcal{M}_{\mathrm{tree}}^{\mathrm{quark}} \times \left[\frac{Q_\chi(1+2Q_{\bar{u}} s_W^2)c_W^2\alpha}{4\pi s_W^4}
    \left(\log c_W - \frac{\pi(1-c_W)M_W}{2c_W m_{{\chi}}}
    \right)\right]\cr
&+\mathcal{M}_{A} \times \frac{Q_\chi(1+2Q_{\bar{u}}s_W^2)\alpha}{12(1+c_W)s_W^2}\frac{c_WM_W}{m_{{\chi}}}+ \order{M_W^2/m_\chi^2}\ , \\
   \label{eq:MbZEW}
  \mathcal{M}_b^{Z(\mathrm{EW})}  =& \mathcal{M}_{\mathrm{tree}}^{\mathrm{quark}} \times \left[\frac{Q_\chi(1+2Q_{d}s_W^2)c_W^2\alpha}{4\pi s_W^4}
    \left(
    \log c_W- \frac{\pi(1-c_W)M_W}{2c_W m_{{\chi}}}\right)
    \right]\cr
&-\mathcal{M}_{A} \times \frac{Q_\chi(1+2Q_{d}s_W^2)\alpha}{12(1+c_W)s_W^2}\frac{c_WM_W}{m_{{\chi}}}
+\order{M_W^2/m_\chi^2}\ .
\end{align}
Here, we have again taken the limit of $r_{1,2}\to 0$ in the loop integration and used the Gordon equations.

By combining the photon and the $Z$-boson box contributions we obtain,
\begin{align}
\label{eq:BOX}
    \frac{\mathcal{M}^{\mathrm{Box(EW)}}}{\mathcal{M}_{\mathrm{tree}}^{\mathrm{quark}}}= \frac{\alpha}{4\pi} \left(
    \log\frac{M_W^2}{m_\gamma^2} -\frac{c_W^4}{s_W^4} \log c_W^2
+ \frac{\pi (1+c_W-c_W^2)}{(1+c_W)s_W^2}
    \frac{M_W}{m_{{\chi}}}\right)
    + \order{M_W^2/m_\chi^2}\ .
\end{align}
Note that this contribution does not depend on $\bar{Q}$.
As we will discuss in Sec.\,\ref{sec:axialWino},
the contributions proportional 
to $\mathcal{M}_A$ are negligibly small, and hence, we do not include them to $\mathcal{M}^{\mathrm{Box(EW)}}$.

\subsubsection{Vertex Corrections}
\begin{figure}[t]
    \centering
    \includegraphics[]{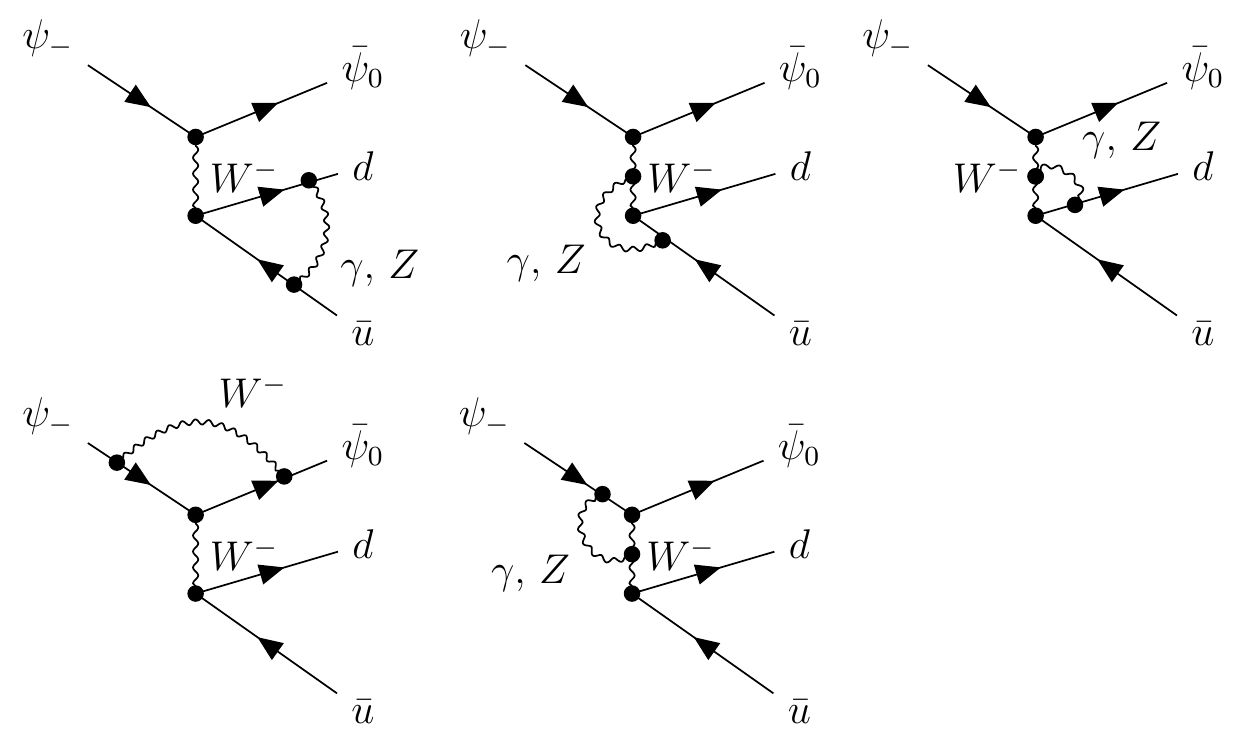}
    \caption{The contributions of the vertex corrections 
    to the Wino decay 
    into the free quarks in the electroweak theory.}
    \label{fig:EW_vertex}
\end{figure}

The vertex corrections are shown in Fig.\,\ref{fig:EW_vertex}.
The $ud$ vertex corrections are given by, 
\begin{align}
    \mathcal{M}^{\gamma}_{ud} &= \mathcal{M}_{\mathrm{tree}}^{\mathrm{quark}}\times \left[-\frac{Q_{d}Q_{\bar{u}}\alpha}{4\pi}
    \left(\frac{1}{\bar{\epsilon}_{\mathrm{EW}}} + \log\frac{\mu_{\mathrm{EW}}^2}{m_\gamma^2}-\frac{1}{2}\right)
    \right]\ , 
    \\
      \mathcal{M}^{Z}_{ud} &=  \mathcal{M}_{\mathrm{tree}}^{\mathrm{quark}}\times \left[\frac{g_L^d g_L^u\alpha}{4\pi c_W^2 s_W^2}
    \left(\frac{1}{\bar{\epsilon}_{\mathrm{EW}}} + \log\frac{\mu_{\mathrm{EW}}^2}{M_Z^2}-\frac{1}{2}\right)
    \right]\ ,
      \\
      \mathcal{M}^{W\gamma}_{ud} &=  \mathcal{M}_{\mathrm{tree}}^{\mathrm{quark}}\times \left[-\frac{(Q_{d}+Q_{\bar{u}})\alpha}{4\pi }
    \left(\frac{3}{\bar{\epsilon}_{\mathrm{EW}}} +3 \log\frac{\mu_{\mathrm{EW}}^2}{M_W^2}+\frac{5}{2}\right)
    \right]\ ,
     \\
      \mathcal{M}^{WZ}_{ud} &=  \mathcal{M}_{\mathrm{tree}}^{\mathrm{quark}}\times \left[-\frac{(g_L^d-g_L^u)\alpha}{4\pi  s_W^2}
    \left(\frac{3}{\bar{\epsilon}_{\mathrm{EW}}} +3 \log\frac{\mu_{\mathrm{EW}}^2}{M_W^2}+\frac{5}{2}
    + \frac{3}{s_W^2}\log c_W^2\right)\right]\ ,
\end{align}
where  $g_L^d-g_L^u = -1 -s_W^2 (Q_{d}+Q_{\bar{u}})$.
Similarly, the Wino vertex corrections are given by,
\begin{align}
     \mathcal{M}^{W}_{\chi} &=  \mathcal{M}_{\mathrm{tree}}^{\mathrm{quark}}\times
     \left[\frac{\alpha}{4\pi s_W^2}
     \left(\frac{1}{\bar{\epsilon}_{\mathrm{EW}}}+ \log\frac{\mu_{\mathrm{EW}}^2}{m_{{\chi}}^2} 
     + 2\log\frac{M_W^2}{m_{{\chi}}^2}
     +4- \frac{3\pi M_W}{m_{{\chi}}}\right)\right]+\order{M_W^2/m_\chi^2}\ ,\\   \mathcal{M}^{W\gamma}_{\chi} &=  \mathcal{M}_{\mathrm{tree}}^{\mathrm{quark}}\times
     \left[\frac{\alpha}{4\pi }
     \left(\frac{3}{\bar{\epsilon}_{\mathrm{EW}}}+ 3\log\frac{\mu_{\mathrm{EW}}^2}{m_{{\chi}}^2} + 4-\frac{2\pi M_W}{m_{{\chi}}}\right)\right]+\order{M_W^2/m_\chi^2}\ ,\\   \mathcal{M}^{WZ}_{\chi} &=  \mathcal{M}_{\mathrm{tree}}^{\mathrm{quark}}\times
     \left[\frac{c_W^2\alpha}{4\pi s_W^2}
     \left(\frac{3}{\bar{\epsilon}_{\mathrm{EW}}}+ 3\log\frac{\mu_{\mathrm{EW}}^2}{m_{{\chi}}^2} + 4
     - \frac{2\pi (1+c_W+c_W^2)M_W}{c_W(1+c_W) m_{{\chi}}} \right)\right]+\order{M_W^2/m_\chi^2}\ .
\end{align}
Here, we have taken the limit of $r_{1,2}\to 0$ in the loop momentum integration and used the Gordon equations.
Note that the vertex corrections are proportional to $\mathcal{M}_{\mathrm{tree}}^{\mathrm{quark}}$ and do not contain the contributions such as $\mathcal{M}_A$.
As in the case of the two-point functions, we have omitted the pure QCD corrections.

The contributions from the vertex counterterms
are given by,
\begin{align}
\label{eq:EWcounter}
    \mathcal{M}_{\mathrm{CT}}^{\mathrm{Vertex}(\mathrm{EW})} = \mathcal{M}_{\mathrm{tree}}^{\mathrm{quark}}\times\left[2(\delta Z_1^W-\delta Z_2^W)+\delta_{\chi} + \delta_{L}\right]\ ,
\end{align}
where $\delta Z_{1}^{W}$
is the renormalization constants of the weak gauge coupling and
$\delta Z_{2}^{W}$ the weak gauge boson wave function factor
(see Eq.\,(4.10) in Ref.\,\cite{Hollik:1988ii}).%
\footnote{In the on-shell scheme of the electroweak theory in Ref.\,\cite{Bohm:1986rj}, 
the renormalized SU(2)$_L$ gauge coupling constant, $g$, is defined by $g_B = Z_{1}^W (Z_2^{W})^{-3/2}g$
where $g_B$ is the bare  SU(2)$_L$ gauge coupling constant.
Accordingly, each fermion--$W$-boson coupling is associated with $\delta Z_{1}^W-\delta Z_2^W+\delta_F$, where $\delta_F$ is the wave function renormalization factor of the fermion.}
Note that $\delta_\chi + \delta_L$ contributions in Eq.\,\eqref{eq:EWcounter} cancels those in Eq.\,\eqref{eq:WFEW}, and hence,
the result does not depend on the choices of the wave function renormalization factors.
In the on-shell renormalization scheme of the electroweak theory in Ref.\,\cite{Bohm:1986rj},
the EW counterterms are set to satisfy
\begin{align}
    \delta Z_1^W-\delta Z_2^W= - \frac{\alpha}{2\pi s_W^2}
    \left(\frac{1}{\bar{\epsilon}_{\mathrm{EW}}}+\log \frac{\mu_{\mathrm{EW}}^2}{M_{W}^2}\right)\ ,
\end{align}
(see e.g., Eq.\,(5.42) of Ref.\,\cite{Bohm:1986rj}).
The resultant vertex correction is given by,
\begin{align}
\label{eq:MVertexEW}
    \mathcal{M}^{\mathrm{Vertex(EW)}}=
    \mathcal{M}_{ud}^{\gamma}
    +\mathcal{M}_{ud}^{Z}
    +\mathcal{M}_{ud}^{W\gamma}
    +\mathcal{M}_{ud}^{WZ}
    +\mathcal{M}_\chi^{W}
    +\mathcal{M}_\chi^{W\gamma}
    +\mathcal{M}_\chi^{WZ}+\mathcal{M}_{\mathrm{CT}}^{\mathrm{Vertex}\mathrm{(EW)}}\ .
\end{align}

\subsubsection{Total Contributions}
Altogether, 
the above contributions result in
\begin{align}
\label{eq:virtual EW}
     \frac{\mathcal{M}^{\mathrm{WF,Box,Vertex(EW)}}}{\mathcal{M}_{\mathrm{tree}}^{\mathrm{quark}}}
     =& \frac{\alpha}{4\pi}
     \left[
     \frac{3}{2}\log\frac{M_Z^2}{m_\gamma^2}
     - \left(
     \frac{1}{s_W^2} 
     -\frac{4}{s_W^4}
     \right)\log c_W
 + \frac{3}{s_W^2}  \right]\cr
 &-\frac{\alpha(4 + 6c_W+ c_W^2)}{4(1+c_W)s_W^2}\frac{M_W}{m_{{\chi}}} + \order{M_W^2/m_\chi^2}
    \ .
\end{align}
Finally, we add the vacuum polarization of the virtual $W$-boson leads to the correction,
\begin{align}  \frac{\mathcal{M}^{\mathrm{VP}}}{\mathcal{M}_{\mathrm{tree}}^{\mathrm{quark}}}=
  \left(1-\displaystyle{\frac{\hat{\Sigma}_{WW}^\mathrm{1PI}(0)}{M_W^2}}\right)^{-1} \simeq 1+\displaystyle{\frac{\hat{\Sigma}_{WW}^\mathrm{1PI}(0)}{M_W^2}}\ ,
\end{align}
where $\hat{\Sigma}_{WW}^\mathrm{1PI}(p^2=0)$ is the renormalized self-energy of the $W$-boson.
Note that renormalized $\hat{\Sigma}_{WW}^\mathrm{1PI}(0)$ includes the Wino contribution. 
Since this contribution is common to all the charged weak interactions, and hence, it does not affect the ratio of the Wino and the pion decay rates.
As a result, we obtain
\begin{align}
\label{eq:virtual total}
\frac{\mathcal{M}^{\mathrm{Virtual(EW)}}}{\mathcal{M}_{\mathrm{tree}}^{\mathrm{quark}}}=
     \frac{\mathcal{M}^{\mathrm{WF,Box,Vertex(EW)}}+\mathcal{M}^{\mathrm{VP}}}{\mathcal{M}_{\mathrm{tree}}^{\mathrm{quark}}}  \ .
\end{align}

So far, we have used the tree-level 
weak interaction with $G_\pi^0 = V_{ud}G_F^0$, where
\begin{align}
G_F^0 = \frac{g^2}{4\sqrt{2}M_W^2}=  \frac{e^2}{4s_W^2\sqrt{2}M_W^2}\ .
\end{align}
As we will discuss, we take the ratio between the charged Wino and the charged pion decay rates where the latter has been calculated by D\&M\,\cite{Descotes-Genon:2005wrq}.
In their analysis, 
the tree-level decay rate of the charged 
pion is not defined by
using $G_\pi^0 = V_{ud}G_F^0$ (see Eq.\,\eqref{eq:GammaPI}) but by ${G}_\pi= V_{ud}{G}_F$ where
${G}_F$ is the Fermi constant determined from the muon lifetime.
In the limit of the vanishing electron mass, $G_F^0$ and $G_F$ are 
related via
\begin{align}
\label{eq:GFmuon}
    G_F^0 = \frac{e^2}{4s_W^2\sqrt{2}M_W^2} ={G}_F
    \times\left[1-\displaystyle{\frac{\hat{\Sigma}_{WW}^\mathrm{1PI}(0)}{M_W^2}}
    -
    \frac{\alpha}{4\pi s_W^2}
    \left(6+\frac{7-4 s_W^2}{2s_W^2} \log c_W^2\right)
    \right]\ ,
\end{align}
at one-loop level
(see e.g. Eq.\,(4.18) of Ref.\,\cite{Hollik:1988ii}).
Accordingly, the ratio to the tree-level 
contribution is slightly shifted by,
\begin{align}
\label{eq:virtual EW final}
     \frac{\hat{\mathcal{M}}^{\mathrm{Virtual(EW)}}}{\hat{\mathcal{M}}_{\mathrm{tree}}^{\mathrm{quark}}}
     =& \frac{\alpha}{4\pi}
     \left[
     \frac{3}{2}\log\frac{M_Z^2}{m_\gamma^2}
    + \left(
     \frac{3}{s_W^2} 
     - \frac{3}{s_W^4}
     \right)\log c_W
     - \frac{3}{s_W^2}
     \right]\cr
 &-\frac{\alpha(4 + 6c_W + c_W^2)}{4(1+c_W)s_W^2}\frac{M_W}{m_{{\chi}}} + \order{M_W^2/m_\chi^2}\ .
\end{align}
Here, $\hat{\mathcal{M}}_{\mathrm{tree}}^{\mathrm{quark}}$ 
denotes to the tree-level amplitude 
in Eq.\,\eqref{eq:quarktree} with $G_\pi^0$
replaced by ${G}_\pi$ and 
\begin{align}
    \hat{\mathcal{M}}^{\mathrm{Virtual(EW)}} ={\mathcal{M}}^{\mathrm{Virtual(EW)}} +{\mathcal{M}}_{\mathrm{tree}}^{\mathrm{quark}}- \hat{\mathcal{M}}_{\mathrm{tree}}^{\mathrm{quark}}\ .
\end{align}
The one-loop correction to the Wino-decay amplitude into the free quarks is UV finite as expected.

Note that the virtual correction does not depend on the arbitrary mass scale $\mu_\mathrm{EW}$, since 
we adopt the on-shell renormalization scheme.
Besides, it depends on the Wino mass
only through its negative power.
This shows that the 
Appelquist-Carazzone theorem holds for the Wino decay into free quarks 
at the one-loop level.

The virtual correction has an 
IR singularity in the limit of $m_\gamma \to 0$.
Since the IR singularity 
of the electroweak theory 
and the Four-Fermi theory 
are common, 
it does not affect the matching procedure.
In fact, we will find that the matching conditions of the counterterms are free from the IR singularity, which provides non-trivial consistency check of our analysis.

\subsubsection{Effects from Wino Axial Current Interactions}
\label{sec:axialWino}
As we have seen in Eqs.\,\eqref{eq:MaPhotonEW}--\eqref{eq:MbZEW}, the box contributions in the electroweak theory lead to the axial current Wino weak interaction, $\mathcal{M}_A$.
In terms of the effective Lagrangian, they amount to,
\begin{align}
\label{eq:LA}
    \mathcal{L}_{A} = \kappa_A G_F^0 
    J_\mu^{\mathrm{quark}}(\bar{\psi}_-\gamma^\mu\gamma_5\psi_0)+h.c. \ ,
\end{align}
where $\kappa_A$ is a dimensionless coefficient suppressed by one-loop factor,
\begin{align}
    \kappa_A =\frac{2Q_\chi(Q_{d}-Q_{\bar{u}})\alpha}{3(1+c_W)}\frac{c_WM_W}{m_{{\chi}}}
    \ .
\end{align}
This operator contributes to the charged Wino decay 
into the pion through Eq.\,\eqref{eq:JmutoPi}.
The axial Wino current contribution is, however, 
the $p$-wave decay, while the tree-level contribution 
in Eq.\,\eqref{eq:iMtree} is the $s$-wave decay. 
Thus, the interference between those contributions 
to the total decay rate vanishes.
As a result, the decay rate through this operator, $\delta\Gamma_A$, is highly suppressed as
\begin{align}
    \frac{\delta\Gamma_A}{\Gamma_\chi} =
    \frac{\kappa_A^2}{8}
    \left(1-\frac{m_\pi^2}{\mathit{\Delta}m^2}\right)\ ,
\end{align}
and hence, it is $\order{\alpha^2M_W^2/m_\chi^2}$ correction.
In the following, we neglect the 
Wino axial current   interactions.

\subsection{Wino Decay into Free Quarks in Four-Fermi Theory}
Let us repeat the same analysis in the Four-Fermi theory.
In the Four-Fermi theory, 
there are no radiative corrections to the neutral Wino.
As stated before,
we take the tree-level mass term of the charged Wino $m_\chi$ to be the physical charged Wino mass.
The neutral Wino mass is also taken to be $m_0=m_\chi - \mathit{\Delta}m$ so that it reproduces the prediction in the electroweak theory.
We also redefine the tree-level Four-Fermi interactions by using ${G}_F$ in Eq.\,\eqref{eq:GFmuon} 
in the following analysis.

\subsubsection{Two-Point Functions}
In the effective Four-Fermi interaction, the relevant two-point functions come from the QED contributions.
In the following, we omit the QCD contributions as in the case of the electroweak theory.
The photon contributions to the two-point functions
are,
\begin{align}
\label{eq:wionotwopoint}
     &\frac{d}{d\slashed{p}}\Sigma_-^\mathrm{\gamma(FF)}(m_{{\chi}})
     = - 
    \frac{Q_\chi^2\alpha}{4\pi}
    \left(\frac{1}{\bar{\epsilon}_{\mathrm{FF}}} + \log \frac{\mu_{\mathrm{FF}}^2}{m_{{\chi}}^2} +2 \log \frac{m_\gamma^2}{m_{{\chi}}^2} + 4\right) \ , \\
     &\frac{d}{d\slashed{p}}\Sigma_0^\mathrm{\gamma(FF)}(m_{{\chi}})= 0\ ,
    \\
\label{eq:quarktwopoint}
     &\frac{d}{d\slashed{p}}\Sigma_{u,d}^\mathrm{\gamma(FF)}(0)
     =-
     \frac{Q_{\bar{u},d}^2\alpha}{4\pi}
     \left(
     \frac{1}{\bar{\epsilon}_{\mathrm{FF}}} + \log\frac{\mu_{\mathrm{FF}}^2}{m_\gamma^2} - \frac{1}{2}
     \right)\ .
\end{align}
These corrections contribute to the decay rate through the amplitude
\begin{align}
\mathcal{M}^{\mathrm{WF(FF)}}&=\hat{\mathcal{M}}_{\mathrm{tree}}^{\mathrm{quark}}\times \left(
\frac{1}{2}\frac{d}{d\slashed{p}}\Sigma_-^\mathrm{\gamma(FF)}(m_{{\chi}})
+\frac{1}{2}\frac{d}{d\slashed{p}}\Sigma_u^\mathrm{\gamma(FF)}(0)
+ \frac{1}{2}\frac{d}{d\slashed{p}}\Sigma_d^\mathrm{\gamma(ff)}(0)
\right) \\
&= \hat{\mathcal{M}}_{\mathrm{tree}}^{\mathrm{quark}}\times
\frac{\alpha}{8\pi}
\left[
- 
Q_\chi^2
    \left(\frac{1}{\bar{\epsilon}_{\mathrm{FF}}} + \log \frac{\mu_{\mathrm{FF}}^2}{m_{{\chi}}^2} +2 \log \frac{m_\gamma^2}{m_{{\chi}}^2} + 4\right) 
-(Q_{d}^2 + Q_{\bar{u}}^2)
     \left(
     \frac{1}{\bar{\epsilon}_{\mathrm{FF}}} + \log\frac{\mu_{\mathrm{FF}}^2}{m_\gamma^2} - \frac{1}{2}\right)
\right]\ .
\end{align}
The counterterm contributions to the same process are given by,
\begin{align}
    \mathcal{M}_{\mathrm{CT}}^{\mathrm{(FF)}} = 
    \hat{\mathcal{M}}_{\mathrm{tree}}^{\mathrm{quark}} \times e^2\left[
    f_{\chi\chi}Q_\chi^2 + f_{d{\bar{u}}}(Q_d+Q_{\bar{u}})^2 + f_{\chi d}Q_\chi Q_d - f_{\chi \bar{u}}Q_\chi Q_{\bar{u}}\right]\ .
\end{align}

\subsubsection{Vertex Corrections}
\begin{figure}[t]
    \centering
    \includegraphics[]{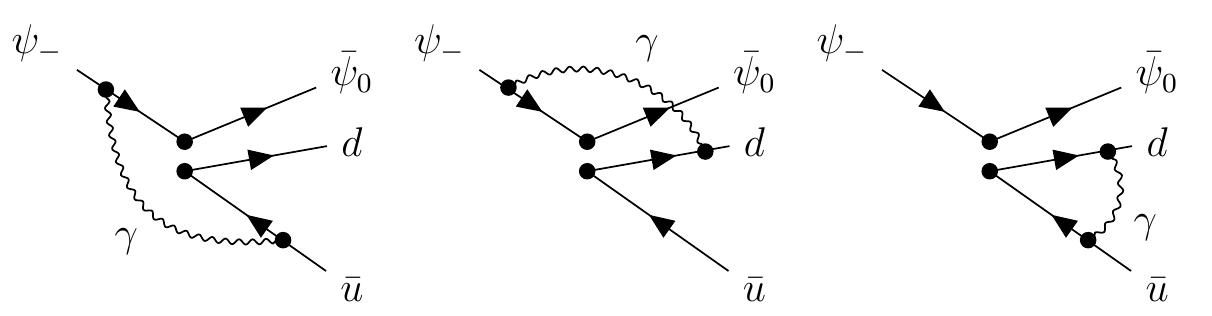}
    \caption{The virtual corrections to the Wino decay 
    into the free quarks in the Four-Fermi theory.}
    \label{fig:FF_virtual}
\end{figure}

The vertex corrections in the Four-Fermi theory are shown in Fig.\,\ref{fig:FF_virtual}.
Those diagrams contribute to the charged Wino decay as,
\begin{align}
\label{eq:MaFF}
 \mathcal{M}_a^{\mathrm{\gamma(FF)}}  =& 
    \hat{\mathcal{M}}_{\mathrm{tree}}^{\mathrm{quark}} \times
    \frac{ Q_\chi Q_{\bar{u}}\alpha}{8\pi}\left(
    \frac{5}{\bar{\epsilon}_{\mathrm{FF}} }
+5\log\frac{\mu_{\mathrm{FF}}^2}{m_{{\chi}}^2}+2\log\frac{m_{{\chi}}^2}{m_\gamma^2}
    +7
    \right) \cr
    -&\mathcal{M}_A \times
    \frac{3 Q_\chi Q_{\bar{u}}\alpha}{8\pi}\left(
    \frac{1}{\bar{\epsilon}_{\mathrm{FF}} }
    +\log\frac{\mu_{\mathrm{FF}}^2}{m_{{\chi}}^2}
    +\frac{5}{3}
    \right)\ ,
    \\
    \label{eq:MbFF}
    \mathcal{M}_b^{\mathrm{\gamma(FF)}}  =& 
    \hat{\mathcal{M}}_{\mathrm{tree}}^{\mathrm{quark}} \times
    \frac{ Q_\chi Q_{d}\alpha}{8\pi}\left(
    \frac{5}{\bar{\epsilon}_{\mathrm{FF}} }
    +5\log\frac{\mu_{\mathrm{FF}}^2}{m_{{\chi}}^2}
    +2\log\frac{m_{{\chi}}^2}{m_\gamma^2}
    +7
    \right) \cr
    +&\mathcal{M}_A \times
    \frac{ 3Q_\chi Q_{d}\alpha}{8\pi}\left(
    \frac{1}{\bar{\epsilon}_{\mathrm{FF}} }
    +\log\frac{\mu_{\mathrm{FF}}^2}{m_{{\chi}}^2}
    +\frac{5}{3}
    \right)\ , \\
\mathcal{M}_c^{\mathrm{\gamma(FF)}}  =& \hat{\mathcal{M}}_{\mathrm{tree}}^{\mathrm{quark}} \times
    \frac{ - Q_{d} Q_{\bar{u}}\alpha}{4\pi}\left(
    \frac{1}{\bar{\epsilon}_{\mathrm{FF}} }
    +\log\frac{\mu_{\mathrm{FF}}^2}{m_\gamma^2}
    -\frac{1}{2}
    \right)\ .
\end{align}
Here, we have again taken the limit of the quark momenta $r_{1,2}\to 0$ in the loop momentum integration and used the Gordon equations. As in the case of the electroweak theory, we have omitted the QCD loop diagrams.

As in the case of the electroweak theory, the radiative corrections lead to the axial Wino current interaction, $\mathcal{M}_A$.
The UV divergences of the $\mathcal{M}_A$ contributions
must be canceled by additional counterterms proportional to Eq.\,\eqref{eq:LA}.
The finite part of the counterterms can be determined by matching $\mathcal{M}_A$ contributions 
in the Four-Fermi theory and the electroweak theory.
As discussed in Sec.\,\ref{sec:axialWino}, 
however,
the Wino decay via the axial Wino current interactions
are negligibly small in the
the electroweak theory.
Thus, we can safely neglect the axial Wino current contributions.

The total virtual correction is given by,
\begin{align}
\hat{\mathcal{M}}^{\mathrm{Virtual(FF)}}&=\mathcal{M}^{\mathrm{WF(FF)}}
 +\mathcal{M}_{\mathrm{CT}}^{\mathrm{(FF)}} 
 +\mathcal{M}_a^{\mathrm{\gamma(FF)}} 
 +\mathcal{M}_b^{\mathrm{\gamma(FF)}} 
 +\mathcal{M}_c^{\mathrm{\gamma(FF)}}\ ,  
\end{align}
and we obtain,
\begin{align}
\label{eq:viretural FF}
\frac{\hat{\mathcal{M}}^{\mathrm{Virtual(FF)}}}{\hat{\mathcal{M}}_{\mathrm{tree}}^{\mathrm{quark}}}
&=
\frac{3\alpha}{8\pi}
\left(
\frac{1}{\bar{\epsilon}_{\mathrm{FF}}}
+\log\frac{\mu_{\mathrm{FF}}^2}{m_\gamma^2}
+ \frac{7}{6}
\right)+
e^2\left(f_{\chi\chi} + f_{d\bar{u}} + \frac{1-\bar{Q}}{2}f_{\chi d}
- \frac{1+\bar{Q}}{2}f_{\chi \bar{u}}\right)\ .
\end{align}
Following D\&M, we take the UV divergent parts of the counterterms,
\begin{align}
\label{eq:divergence of f}
f_{\chi\chi}&= \frac{1}{32\pi^2} \frac{1}{\bar{\epsilon}_{\mathrm{FF}}} + f_{\chi\chi}^{r}(\mu_{\mathrm{FF}})\ , \\
\frac{1}{2}(f_{\chi d}-f_{\chi \bar{u}})&= -\frac{5}{32\pi^2} \frac{1}{\bar{\epsilon}_{\mathrm{FF}}} + \frac{1}{2}(f^r_{\chi d}(\mu_{\mathrm{FF}})-f^r_{\chi\bar{u}}(\mu_{\mathrm{FF}})) \ , \\
f_{d\bar{u}}&=\frac{1}{32\pi^2} \frac{1}{\bar{\epsilon}_{\mathrm{FF}}}+f_{d\bar{u}}^r(\mu_{\mathrm{FF}})\ ,
\end{align}
with which the UV divergence in the virtual correction is cancelled.
Here, $f^r(\mu_{\mathrm{FF}})$'s are the finite parts of the counterterms in the $\overline{\mathrm{MS}}$,
which will be determined shortly.
The UV divergences
coincide with those of the counterterms for the pion decay given by D\&M\,\cite{Descotes-Genon:2005wrq}.

\subsection{Electroweak and Four-Fermi Matching Conditions}
By comparing Eqs.\,\eqref{eq:virtual EW final} and \eqref{eq:viretural FF}, we obtain two conditions 
\begin{align}
\label{eq:fwd+fwbaru}
    f_{\chi d} + f_{\chi \bar{u}} =&\, 0 \ , \\
    e^2(f_{\chi\chi} + f_{d\bar{u}} + f_{\chi d}/2-f_{\chi \bar{u}}/2) =& \frac{\alpha}{4\pi}
    \left[-\frac{3}{2}\frac{1}{\bar{\epsilon}_{\mathrm{FF}}} -\frac{3}{2} \log\frac{\mu_{\mathrm{FF}}^2}{M_Z^2}
     +\left(
      \frac{3}{s_W^2} 
     - \frac{3}{s_W^4}
     \right)\log c_W
     -\frac{3}{s_W^2}
     -\frac{7}{4}
   \right] \cr
   \label{eq:FFEWmatching}
 &-\frac{\alpha(4 + 6 c_W + c_W^2)}{4(1+c_W)s_W^2}\frac{M_W}{m_{{\chi}}} + \order{M_W^2/m_\chi^2}\ .
\end{align}
Notice that the matching conditions do not depend on $m_{\gamma}$,
which provides a non-trivial consistency check of the matching procedure.
The full $M_W/m_\chi$
dependence is given in  Appendix\,\ref{sec:FullMW}.

\section{Matching Between Four-Fermi Theory and ChPT}
\label{sec:FFtoChPT}
As the second step of the matching procedure, we match the Four-Fermi theory to the ChPT
extended with QED and the Winos.
\subsection{Chiral Lagrangian}
\label{sec:ChiralLagrangian}
Following D\&M analysis, let us first construct the three-flavor ChPT extended with QED and the Winos.
In the ChPT, the pions and Kaons are introduced as the
coordinates 
of the coset space of
the spontaneously broken chiral flavor symmetry, SU(3)$_L\times$SU(3)$_R\to$SU(3)$_V$.
The leading order terms, i.e., $\order{p^2}$ terms are,
\begin{align}
\label{eq:chptleading}
    \mathcal{L}_{p^2}=&\frac{F_0^2}{4}
    \langle u_\mu u^\mu + \chi^{\mathrm{(sp)}}_+\rangle + e^2 F_0^2 Z_\mathrm{ChPT} \langle\mathcal{Q}_L\mathcal{Q}_R\rangle \cr
    &-\frac{1}{4} F_{\mu\nu}F^{\mu\nu}
    + \sum_{\ell}\left[ \bar{\ell}(i\slashed{D}-m_\ell)\ell + \bar{\nu}_{\ell L}i\slashed{\partial}\nu_{\ell L}\right]\cr
    &+ \bar{\psi}_-(i\slashed{D}-m_{{\chi}})\psi_- 
     + \bar{\psi}_0(i\slashed{\partial}-m_{0})\psi_0\ .
\end{align}
The angle bracket $\langle\rangle$ denotes the trace of the flavor index.
The definitions of the building blocks, $u_\mu$, $\mathcal{Q}_L$, $\mathcal{Q}_R$ are given below  (see also Refs.\,\cite{Knecht:1999ag,Descotes-Genon:2005wrq}).
The spurion field
$\chi_+^{\mathrm{(sp)}}$
represents the 
explicit chiral symmetry breaking
due to the quark masses.%
\footnote{The spurion fields $\chi_+^{\mathrm{(sp)}}$ and 
$\chi_-^{\mathrm{(sp)}}$ are denoted by $\chi_+$ and $\chi_-$ in Ref.\,\cite{Knecht:1999ag}, respectively.}
The term proportional to $Z_\mathrm{ChPT} = \order{1/16\pi^2}$ is responsible for the mass difference between the charged and the neutral pions.

The QED and the weak interactions (including the Wino interactions) also break the chiral symmetry explicitly.
To treat those explicit breaking effects in a systematic manner, 
we use the spurion formalism in Refs.\,\cite{Urech:1994hd,Knecht:1999ag}. 
In this formalism, 
we introduce spurion fields $\mathbf{q}_{L,R}$ and $\mathbf{q}_W$, 
which transform under the chiral 
SU(3)$_L\times$SU(3)$_R$ symmetry as 
\begin{align}
    \mathbf{q}_{L}\to 
    \hat{g}_L \mathbf{q}_{L} \hat{g}_L^\dagger
    \ , 
    \,
     \mathbf{q}_{R}\to 
    \hat{g}_R \mathbf{q}_{R} \hat{g}_R^\dagger\ , \,
     \mathbf{q}_{W}\to 
    \hat{g}_L \mathbf{q}_{W} \hat{g}_L^\dagger
    \ ,
\end{align}
where $\hat{g}_{L,R}\in \mbox{SU(3)}_{L,R}$.
The spurions $\mathbf{q}_{L,R}$ are hermitian.
The physical values of the spurions $\mathbf{q}_{L,R}$ and $\mathbf{q}_{W}$ are 
\begin{align}\label{eq:charges}
Q_{\mathrm{QED}}=\left(
\begin{array}{ccc}
\frac{2}{3} &0&0\\
0&-\frac{1}{3}&0\\
0&0&-\frac{1}{3}
\end{array}\right)\ ,
\quad
Q_W= -2\sqrt2  \left(
\begin{array}{ccc}
0&V_{ud}&V_{us}\\
0&0&0\\
0&0&0
\end{array}\right)\ ,
\end{align}
respectively.
Here, $V_{ud}$ and $V_{us}$ are 
the elements of the CKM matrix.

In terms of the spurions, the QED interaction and the weak interaction of the pions are described by the couplings to the left/right vector fields, which are defined by,
\begin{align}
\label{eq:lmu}
    &l_\mu  = e \mathbf{q}_L A_\mu + {G}_F
    \left(\mathbf{q}_W \bar{\ell}_L \gamma_\mu \nu_L 
    +\sqrt{2}\mathbf{q}_W  \bar{\psi}_-\gamma_\mu \psi_0
+\bar{\nu}_L \gamma_\mu \ell_L \mathbf{q}_W^\dagger 
    +\sqrt{2}
    \bar{\psi}_0\gamma_\mu \psi_-
    \mathbf{q}_W^\dagger  
    \right)\ ,\\
    \label{eq:rmu}
    &r_\mu = e \mathbf{q}_R A_\mu\ .
\end{align}
Note that we use ${G}_F$ in Eq.\,\eqref{eq:GFmuon}.
The sign convention of $A_\mu$ terms are opposite to those in
Refs.\,\cite{Knecht:1999ag,Descotes-Genon:2005wrq}.
The coset variables $u_{R,L}$ are 
expressed by the coset coordinates, $\pi^a(a=1\mbox{--}8)$, as
\begin{align}
    &u_{R} = e^{i\pi^at^a}\ , \cr
    &u_{L} = e^{-i\pi^at^a} \ ,
\end{align}
where $t^a = \lambda^a/2$ with $\lambda^a$ being the Gell-Mann matrices.
The coset variables couple to 
the vector fields $l_\mu$ and $r_\mu$ via
\begin{align}
\label{eq:umu}
    u_\mu = i 
    \left[ u_R^\dagger (\partial_\mu - ir_\mu)u_R 
    - u_L^\dagger(\partial_\mu - il_\mu)u_L\right]\ .
\end{align}
The dressed spurion fields are defined by,
\begin{align}
    \mathcal{Q}_L = u_L^\dagger \mathbf{q}_L u_L \ , \quad \mathcal{Q}_R = u_R^\dagger \mathbf{q}_R u_R\ , \quad \mathcal{Q}_W = u_L^\dagger \mathbf{q}_W u_L \ .
\end{align}
We will also use 
\begin{align}
    \mathcal{Q}_L^\mu = u_L^\dagger (D^\mu\mathbf{q}_L) u_L \ , \quad \mathcal{Q}_R^\mu = u_R^\dagger 
    (D^\mu\mathbf{q}_R) u_R\ ,
\end{align}
where
\begin{align}
    D_\mu\mathbf{q}_L = \partial_\mu \mathbf{q}_L 
    - i [\ell_{\mu},\mathbf{q}_L]\ , \quad
    D_\mu\mathbf{q}_R = \partial_\mu \mathbf{q}_R 
    - i [r_{\mu},\mathbf{q}_R]\ .
\end{align}

With the above definitions,
the kinetic term of the pions and
the tree-level QED and weak interactions can be drawn from the first term of Eq.\,\eqref{eq:chptleading},
\begin{align}
\label{eq:TreeLevel}
    \mathcal{L}_{p^2} \supset\,\, &|D_\mu \pi^-|^2  - m_\pi^2 |\pi^-|^2  - 2 F_0 G_F V_{ud}
    D_\mu \pi^- 
    \left(\bar{\ell}_L\gamma^\mu\nu_{\ell L} + \sqrt{2}
     \bar{\psi}_-
\gamma^\mu\psi_0\right) + h.c.\ ,
\end{align}
which reproduces the Wino-pion interaction in Eq.\,\eqref{eq:WinoPion}
if we replace $G_F \to G_F^0$ and $F_\pi \to F_0$, respectively.%
\footnote{By field redefinitions of the Winos, the Wino-pion derivative interaction can be rewritten as Yukawa interaction~\cite{Kinoshita:1959ha} (see Appendix~\ref{sec:Wino-Pion Interaction}).}
The physical decay constant
$F_\pi$
is related to $F_0$ at the one-loop level, via,
\begin{align}
\label{eq:Fpi}
    F_\pi \equiv\,& F_0\left\{
    1+\frac{4}{F_0^2}
    \left[L_4^r(\mu_\mathrm{ChPT})(M_\pi^2 + M_K^2)
    +L_5^r(\mu_\mathrm{ChPT})M_\pi^2
    \right]\right. \cr 
    & \left.
    -\frac{1}{2F_0^2} 
    \left[2M_\pi^2 \log\frac{M_\pi^2}{\mu_\mathrm{ChPT}^2} +2M_K^2 \log\frac{M_K^2}{\mu_\mathrm{ChPT}^2}\right]
    \right\}\ ,
\end{align}
where $\mu_{\mathrm{ChPT}}$ is the $\overline{\mathrm{MS}}$
renormalization scale and  $L_{4,5}^{r}(\mu_\mathrm{ChPT})$ 
and the mass parameters $M^2_{\pi,K}$ 
are given in 
Refs.\,\cite{Gasser:1984gg, Knecht:1999ag}.
The $\mu_\mathrm{ChPT}$ dependence in 
Eq.\,\eqref{eq:Fpi} is 
canceled by those of  $L_{4,5}^{r}(\mu_\mathrm{ChPT})$.

Note that we use $G_F$ and 
$F_0$ for the tree-level parameters in the ChPT as a convention.
This convention is useful
since the ratio between the tree-level Wino decay rate and the tree-level pion decay rate by D\&M coincides with Eq.\,\eqref{eq:treeRatio}.

To discuss the virtual photon corrections in the ChPT, we consider the $\order{e^2 p^2}$ terms
introduced in Refs.\,\cite{Urech:1994hd,Knecht:1999ag},
\begin{align} 
\mathcal{L}_{K} &=
e^2 {F_0}^2 \bigg\{ \frac{1}{2} K_1 \; \langle (\mathcal{Q}_L)^2 +
(\mathcal{Q}_R)^2\rangle
\langle u_\mu  u^\mu\rangle  + K_2 \; \langle \mathcal{Q}_L \mathcal{Q}_R\rangle 
  \langle u_\mu u^\mu
 \rangle \cr
 &\mbox{} - K_3 \; [\langle \mathcal{Q}_L u_\mu\rangle 
\langle \mathcal{Q}_L u^\mu
 \rangle 
 + \langle \mathcal{Q}_R u_\mu\rangle 
 \langle \mathcal{Q}_R u^\mu\rangle ] + K_4 \; \langle \mathcal{Q}_L u_\mu\rangle 
 \; \langle \mathcal{Q}_R u^\mu \rangle
  \cr
& \mbox{} + K_5 \; \langle[(\mathcal{Q}_L)^2 + (\mathcal{Q}_R)^2] 
 u_\mu u^\mu\rangle  + K_6 \; \langle (\mathcal{Q}_L \mathcal{Q}_R + 
 \mathcal{Q}_R \mathcal{Q}_L) u_\mu u^\mu\rangle \cr
& \mbox{} + \frac{1}{2} K_7 \; \langle (\mathcal{Q}_L)^2 
 + (\mathcal{Q}_R)^2\rangle \; \langle \chi_+^{\mathrm{(sp)}}\rangle + K_8\; \langle \mathcal{Q}_L \mathcal{Q}_R\rangle 
 \; \langle \chi^{\mathrm{(sp)}}_+\rangle \cr
& \mbox{}+ K_9 \; \langle [(\mathcal{Q}_L)^2 + (\mathcal{Q}_R)^2] 
 \chi^{\mathrm{(sp)}}_+\rangle+ K_{10}\; \langle(\mathcal{Q}_L \mathcal{Q}_R 
 + \mathcal{Q}_R \mathcal{Q}_L) \chi^{\mathrm{(sp)}}_+\rangle \cr
& \mbox{} - K_{11} \; \langle(\mathcal{Q}_L \mathcal{Q}_R - \mathcal{Q}_R \mathcal{Q}_L) \chi^{\mathrm{(sp)}}_-\rangle \cr
& \mbox{}- iK_{12}\; \langle[ (\mathcal{Q}_{L\mu} \mathcal{Q}_L -
\mathcal{Q}_L  \mathcal{Q}_{L\mu}) 
 - (\mathcal{Q}_{R\mu} \mathcal{Q}_R - 
 \mathcal{Q}_R \mathcal{Q}_{R\mu})] u^\mu\rangle \cr
& \mbox{}+ K_{13} \; \langle  \mathcal{Q}_{L\mu} 
 \mathcal{Q}_R^\mu
 \rangle + K_{14} \; \langle ( \mathcal{Q}_{L\mu} \mathcal{Q}_L^\mu) +
 (\mathcal{Q}_{R\mu}\mathcal{Q}_{R}^\mu)\rangle  
 \bigg\}\ .
\end{align}
The $K$-terms provide the counterterms to cancel the divergences due to the QED corrections to the ChPT.
All the counterterms are proportional to the spurion fields.
The values of the $K$'s have been calculated in Ref.\,\cite{Moussallam:1997xx,Ananthanarayan:2004qk}.

In addition to the $K$-terms, we introduce the counterterms for the pion-Wino weak interactions:
\begin{align}
\label{eq:countertermLY}
&{\mathcal{L}}_{\mathrm{Y}}= 
e^2 \Big\{ \sqrt{2}F_0^2 G_F\Big[
{Y_1}\bar{\psi}_-\gamma_\mu \psi_0 \langle u^\mu
\{\mathcal{Q}_R,\mathcal{Q}_W\} 
\rangle 
 +{\hat Y_1}\bar{\psi}_-\gamma_\mu\psi_0 \langle{u^\mu\{\mathcal{Q}_L,\mathcal{Q}_W\} } \rangle
 \cr
 &\phantom{{\mathcal{L}}_{Y}= e^2 \sum_l \Big\{ F_0^2 \Big[}
 +{Y_2}\bar{\psi}_-\gamma_\mu\psi_0
 \langle{u^\mu[\mathcal{Q}_R,\mathcal{Q}_W]}\rangle
+{\hat Y}_2\bar{\psi}_-\gamma_\mu\psi_0\langle{u^\mu[\mathcal{Q}_L,\mathcal{Q}_W]}\rangle
 \cr
&\phantom{{\mathcal{L}}_{Y}= e^2 \sum_l \Big\{ F_0^2 \Big[}
 +{Y_3}m_{{\chi}}\bar{\psi}_- \psi_0\langle{\mathcal{Q}_R\mathcal{Q}_W}\rangle
 \cr
&\phantom{{\mathcal{L}}_{Y}= e^2 \sum_l \Big\{ F_0^2 \Big[}
+i{Y_4}\bar{\psi}_-\gamma_\mu\psi_0
\langle{\mathcal{Q}_{L}^\mu \mathcal{Q}_W }\rangle
 +i{Y_5}\bar{\psi}_-\gamma_\mu\psi_0\langle{\mathcal{Q}_{R}^\mu \mathcal{Q}_W}\rangle +h.c.\Big]
 \cr
&\phantom{{\mathcal{L}}_{Y}= e^2 \sum_l \Big\{ F_0^2 \Big[}
 + {\hat Y_6}\bar{\psi}_- (i\slashed{\partial}-e\slashed{A})\psi_-
 + {\hat Y_7} m_{{\chi}}\bar{\psi}_-\psi_-\Big\} \ .
\end{align}
These terms are in parallel to the $X$-terms for the pion-lepton weak interactions in Refs.\,\cite{Knecht:1999ag,Descotes-Genon:2005wrq}.
The $\hat{Y}_6$ and $\hat{Y}_7$ terms correspond to the renormalization factors 
for the wave function and 
the mass of the charged Wino, respectively.
We take the Wino mass counterterm so that $m_\chi$ becomes the physical charged Wino mass by adjusting $\hat{Y}_7$.

In this paper, we define 
the constants $K$'s and $Y$'s so that they are dimensionless even when the spacetime dimension is $d(\neq4)$ in the dimensional regularization. 
Thus, $K$'s in Ref.\,\cite{Urech:1994hd}
are $\mu_{\mathrm{ChPT}}^{d-4}$ times $K$'s in this paper.
Accordingly $K$'s and $Y$'s in this paper depend 
on the renormalization scale $\mu_{\mathrm{ChPT}}$ in the $\overline{\mathrm{MS}}$, 
while  $K$'s in Ref.\,\cite{Urech:1994hd} do not depend on $\mu_{\mathrm{ChPT}}$.
The definition for the finite part of $K$'s, i.e., $K^r(\mu_{\mathrm{ChPT}})$, is the same with those in Ref.\,\cite{Urech:1994hd}.

The constants $K$'s and $Y$'s contribute to the Wino decay rate through,
\begin{align}
\label{eq:pionK}
    \mathcal{L}_K \supset\,\,&
    e^2\Bigg\{\left[\frac{8}{3} (K_1+
    K_2 ) + \frac{20}{9} (K_5+K_6)
    \right]
    |D_\mu \pi^-|^2 \cr 
    &-
    2\sqrt{2}F_0{G}_F V_{ud}\!\left[\frac{8}{3} (K_1+
    K_2) + \frac{20}{9} (K_5+K_6)+2 K_{12} 
\right]\!\! 
\left(
\bar{\psi}_{-}\gamma^\mu \psi_0 D_\mu \pi^- + h.c.\right)\!\!\Bigg\}\ ,
\end{align}
and 
\begin{align}
\label{eq:LY}
    \mathcal{L}_Y \supset&2\sqrt{2}e^2F_0{G}_F V_{ud}
    \Bigg\{\left[\frac{2}{3} (Y_1+\hat{Y}_1) 
    +2 (Y_{2}+\hat{Y}_{2})  
\right]   \bar{\psi}_- \gamma^\mu \psi_0D_\mu \pi^-+ \left(2i Y_3 m_{{\chi}}\bar{\psi}_-\psi_0
    \pi^-
+ h.c.\right)\Bigg\} \cr
& +e^2\hat Y_6\bar{\psi}_- i\slashed{\partial}\psi_- \ .
\end{align}
Note that the $Y_3$ term induces the Yukawa interaction of the pion which is not suppressed by the pion momentum in contrast to the tree-level interaction in Eq.\,\eqref{eq:TreeLevel}.

Altogether, the counterterm contribution to the decay rate is given by,
\begin{align}
\label{eq:structure dependent Wino}
    \frac{\delta\Gamma_\chi}
   {\Gamma_{\chi}}\bigg|_{K,Y} &= 
    e^2 
    \Bigg[\frac{8}{3} (K_1+K_2) + \frac{20}{9} (K_5+K_6)+4 K_{12}\cr
    &\hspace{3cm}
    -\hat{Y}_6 -\frac{4}{3} 
    (Y_1+\hat{Y}_1)
    -4 \left(Y_2+\hat{Y}_2
    - \frac{m_{{\chi}}}{\mathit{\Delta} m} Y_3\right)
    \Bigg]\ .
\end{align}
Due to the non-derivative nature of the $Y_3$ interaction, the $Y_3$ contribution is relatively enhanced by $m_\chi/\mathit{\Delta}m$.

\subsection{Spurions in Four-Fermi theory}
To determine the constants $Y$'s,
we compare the current correlators in the ChPT and the Four-Fermi theory.
For that purpose, we rewrite the Four-Fermi theory
with the spurion fields $\mathbf{q}_{L,R,W}$ so that the Four-Fermi theory respects the SU(3)$_L\times$SU(3)$_R$ flavor symmetry.
The tree-level Lagrangian is given by,
\begin{align}
\label{eq:quarkpicture}
    \mathcal{L}^{\mathrm{FF}+\mathbf{q}}_{\mathrm{tree}} = 
    &\bar{\psi}_Li\slashed{D}_L \psi_L
    +\bar{\psi}_Ri\slashed{D}_R \psi_R+
    \sqrt{2}G_F   \left(\bar{\psi}_L\gamma_\mu \mathbf{q}_W \psi_L
    \times 
 \bar{\psi}_-\gamma^\mu \psi_0 + h.c. \right)
  \ ,
\end{align}
where $\psi_{L,R}$ are the left/right projections of the three-flavor Dirac quarks $(u,d,s)$. 
The covariant derivatives are $D_{L,R,\mu}=\partial_\mu - ie \mathbf{q}_{L,R}A_\mu$. The normalization of the weak interaction 
is due to Eq.\,\eqref{eq:WinoWeak} and Eq.\,\eqref{eq:charges}.

In Eq.\,\eqref{eq:FF counterterms}, we have introduced 
the counterterms in the Four-Fermi theory.
Here, we rewrite them as
\begin{align}
\label{eq:quark counterterms}
\mathcal{L}_{\mathrm{CT}}
^{\mathrm{FF}+\mathbf{q}}
= &
-2 e^2 f_{\chi\chi}\bar{\psi}_-(i \slashed{D}-m_{{\chi}}) \psi_-  -ie^2 f_{d\bar{u}}
\left[\bar{\psi}_L[\mathbf{q}_L,D^\mu\mathbf{q}_L]\gamma_\mu \psi_L + (L \leftrightarrow R) \right]
\cr
&- \sqrt{2} e^2 G_F
\left[\bar{\psi}_-\gamma_\lambda \psi_0
\times
(f_{\chi d}
\bar{\psi}_L \gamma^\lambda
\mathbf{q}_W \mathbf{q}_L \psi_L
+f_{\chi\bar{u}}
\bar{\psi}_L \gamma^\lambda
\mathbf{q}_L\mathbf{q}_W \psi_L
)
+ h.c.\right]\ .
\end{align}
By setting the $\mathbf{q}_{L,R} = \mbox{diag}(-Q_{\bar{u}},Q_d,Q_s)$ and $\mathbf{q}_W = Q_W$, and re-scaling the Wino field by 
\begin{align} 
    \psi_- \to \left(1+e^2 f_{\chi\chi}\right) \psi_- \ ,
\end{align}
the counterterms in Eq.\,\eqref{eq:quark counterterms} reproduce those in 
Eq.\,\eqref{eq:FF counterterms},
where no counterterms for the kinetic terms appear.

From Eqs.\,\eqref{eq:countertermLY}
and \eqref{eq:quark counterterms}, we find that both
$e^2\hat{Y}_6$ and $-2e^2 f_{\chi\chi}$ 
play the role of the counterterm to the kinetic term in each theory.
Incidentally, $f_{d\bar{u}}$ term is equivalent to the counterterm to the quark kinetic term,
\begin{align}
\label{eq:kinetic counterterm}
    \mathcal{L}'^{\mathrm{FF}+\mathbf{q}}_{\mathrm{CT}} = &- 2e^2 f_{d\bar{u}}\times\frac{1}{2}
    \left[
    \bar{\psi}_L \mathbf{q}_L i\overleftrightarrow{\slashed{D}}\mathbf{q}_L \psi_L
    + (L \leftrightarrow R)
    \right]\ ,
\end{align}
and hence, $f_{d\bar{u}}$ term is
related to $g_{23}$ term in D\&M's paper 
(see Appendix\,\ref{sec:quark counterterm}).

\subsection{Current Correlators}
We have introduced the charge spurion fields
$\mathbf{q}_{L,R,W}$ in both the ChPT and the Four-Fermi theory.
Following D\&M analysis (see also Ref.\,\cite{Moussallam:1997xx}),
we calculate
the current correlators 
in two theories 
by taking derivatives of the generating functional with respect to the spurions.
By comparing them, we can derive the matching conditions for $Y$'s.

Following D\&M analysis, we consider current correlators
\begin{align}
i f^{abc}F_{VW}
+ d^{abc} D_{VW}
= i
\int d^4 x 
\langle \psi_0(q)  \pi^a(r)|
\frac{\delta^2 W(\mathbf{q})}{\delta \mathbf{q}_{V}^b(x)\delta\mathbf{q}_W^{*c}(0)}|\psi_-(p)\rangle\ .
\end{align}
Here, $W(\mathbf{q})$ 
corresponds to the generating functional
of the ChPT, $W^\mathrm{ChPT}(\mathbf{q})$, 
or the Four-Fermi theory,
$W^{\mathrm{FF}}(\mathbf{q})$.
They depend on the spurion fields $\mathbf{q}_{L,R,W}$. 
We have also defined the vector and the axial spurions via, $\mathbf{q}_V \equiv\mathbf{q}_R+\mathbf{q}_L$  and 
$\mathbf{q}_A \equiv\mathbf{q}_R-\mathbf{q}_L$.
The spurions $\mathbf{q}^a$'s  ($a=1\mbox{--}8$)
are given by 
$\mathbf{q}^a = \tr[\lambda^a \mathbf{q}]$.
The structure constants $f^{abc}$ and $d^{abc}$ are given by,
\begin{align}
    f^{abc} = -2i
    \tr[t^a,[t^b,t^c]]\ , \quad d^{abc} =2\tr[t^a,\{t^b,t^c\}]\ .
\end{align}
Note that we set $\mathbf{q}_{L,R,W}=0$ after taking 
derivatives with respect to them.

In the following, we take
$(a,b,c)=(3,1,2)$ for $F_{VW}$ ($f^{312}=1$, $d^{312}=0$), 
and $(a,b,c)=(1,1,8)$ for $D_{VW}$ ($f^{118}=0$, $d^{118}=1/\sqrt{3}$).
As we will see,
$F_{VW}$ and $D_{VW}$ determine  combinations $Y_2 + \hat{Y}_2 - (m_\chi/{\mathit{\Delta}m})Y_3$
and  $Y_1+\hat{Y}_1$, respectively.
In the analysis of the current correlators, 
we assume the kinematics with $p=r+q$, $p^2 = m_\chi^2$, and $q^2=m_0^2$. 
We also 
take the chiral limit, i.e., $m_\pi^2 = 0$.

\subsection{Calculation of \texorpdfstring{$F_{VW}$}{}}
\subsubsection{\texorpdfstring{$F_{VW}$}{} in ChPT}
\begin{figure}[t]
    \centering
    \includegraphics[]{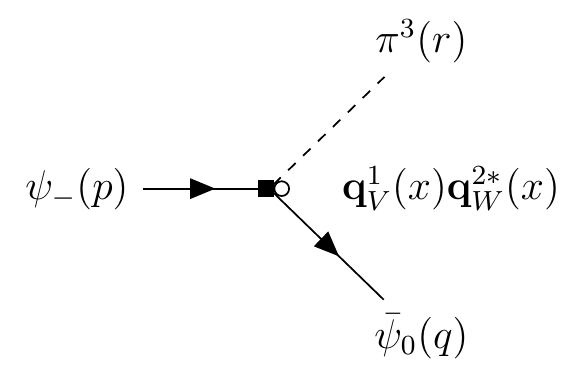}
    \caption{The tree-level contribution to $F_{VW}$ 
    in the ChPT.
    The point $\shikakumaru$
    is the vertex proportional to $\mathbf{q}_V^1(x) \mathbf{q}_W^{2*}(x)$.
    }
    \label{fig:FVW_ChPT_tree}
\end{figure}

In the ChPT, the tree-level contribution 
to $F_{VW}$ (Fig.\,\ref{fig:FVW_ChPT_tree}) is given by,
\begin{align}
\label{eq:FVWtree}
    F_{VW}^{(\mathrm{tree,ChPT})}
       =&
    -\frac{\sqrt{2}}{2}e^2 F_0 G_F  \left(Y_2+\hat{Y}_2-\frac{m_\chi}{\mathit{\Delta}m}Y_3\right)
\bar{u}_{0}(q)\slashed{r}
      u_{-}(p)\ .
\end{align}
Here, we have used the relation, 
$\mathit{\Delta}m\bar{u}_{0}(q)u_{-}(p)= \bar{u}_{0}(q)\slashed{r} u_{-}(p)$.
As the constants $Y_{2}$, $\hat{Y}_{2}$ and  $Y_3$
appear in the form of the above combination in the Wino decay rate (see Eq.\,\eqref{eq:structure dependent Wino}), 
we do not need determine these constants individually.

\begin{figure}[t]
    \centering
    \includegraphics[]{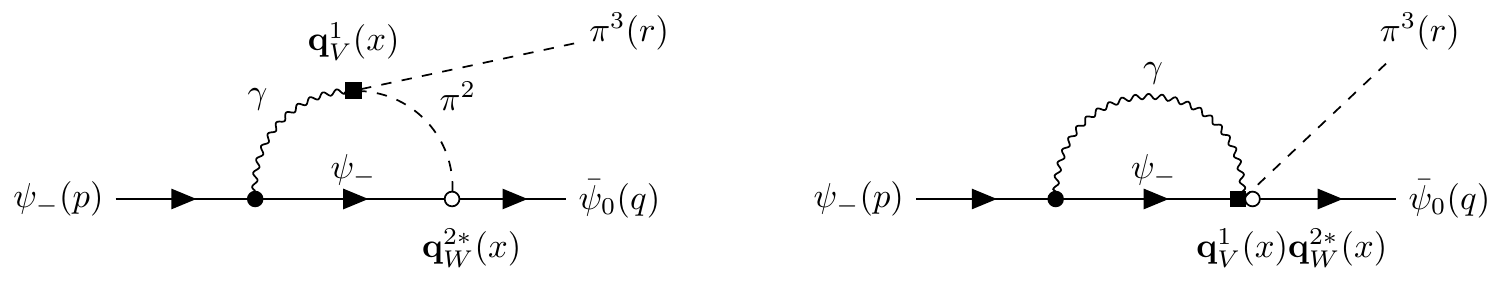}
    \caption{The one-loop contribution to $F_{VW}$ 
    in the ChPT. The points $\blacksquare$ and $\maru$
    are the separated vertices proportional to $\mathbf{q}_V^1(x)$ and $\mathbf{q}_W^{2*}(x)$, respectively.}
    \label{fig:FVW_ChPT_loop}
\end{figure}
The one-loop contribution (Fig.\,\ref{fig:FVW_ChPT_loop}) is given by
\begin{align}
    F_{VW}^{\mathrm{(loop,ChPT)}}     &= -i\frac{\sqrt{2}}{4}e^2F_0G_F
     \int\frac{d^4 \ell}{(2\pi)^4}
     \frac{1}{\ell^2-m_{\gamma}^2}
     \frac{\bar{u}_{0}(q)\gamma_\nu (\slashed{p}-\slashed{\ell}+m_{{\chi}})\gamma_\mu u_{-}(p)}{(p-\ell)^2 - m_{{\chi}}^2} {H_\pi^{\mu\nu}}(\ell-r) \ , \cr
     H_\pi^{\mu\nu}(\ell) &= \frac{\ell^\mu\ell^\nu-r^\mu \ell^\nu}{\ell^2} - g^{\mu\nu}\ .
\end{align}
By performing loop integration and combining with the tree-level contribution, we obtain,
\begin{align}
\label{eq:FVW_V_ChPT}
   F_{VW}^{\mathrm{(ChPT)}} =& \frac{\sqrt{2}}{2} F_0 G_F  \Bigg[
    -e^2 \left(Y_2+\hat{Y}_2 -\frac{m_{{\chi}}}{\mathit{\Delta} m} Y_3\right) -
   \frac{\alpha}{8\pi}
   \frac{m_{{\chi}}}{\mathit{\Delta} m} \left(
   \frac{3}{\bar{\epsilon}_{\mathrm{ChPT}}} + 
   3 \log\frac{\mu_\mathrm{ChPT}^2}{m_{{\chi}}^2}+4
   \right)\cr
    &+
   \frac{\alpha}{16\pi}
  \left(
   \frac{5}{\bar{\epsilon}_{\mathrm{ChPT}}} + 
    5\log\frac{\mu_\mathrm{ChPT}^2}{m_{{\chi}}^2}
   + \log\frac{m_{{\chi}}^2}{m_\gamma^2}
   +\frac{9}{2}
   -\frac{5}{3}\pi^2 
   -4\log\frac{m_\gamma^2} {4\mathit{\Delta}m^2}
   -
   \log^2\frac{m_\gamma^2}{4 \mathit{\Delta}m^2}
   \right) \cr
   &+\frac{\alpha}{4\pi}
   \log\frac{m_{{\chi}}^2}{4\mathit{\Delta}m^2}
   \Bigg]\times \bar{u}_{0}(q)\slashed{r}u_{-}(p)\ .
\end{align}

\subsubsection{\texorpdfstring{$F_{VW}$}{} in Four-Fermi theory}
\begin{figure}[t]
    \centering
    \includegraphics[]{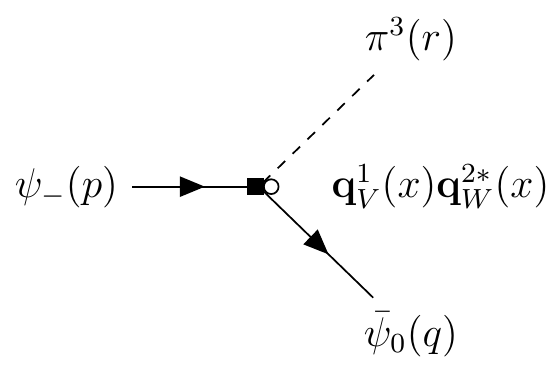}
    \caption{The tree-level contribution to $F_{VW}$ 
    in the Four-Fermi theory.
     The point $\shikakumaru$
    is the vertex proportional to
     $\mathbf{q}_V^1(x)\mathbf{q}_W^{2*}(x)$.
    Note that the vertices are provided by the quark current. 
    }
    \label{fig:FVW_FF_tree}
\end{figure}

In the Four-Fermi theory, the tree-level contribution to $F_{VW}$ (Fig. \ref{fig:FVW_FF_tree}) is given by,
\begin{align}
\label{eq:FVW_FF_tree}
    F_{VW}^{(\mathrm{tree,FF})} 
=
    \frac{\sqrt{2}}{8} e^2 F_0 G_F (f_{\chi d}-f_{\chi \bar{u}})  \times
\bar{u}_{0}(q)\slashed{r}
      u_{-}(p)\ .
\end{align}

\begin{figure}[t]
    \centering
    \includegraphics[]{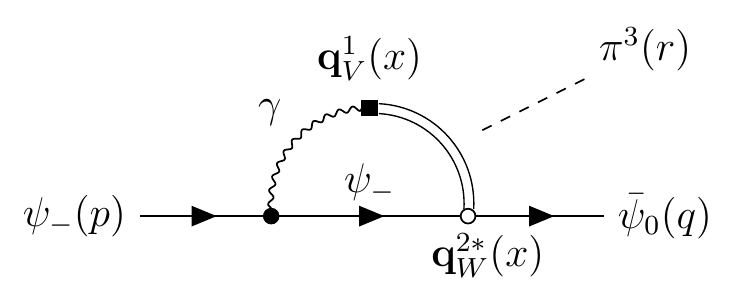}
    \caption{The one-loop contribution to $F_{VW}$ 
    in the Four-Fermi theory.
    The double line denotes the current correlator $\Gamma_{VA}^{\mu\nu}$.
    The points $\blacksquare$ and $\maru$
    are the separated vertices proportional to $\mathbf{q}_V^1(x)$ and $\mathbf{q}_W^{2*}(x)$, respectively.}
    \label{fig:FVW_FF_loop}
\end{figure}
At the one-loop level, $F_{VW}$ is given by,
\begin{align}
\label{eq:FVWQCDloop}
     F_{VW}^{(\mathrm{loop,FF})} 
=-i\frac{\sqrt{2}}{4} e^2 G_F\int \frac{d^4\ell}{(2\pi)^4}
 \Gamma_{VA}^{\mu\nu}(\ell,r) \times \frac{-i}{\ell^2 - m_\gamma^2} 
 \bar{u}_{0}(q)\gamma_\nu 
     \frac{i(\slashed{p} - \ell + m_{{\chi}})}{(p-\ell)^2-m_{{\chi}}^2}
     \gamma_\mu
     u_{-}(p)
        \ .
\end{align}
Here,
following D\&M, we have defined,
\begin{align}
 i f^{abc}\Gamma_{VA}^{\mu\nu}(
 \ell,r)\equiv\int d^4x e^{-i\ell x} 
    \langle \pi^a(r)|T J_{V}^{b\mu}(x) J_{A}^{c\nu}(0) | 0\rangle\ ,
\end{align}
where $J_A^a$ $(a=1\mbox{--}8)$ are the axial currents
in Eq.\,\eqref{eq:JA} extended to the three flavor quarks, while $J_V^a$ are the corresponding vector currents.

The evaluation of $\Gamma_{VA}^{\mu\nu}$ 
involves the strong dynamics of QCD which requires non-perturvative calculation.
Instead, we 
rely on the MRM in Ref.\,\cite{Weinberg:1967kj}
as in D\&M analysis.
As emphasized earlier, the MRM satisfies the Weinberg sum rule and the leading QCD asymptotic constraints.
In this model, $\Gamma_{VA}^{\mu\nu}$, is given by
\begin{align}
    \Gamma_{VA}^{\mu\nu}(k,p) = 
    F_0\Bigg[
    \frac{(k^\mu+2 q^\mu)q^\nu}{q^2} - g^{\mu\nu} + F(k^2,q^2) P^{\mu\nu}
    + G(k^2,q^2) Q^{\mu\nu}
    \Bigg] \ ,
\end{align}
where $q= p-k$ with
\begin{align}
    P^{\mu\nu} = q^{\mu} k^{\nu} 
    - (k\cdot q)g^{\mu\nu}\ , 
    \quad
    Q^{\mu\nu} =k^2 q^\mu q^\nu
    +q^2 k^\mu k^\nu - (k\cdot q)
    k^\mu q^\nu - k^2 q^2 g^{\mu\nu} \ ,
\end{align}
and 
\begin{align}
    F(k^2,q^2) = \frac{k^2-q^2+2(M_A^2-M_V^2)}{2(k^2-M_V^2)(q^2 - M_A^2)} 
    \ ,  \quad
     G(k^2,q^2) = \frac{-q^2+2M_A^2}{(k^2-M_V^2)(q^2 - M_A^2)q^2} \ .
\end{align}
Here, $M_V$ and $M_A = \sqrt{2}M_V$ are the mass parameters with $M_V$ being around the $\rho$ meson mass.

By performing loop momentum integration, we obtain,
\begin{align}
\label{eq:FVW_V_FF_loop}
   F_{VW}^{(\mathrm{loop,FF})} =& \frac{\alpha}{16\pi\sqrt{2}} F_0 G_F  \Bigg[
    \frac{5}{\bar{\epsilon}_{\mathrm{ChPT}}}
    + 5 \log\frac{\mu_{\mathrm{FF}}^2}{m_{{\chi}}^2}
    + \log\frac{m_{{\chi}}^2}{m_\gamma^2}
    + \log\frac{m_{{\chi}}^2}{M_V^2}\cr
    &-\frac{5\pi^2}{3} - 4
    \log\frac{m_\gamma^2}{M_V^2}
- \log^2\frac{m_\gamma^2}{4 \mathit{\Delta}m^2}
 \cr 
    & + \frac{6 M_A^2 - 9 M_V^2}{M_A^2 - M_V^2}
+3 \frac{M_V^4}{(M_A^2 -M_V^2)^2}
\log\frac{M_A^2}{M_V^2}\cr
&-\frac{4}{\mathit{\Delta}m}\frac{\pi M_AM_V}{M_A+M_V}    
     \Bigg]\times \bar{u}_{0}(q)\slashed{r}u_{-}(p)\ .
\end{align}
The total contribution to $F_{VW}$ in the Four-Fermi theory is given by
\begin{align}
F^\mathrm{(FF)}_{VW} = F^\mathrm{(tree,FF)}_{VW}+F^\mathrm{(loop,FF)}_{VW}\ .
\end{align}

Note that the UV divergence and the $\log m_{{\chi}}$ dependences are determined by the asymptotic behavior of $\Gamma_{VA}^{\mu\nu}$ at $\ell^2 \to -\infty$,
\begin{align}
     \Gamma_{VA}^{\mu\nu}(\ell,p) =
     F_0 \times \frac{(\ell \cdot p) g^{\mu\nu} -\ell^\mu p^\nu - \ell^\nu p^\mu }{\ell^2} + O(\ell^{-2})\ .
\end{align}
This behavior is consistent with the operator product expansion of QCD  (see Eq.\,(58) of Ref.\,\cite{Moussallam:1997xx}).

\subsection{Calculation of \texorpdfstring{$D_{VW}$}{}}
\subsubsection{\texorpdfstring{$D_{VW}$}{} in ChPT}
\begin{figure}[t]
    \centering
    \includegraphics[]{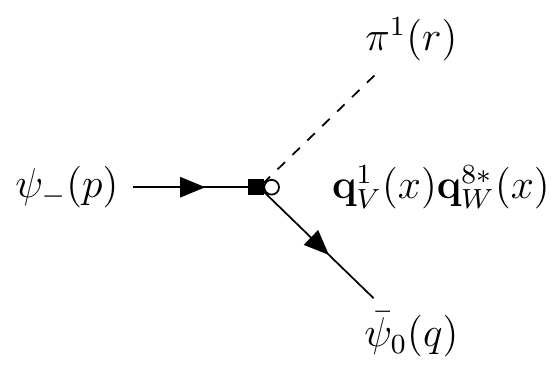}
    \caption{The tree-level contribution to $D_{VW}$ 
    in the ChPT. 
    The point $\shikakumaru$
    is the vertice proportional to $\mathbf{q}_V^1(x) \mathbf{q}_W^{8*}(x)$.}
    \label{fig:DVW_ChPT_tree}
\end{figure}
The tree-level contribution to $D_{VW}$ (Fig.\,\ref{fig:DVW_ChPT_tree}) is given by,
\begin{align}
\label{eq:DVW_ChPT_tree}
    D_{VW}^{\mathrm{(tree,ChPT)}} =   \frac{\sqrt{2}}{2}e^2 F_0 G_F  (Y_1+\hat{Y}_1)\ .
\end{align}
In the ChPT, there is no one-loop level contributions to $D_{VW}$, 
\begin{align}
\label{eq:DVW_ChPT_loop}
    D_{VW}^{\mathrm{(loop,ChPT)}} =  0 \ .
\end{align}

\subsubsection{\texorpdfstring{$D_{VW}$}{} in Four-Fermi theory}

\begin{figure}[t]
    \centering
    \includegraphics[]{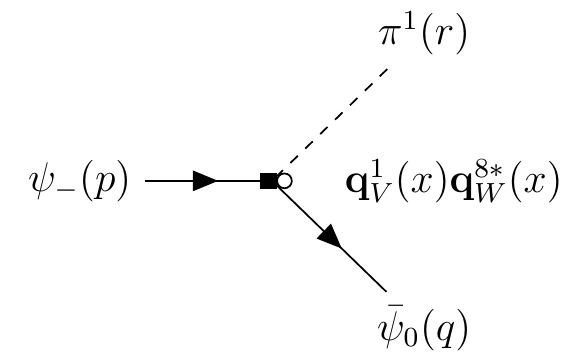}
    \caption{The tree-level contribution to $D_{VW}$ 
    in the Four-Fermi theory.  The point $\shikakumaru$
    is the vertice proportional to $\mathbf{q}_V^1(x) \mathbf{q}_W^{8*}(x)$.
    Note that the vertices are provided by the quark current.
    }
    \label{fig:DVW_FF_tree}
\end{figure}
In the Four-Fermi theory, the tree-level contribution to $D_{VW}$ 
(Fig.\,\ref{fig:DVW_FF_tree}) is given by,
\begin{align}
\label{eq:DVW_FF_tree}
    D_{VW}^{(\mathrm{tree,FF})} 
=
    \frac{\sqrt{2}}{8} e^2 F_0 G_F (f_{\chi d}+f_{\chi \bar{u}})  \times
\bar{u}_{0}(q)\slashed{r}
      u_{-}(p)\ .
\end{align}
\begin{figure}[t]
    \centering
    \includegraphics[]{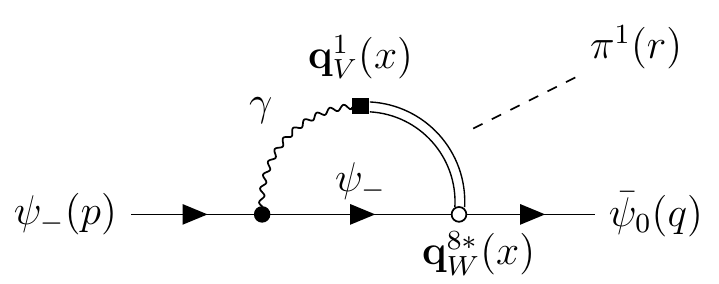}
    \caption{The one-loop contribution to $D_{VW}$ 
    in the Four-Fermi theory.
    The double line denotes the current correlator $\Gamma_{VV}^{\mu\nu}$. 
    The points $\blacksquare$ and $\maru$
    are the separated vertices proportional to $\mathbf{q}_V^1(x)$ and $\mathbf{q}_W^{8*}(x)$, respectively.}
    \label{fig:DVW_FF_loop}
\end{figure}
At the one-loop level, $D_{VW}$ (Fig.\,\ref{fig:DVW_FF_loop}) is given by,
\begin{align}
\label{eq:DVWQCDloop}
     D_{VW}^{(\mathrm{loop,FF})} 
=&\frac{\sqrt{2}}{4} e^2 G_F\int \frac{d^4\ell}{(2\pi)^4}
 \Gamma_{VV}^{\mu\nu}(\ell,r) \times \frac{-i}{\ell^2 - m_\gamma^2} 
 \bar{u}_{0}(q)\gamma_\nu 
     \frac{i(\slashed{p} - \slashed{\ell} + m_{{\chi}})}{(p-\ell)^2-m_{{\chi}}^2}
     \gamma_\mu
     u_{-}(p)
        \ .
\end{align}
Here, following D\&M, we have defined,
\begin{align}
  d^{abc}\Gamma_{VV}^{\mu\nu}(
 \ell,r)=\int d^4x e^{-i\ell x} 
    \langle \pi^a(r)|T J_{V}^{b\mu}(x) J_{V}^{c\nu}(0) | 0\rangle\ .
\end{align}

To evaluate $\Gamma_{VV}^{\mu\nu}$, we again rely on the MRM.
There, $\Gamma_{VV}^{\mu\nu}$ is given in
Ref.\,\cite{Knecht:1999ag},
\begin{align}
    \Gamma_{VV}^{\mu\nu}(k, p) = i F_0 \varepsilon^{\mu\nu\rho\sigma}k_{\rho}
    p_\sigma \Gamma_{VV}(k, p)\ ,
\end{align}
with 
\begin{align}
    \Gamma_{VV}(k, p) = 
    \frac{2k^2-2k\cdot p - c_V}{2(k^2 - M_V^2)((p-k)^2 - M_V^2)}\ .
\end{align}
The value of $c_V$ is given by
\begin{align}
\label{eq:cV}
    c_V = \frac{N_c M_V^4}{4\pi^2 F_0^2}\ ,
\end{align}
in the large $N_c$ limit~\cite{Witten:1983tw}.

By performing the loop integration, we find
\begin{align}
    D_{VW}^{\mathrm{(loop,FF)}} = -\frac{i\alpha}{16\sqrt{2}\pi}F_0G_F\left(
    \frac{3}{\bar{\epsilon}_{\mathrm{FF}}} + 3 \log \frac{\mu_{\mathrm{FF}}^2}{m_{{\chi}}^2} +2
    \right)\times \bar{u}_{0}(q)
    \slashed{r} \gamma_5 u_{-}(p)\ .
\end{align}
This result indicates that low energy constants $Y$'s and counterterms $f$'s are not enough and we need to introduce the 
terms where $\gamma_5$'s are inserted between the charged and the neutral Wino.
As we have discussed in Sec.\,\ref{sec:axialWino}, 
the electroweawk theory predicts that 
the contributions of the axial Wino current interaction 
are of $\order{\alpha^2}$ compared with the tree-level  contribution.
Thus, in our analysis, we do not introduce additional low energy constants, and simply neglect those contributions and take
\begin{align}
\label{eq:DVW_FF_loop}
    D_{VW}^{(\mathrm{loop,FF})} = 0 \ .
\end{align}

\subsection{Matching Conditions}
Now, let us derive the matching conditions
by requiring that the current correlators 
in the ChPT reproduce those in the Four-Fermi theory.
 The matching condition for $Y_1+\hat{Y}_1$ is given by
 matching $D_{VW}$, which results in 
\begin{align}
\label{eq:Y1condition}
   e^2(Y_1 + \hat{Y}_1) = \frac{1}{4}e^2 (f_{\chi d}+f_{\chi \bar{u}}) = 0\ ,
\end{align}
where we have used Eq.\,\eqref{eq:fwd+fwbaru} in the second equality.
The matching condition for $Y_2+\hat{Y}_2-(m_\chi/\mathit{\Delta}m) Y_3$ is given by matching $F_{VW}$, and written as
\begin{align}
\label{eq:Y2 and Y3 condition}
   -&e^2 \left(Y_2+\hat{Y}_2 -\frac{m_{{\chi}}}{\mathit{\Delta} m} Y_3\right) -
   \frac{\alpha}{8\pi}
   \frac{m_{{\chi}}}{\mathit{\Delta} m} \left(
   \frac{3}{\bar{\epsilon}_\mathrm{ChPT}} + 
   3 \log\frac{\mu_\mathrm{ChPT}^2}{m_{{\chi}}^2}+4
   \right)
   \cr
    +&
   \frac{\alpha}{16\pi}
  \left(
   \frac{5}{\bar{\epsilon}_\mathrm{ChPT}} + 
    5\log\frac{\mu_\mathrm{ChPT}^2}{m_{{\chi}}^2}
   + \log\frac{m_{{\chi}}^2}{m_\gamma^2}
   +\frac{9}{2}
   -\frac{5}{3}\pi^2 
   -4\log\frac{m_\gamma^2} {m_\chi^2}
   -
   \log^2\frac{m_\gamma^2}{4 \mathit{\Delta}m^2}
   \right) 
   \cr
   =\, &\frac{1}{4} e^2 (f_{\chi d}-f_{\chi \bar{u}}) 
      \cr
  &+ \frac{\alpha}{16\pi}\Bigg(
    \frac{5}{\bar{\epsilon}_\mathrm{FF}}
    + 5 \log\frac{\mu_{\mathrm{FF}}^2}{m_{{\chi}}^2}
    + \log\frac{m_{{\chi}}^2}{m_\gamma^2}
    + \log\frac{m_{{\chi}}^2}{M_V^2}
    -\frac{5\pi^2}{3} - 4
    \log\frac{m_\gamma^2}{M_V^2}
- \log^2\frac{m_\gamma^2}{4 \mathit{\Delta}m^2}\cr
    &\hspace{40pt} + \frac{6 M_A^2 - 9 M_V^2}{M_A^2 - M_V^2}
+3 \frac{M_V^4}{(M_A^2 -M_V^2)^2}
\log\frac{M_A^2}{M_V^2}
-\frac{4}{\mathit{\Delta}m}\frac{\pi M_AM_V}{M_A+M_V}    
     \Bigg)\ .
\end{align}
Note that the both side have the identical $m_\gamma$ dependence and the matching condition is free from the IR divergences.

The matching condition for $\hat{Y}_6$ 
is simply given by,
\begin{align}
    \label{eq:Y6condition}
    e^2\hat{Y}_6 + \frac{\alpha}{4\pi}\frac{1}{\bar{\epsilon}_{\mathrm{ChPT}}} = -2e^2 f_{\chi\chi}
     + \frac{\alpha}{4\pi}\frac{1}{\bar{\epsilon}_{\mathrm{FF}}} + \frac{\alpha}{4\pi} \log\frac{\mu_{\mathrm{FF}}^2}{\mu_\mathrm{ChPT}^2} \ .
\end{align}
In addition, we will use the following expression for, $K_{12}$,%
\footnote{
We obtain this expression by combining 
Eqs.\,(91) and (102) in Ref.\,\cite{Descotes-Genon:2005wrq}.
Here, we take account of the difference between the Pauli-Villars regularization in Ref.\,\cite{Descotes-Genon:2005wrq}
and the $\overline{\mathrm{MS}}$ scheme in our analysis.}
\begin{align}
\label{eq:K12}
 e^2K_{12} =& -\frac{1}{8}
 \frac{\alpha}{4\pi}\frac{1}{\bar{\epsilon}_{\mathrm{ChPT}}}   +\frac{1}{2}e^2f_{d\bar{u}}^r(\mu_\mathrm{FF}) + \frac{\alpha}{16\pi}
 \log \frac{\mu_\mathrm{ChPT}^2}{\mu_{\mathrm{FF}}^2}
 \cr
 &+ \frac{1}{8}
 \frac{\alpha}{4\pi}\left(
 -3 \log \frac{\mu_\mathrm{ChPT}^2}{M_V^2}
 + \frac{3(M_A^2 + M_V^2)M_V^2}{(M_A^2-M_V^2)^2}
 \log\frac{M_A^2}{M_V^2} - \frac{6 M_A^2}{M_A^2 - M_V^2}
 + \frac{3}{2}\right)
\ ,
\end{align}
which is obtained by using the MRM in Ref.\,\cite{Moussallam:1997xx}.

\section{Radiative Correction to  Charged Wino Decay}
\label{sec:QED}
Finally, let us calculate the radiative correction to the Wino decay rate in the  ChPT.
In this section, 
the tree-level amplitude is 
taken to be
\begin{align}
\label{eq:iMtreehat}
    i\hat{\mathcal{M}}_{{\mathrm{tree}}} &=  2\sqrt{2}V_{ud} F_0 G_F \mathit{\Delta} m\,\bar{u}_{0}(q)  u_{-}(p)\ ,
\end{align}
since we have redefined the tree-level Lagrangian by Eq.\,\eqref{eq:TreeLevel}.

\subsection{Two-Point Functions}
The derivative of the self-energy of the Wino at $\slashed{p} = m_\chi$ in the ChPT is,
\begin{align}
\label{eq:dpdSigwino}
    \frac{d}{d\slashed{p}}\Sigma_-^\mathrm{(ChPT)}(m_{{\chi}})
     =&
       - 
    \frac{\alpha}{4\pi}
    \left(\frac{1}{\bar{\epsilon}_{\mathrm{ChPT}}} + \log \frac{\mu_\mathrm{ChPT}^2}{m_{{\chi}}^2} +2 \log \frac{m_\gamma^2}{m_{{\chi}}^2} + 4\right)\ .
\end{align}
And that of pion is,
\begin{align}
\frac{d}{dp^2}\Sigma_\pi^\mathrm{(ChPT, QED)}(m_\pi^2) = \frac{\alpha}{2\pi}
    \left(\frac{1}{\bar{\epsilon}_{\mathrm{ChPT}}} + \log\frac{\mu_\mathrm{ChPT}^2}{m_\gamma^2}
\right)\ .
\end{align}
Here, we only consider the QED correction to the pion, since the pion self-energy from the pion loops cancels 
when we take the ratio between the Wino and the pion decay rates.

The above self-energy contributes to the NLO decay amplitude as,
\begin{align}
   \mathcal{M}^{\mathrm{WF}} =  \hat{\mathcal{M}}_{\mathrm{tree}} \times
    \left(\frac{1}{2} \frac{d}{d\slashed{p}}\Sigma_-^\mathrm{(ChPT)}(m_{{\chi}})+\frac{1}{2} \frac{d}{dp^2}\Sigma_\pi^\mathrm{(ChPT,QED)}(m_\pi^2)\right)\ .
\end{align}
Note that we have already taken the wave function renormalization counterterms into account in Eq.\,\eqref{eq:structure dependent Wino},
and therefore we do not need to include them here.

\subsection{Vertex Corrections}
\begin{figure}[t]
    \centering
    \includegraphics[]{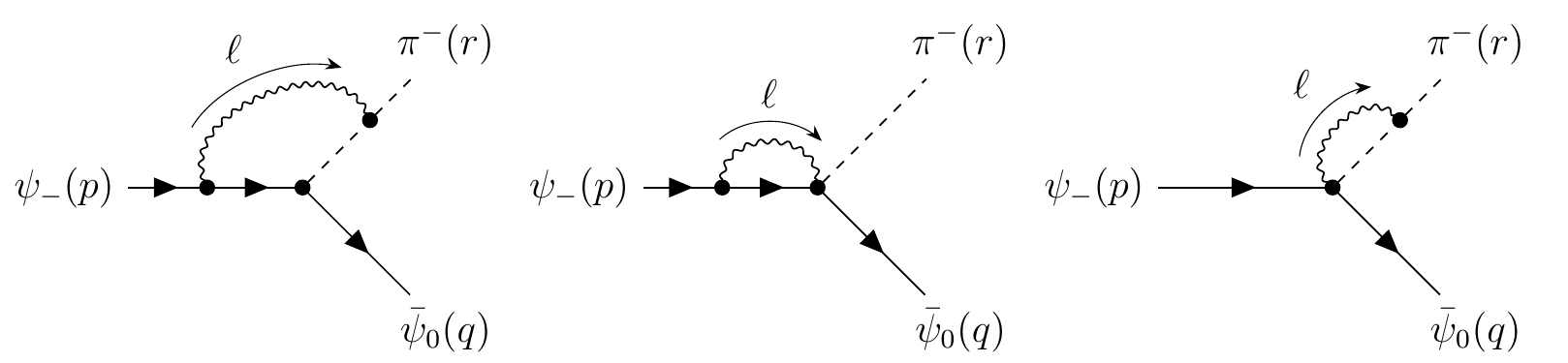}
    \caption{The virtual QED correction to the Wino decay.}
    \label{fig:WinoQEDVirtual}
\end{figure}
The vertex corrections to the Wino decay process are given in Fig.\,\ref{fig:WinoQEDVirtual}.
To estimate these corrections, we take the limits of  $\mathit{\Delta} m \ll m_{{\chi}}$, $m_\pi \ll m_{{\chi}}$ and $m_\gamma \to 0$.
We also use $\mathit{\Delta}m\bar{u}_{0}(q)u_{-}(p)= \bar{u}_{0}(q)\slashed{r} u_{-}(p)$.
The first contribution is,
\begin{align}
    \frac{\mathcal{M}^{\mathrm{Vertex,1}}}{\hat{\mathcal{M}}_{\mathrm{tree}}} =& \frac{\alpha}{4\pi} \Bigg\{\frac{5}{2}\left(\frac{1}{\bar{\epsilon}_{\mathrm{ChPT}}}+\log \frac{\mu_\mathrm{ChPT}^2}{m_\pi^2}\right)
     -
    \frac{2}{\sqrt{1-z^2}}
   \log\frac{1-\sqrt{1-z^2}}{1+\sqrt{1-z^2}}\log\frac{m_\gamma}{m_\pi}
   +\frac{11}{2} \cr
    &-\frac{1}{3\sqrt{1-z^2}}\Bigg[
    \pi^2 -6\log^2 z - 3\log^2\left(2(\sqrt{1-z^2}-1+z^2)\right)+ 6 \log^2(2\sqrt{1-z^2})   \cr
    &+12\log z \log (1-\sqrt{1-z^2}) + 6 \mathrm{Li}_2\left(\frac{1}{2}-\frac{1}{2\sqrt{1-z^2}}\right)
    \Bigg]\Bigg\}\ ,
\end{align}
where $z = m_\pi/\mathit{\Delta}m$.

The second contribution in Fig.\,\ref{fig:WinoQEDVirtual} is given by,
\begin{align}
\label{eq:unwanted} \frac{\mathcal{M}^{\mathrm{Vertex,2}}}{\hat{\mathcal{M}}_{\mathrm{tree}}} = -\frac{\alpha}{4\pi} \frac{m_{{\chi}}}{\mathit{\Delta} m}
\left(
\frac{3}{\bar{\epsilon}_{\mathrm{ChPT}}}
+3\log\frac{\mu_\mathrm{ChPT}^2}{m_\chi^2}  + 4
\right)\ .
\end{align}
Note that this contribution is enhanced by a factor of $m_{\chi}/\mathit{\Delta} m$
and can even exceed the tree-level contribution for a large limit of $m_\chi$. 
As we will see, however, the $m_\chi/\mathit{\Delta} m$ enhanced contribution 
is completely cancelled by the constants $Y$'s which are obtained by matching from the electroweak theory.

Finally, the third contribution in Fig.\,\ref{fig:WinoQEDVirtual}
is given by,
\begin{align}
\frac{\mathcal{M}^{\mathrm{Vertex,3}}}{\hat{\mathcal{M}}_{\mathrm{tree}}} = -\frac{\alpha}{8\pi}
    \left(
    \frac{3}{\bar{\epsilon}_{\mathrm{ChPT}}} + 3\log\frac{\mu_\mathrm{ChPT}^2}{m_\pi^2} + 7
    \right)\ .
\end{align}
By combining all the vertex corrections, $\mathcal{M}^{\mathrm{Vertex},1}$,
$\mathcal{M}^{\mathrm{Vertex},2}$,
$\mathcal{M}^{\mathrm{Vertex},3}$, 
and $\mathcal{M}^\mathrm{WF}$,
we obtain
\begin{align}
\label{eq:WinoVirtual}
   \frac{\mathcal{M}^{\mathrm{Virtual}}}{\hat{\mathcal{M}}_{\mathrm{tree}}} =& \frac{\alpha}{8\pi}\Bigg\{ 3\left(\frac{1}{\bar{\epsilon}_\mathrm{ChPT}}+\log\frac{\mu_\mathrm{ChPT}^2}{m_\pi^2} \right)-6\frac{m_\chi}{\mathit{\Delta}m}\left( \frac{1}{\bar{\epsilon}_\mathrm{ChPT}}+\log\frac{\mu_\mathrm{ChPT}^2}{m_\chi^2} + \frac{4}{3}\right) + 3 \log \frac{m_\chi^2}{m_\pi^2}\cr
    &\quad -8\left(1+\frac{1}{2\sqrt{1-z^2}}\log \frac{1-\sqrt{1-z^2}}{1+\sqrt{1-z^2}}
    \right)\log \frac{m_\gamma}{m_\pi}\cr
    &\quad -\frac{2}{3\sqrt{1-z^2}}\Bigg[
    \pi^2 - 6\log^2 z -3\log^2 \left(2(\sqrt{1-z^2}-1+z^2)\right)+6\log^2(2\sqrt{1-z^2})\cr
    &\quad +12\log z\log(1-\sqrt{1-z^2}) + 6\mathrm{Li}_2 \left(\frac{1}{2}-\frac{1}{2\sqrt{1-z^2}}\right)\Bigg]\Bigg\}\ .
\end{align}

\subsection{Real Photon Emission in Wino Decay}
\begin{figure}[t]
    \centering
    \includegraphics[]{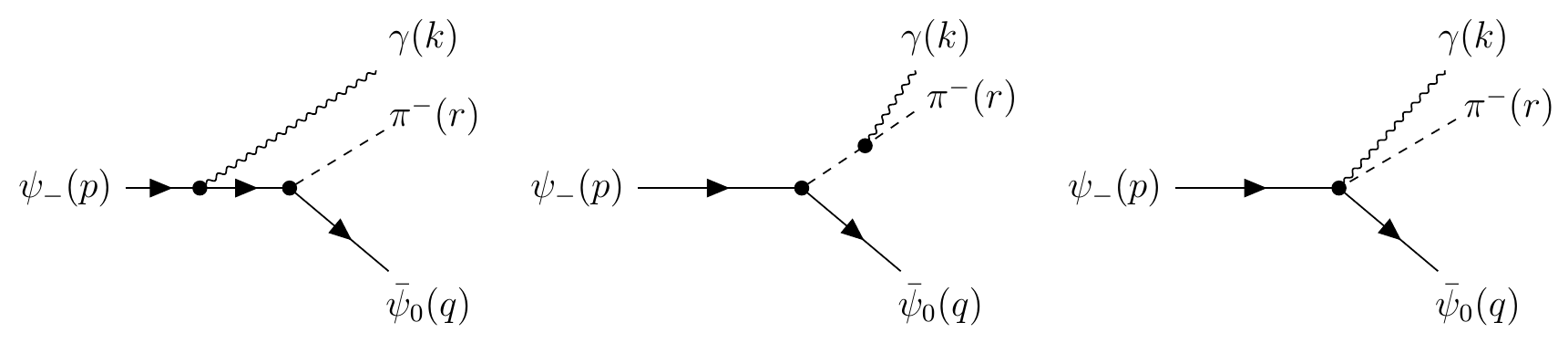}
    \caption{The real photon emission associated to the Wino decay.}
    \label{fig:WinoQEDSoft}
\end{figure}
To cancel the IR singularity in the virtual correction, we need to take into account the real photon emission.
In this subsection, we show the Wino decay rate with real photon emission. 
The details are given in  Appendix~\ref{sec:Details of Real Emission}.
The relevant diagrams are given in Fig.\,\ref{fig:WinoQEDSoft}.
We calculate the real emission rate by dividing the energy range of the photon 
$E_\gamma<E_\mathrm{soft}$ (soft part) and 
$E_\mathrm{soft} < E_\gamma$ (hard part) where $E_\mathrm{soft}$ is taken in between
\begin{align}
    m_\gamma \ll E_\mathrm{soft} \ll \frac{m_{{\chi}}^2 + m_\gamma^2-(m_0+m_\pi)^2}{2m_{{\chi}}} \simeq \mathit{\Delta} m - m_\pi \ .
\end{align}

The resultant soft part
is given by,
\begin{align}
\label{eq:Wino Emit soft Final}
    \frac{\delta\Gamma_\chi^{\mathrm{emit,soft}}}
   {\Gamma_{\chi}} 
= \frac{\alpha}{\pi}
    \Bigg\{&
    1+2 
    \left( 1+\frac{1}{2\sqrt{1-z^2}}
    \log
    \frac{1-\sqrt{1-z^2}}{1+\sqrt{1-z^2}}\right)\log\frac{m_{\gamma}}{2 E_\mathrm{soft}}\cr
    &+ \frac{1}{2\sqrt{1-z^2}}\bigg[-
    \frac{\pi^2}{3}-
   \left(
   1-\log
    \frac{4(1-z^2)}{z^2}
   \right)
   \log
    \frac{1-\sqrt{1-z^2}}{1+\sqrt{1-z^2}} \cr
    & + \frac{1}{2}\log^2\frac{1-\sqrt{1-z^2}}{1+\sqrt{1-z^2}}+ 2\,\mathrm{Li_2}\left(
    \frac{1-\sqrt{1-z^2}}{1+\sqrt{1-z^2}}
    \right)\bigg]\Bigg\}\ ,
\end{align}
in the limit of $m_{{\chi}}\gg \mathit{\Delta} m> m_\pi $.
By comparing Eqs.\,\eqref{eq:WinoVirtual} and \eqref{eq:Wino Emit soft Final}, we find that 
the IR divergence does not appear 
in the sum of the virtual correction and the soft real emission.

The hard part of the real emission is given by
\begin{align}
\label{eq:Wino Emit Hard}
      \frac{\delta\Gamma_\chi^{\mathrm{emit,hard}}}
   {\Gamma_{\chi}}
    =\frac{\alpha}{\pi}
    \Bigg\{4&
    -2\log
    \frac{4(1-z^2)}{z}
    +2 
    \left( 1+\frac{1}{2\sqrt{1-z^2}}\log
    \frac{1-\sqrt{1-z^2}}{1+\sqrt{1-z^2}}
    \right)\log \frac{2E_\mathrm{soft}}{\mathit{\Delta}m} \cr
&+\frac{1}{\sqrt{1-z^2}}
    \bigg[
    -\frac{\pi^2}{3} - \log z \log 
    \frac{1-\sqrt{1-z^2}}{1+\sqrt{1-z^2}}
    \cr
    &
    + \frac{1}{2}\log^2\frac{1-\sqrt{1-z^2}}{1+\sqrt{1-z^2}} +2\,\mathrm{Li}_2\left(\frac{1-\sqrt{1-z^2}}{1+\sqrt{1-z^2}}    \right)\bigg]\Bigg\}\ ,
\end{align}
 in the limit of the large Wino mass.
 The
 sum of $\Gamma^{\mathrm{emit,hard}}_\chi$ and $\Gamma^{\mathrm{emit,soft}}_\chi$ does not depend on  $E_\mathrm{soft}$.
 
\subsection{Virtual Correction and Real Photon Emission}
By combining the virtual photon correction and the real emission contributions, we obtain
\begin{align}
\label{eq:dGamphoton}
 \frac{\delta\Gamma_\chi}
   {\Gamma_{\chi}}\bigg|_{\gamma} 
     &=\frac{\alpha}{4\pi}
    \Bigg[3 \left(\frac{1}{\bar{\epsilon}_{\mathrm{ChPT}}}+\log\frac{\mu_\mathrm{ChPT}^2}{m_\pi^2}\right)
    - \frac{6m_{{\chi}}}{\mathit{\Delta} m}\left(\frac{1}{\bar{\epsilon}_{\mathrm{ChPT}}}+\log\frac{\mu_\mathrm{ChPT}^2}{m_{{\chi}}^2}
    + \frac{4}{3}\right)+ 3\log\frac{m_{{\chi}}^2} {\mathit{\Delta} m^2}+ 4f_{{\chi}}\left(\frac{m_\pi}{\mathit{\Delta} m}\right)\Bigg]\ ,
\end{align}
where 
\begin{align}
\label{eq:fchi}
    f_{{\chi}}(z) = &
    5
+\frac{5}{2}\log z
    -2 \log\left(4(1-z^2)\right)\cr
    &+\frac{1}{\sqrt{1-z^2}}
    \Bigg[
-\frac{2\pi^2}{3}
-\frac{1}{2}\log^2 2
+(1-2\log 2) \log z
+6\log^2 z
\cr
&
-\frac{1}{2} 
(\log 2 + 2\log z) \log\left(1-z^2\right)
-\frac{1}{8} \log^2\left(1-z^2\right)
\cr
&+\frac{1}{2}
\left(
-2 + 6\log 2 
+3 \log\left(1-z^2\right)
-20\log z
\right)\log\left(
1-\sqrt{1-z^2}\right)
    \cr
& +\frac{7}{2}
\log^2
\left(
1-\sqrt{1-z^2}
\right)
-\mathrm{Li}_2\left(\frac{1}{2}-\frac{1}{2\sqrt{1-z^2}}\right)
+3\mathrm{Li}_2\left(
\frac{1-\sqrt{1-z^2}}{1+\sqrt{1-z^2}}\right)\Bigg]\ .
\end{align}
The final expression is free from the IR divergence as expected, while 
it is UV divergent.
In the next section, we combine Eqs.\,\eqref{eq:dGamphoton}
with the counterterm contributions in Eq.\,\eqref{eq:structure dependent Wino}.

\section{Results}
\label{sec:Results}
\subsection{Wino Decay Rate @ NLO}
By combining Eqs.\,\eqref{eq:structure dependent Wino} and \eqref{eq:dGamphoton},
the Wino decay rate at the NLO is given by,
\begin{align}
\label{eq:dGammachi}
 \frac{\delta\Gamma_\chi}
   {\Gamma_{\chi}}
     =&  \frac{\alpha}{4\pi}
    \Bigg[3 \left(\frac{1}{\bar{\epsilon}_{\mathrm{ChPT}}}+\log\frac{\mu_\mathrm{ChPT}^2}{m_\pi^2}\right)
    - \frac{6m_{{\chi}}}{\mathit{\Delta} m}\left(\frac{1}{\bar{\epsilon}_\mathrm{ChPT}}+\log\frac{\mu_\mathrm{ChPT}^2}{m_{{\chi}}^2}
    + \frac{4}{3}\right)+ 3\log\frac{m_{{\chi}}^2} {\mathit{\Delta} m^2}+ 4f_{{\chi}}\left(\frac{m_\pi}{\mathit{\Delta} m}\right)\Bigg]
    \cr
    & + 
    e^2 
    \Bigg[\frac{8}{3} (K_1+K_2) + \frac{20}{9} (K_5+K_6)+4 K_{12}-\hat{Y}_6 -\frac{4}{3} (Y_1+\hat{Y}_1)
    -4 \left(Y_2+ \hat{Y}_2 -
     \frac{m_{{\chi}}}{\mathit{\Delta} m} Y_3\right)
    \Bigg]\ .
\end{align}
The relevant constants, $Y$'s, 
are given in
Eqs.\,\eqref{eq:Y1condition}, 
\eqref{eq:Y2 and Y3 condition} and \eqref{eq:Y6condition},
\begin{align}
\label{eq:Yr}
e^2(Y_1+\hat{Y}_1) =& 0\ , \\
e^2 \left(Y_2+\hat{Y}_2 -\frac{m_{{\chi}}}{\mathit{\Delta} m} Y_3\right) 
=& \frac{\alpha}{16\pi}
  \Bigg[5\left(\frac{1}{\bar{\epsilon}_\mathrm{ChPT}} + 
    \log\frac{\mu_\mathrm{ChPT}^2}{\mu_{\mathrm{FF}}^2}\right)
    -
   \frac{6 m_{{\chi}}}{\mathit{\Delta} m} \left(
   \frac{1}{\bar{\epsilon}_\mathrm{ChPT}} + 
   \log\frac{\mu_\mathrm{ChPT}^2}{m_{{\chi}}^2}+\frac{4}{3}
   \right)
   +3\log\frac{m_\chi^2}{M_V^2}
   \cr
&\hspace{30pt}
   +\frac{9}{2}
   - \frac{6 M_A^2 - 9 M_V^2}{M_A^2 - M_V^2}
   -3 \frac{M_V^4}{(M_A^2 -M_V^2)^2}
   \log\frac{M_A^2}{M_V^2}
   +\frac{4}{\mathit{\Delta}m}\frac{\pi M_AM_V}{M_A+M_V}    
     \Bigg]\cr
&-\frac{1}{4} e^2 (f_{\chi d}^r(\mu_{\mathrm{FF}})-f_{\chi \bar{u}}^r(\mu_{\mathrm{FF}})) \ ,\\
e^2\hat{Y}_6 =&   - \frac{\alpha}{4\pi}
        \frac{1}{\bar{\epsilon}_{\mathrm{ChPT}}}
    -2e^2f^{r}_{\chi\chi}(\mu_{\mathrm{FF}}) + \frac{\alpha}{4\pi} \log \frac{\mu_\mathrm{FF}^2}{\mu_\mathrm{ChPT}^2} \ , 
\end{align}
where $f^r(\mu_{\mathrm{FF}})$'s denote the finite parts of the counterterms in the $\overline{\mathrm{MS}}$ scheme.
The UV divergences of 
$K_{1,5}$ are given in
Ref.\,\cite{Urech:1994hd,Neufeld:1995mu};
\begin{align}
    &e^2K_{1} = -\frac{3}{8}\frac{\alpha}{4\pi}
    \frac{1}{\bar{\epsilon}_{\mathrm{ChPT}}} 
    + e^2K_1^r(\mu_\mathrm{ChPT})
   \ ,\ \\
    &e^2K_{5} = \frac{9}{8}\frac{\alpha}{4\pi}
    \frac{1}{\bar{\epsilon}_{\mathrm{ChPT}}} 
   + e^2K^r_5(\mu_\mathrm{ChPT})  \ , 
  \end{align}
where $K^r_{1,5}(\mu_\mathrm{ChPT})$s are the finite part in the $\overline{\mathrm{MS}}$ scheme.
The constants $e^2K_{2,6}$ are finite corrections of $\order{e^2}$, and hence,
$\mu_{\mathrm{ChPT}}$ independent.
We do not need explicit expressions of $K_{1,5}^r$ and $K_{2,6}$ as they are canceled when we take the ratio between the Wino and the pion decay rates.
We also use $K_{12}$ in Eq.\,\eqref{eq:K12}.

Altogether, we obtain the NLO decay rate of the charged Wino, 
\begin{align}
\label{eq:dGamchi}
  \frac{\delta\Gamma_\chi}
   {\Gamma_{\chi}}=&-\frac{\alpha M_AM_V}{\mathit{\Delta}m (M_A+M_V)}
   + \frac{\alpha}{16\pi} g_{{\chi}}\!\left(\frac{M_V}{M_A},\frac{\mathit{\Delta}m}{M_A}\right)+
    \frac{\alpha}{\pi} f_{{\chi}}\left(\frac{m_\pi}{\mathit{\Delta}m}\right)
    \cr
    &+ e^2(2 f_{\chi\chi}^r(\mu_{\mathrm{FF}}) + f_{\chi d}^r(\mu_{\mathrm{FF}})- f_{\chi\bar{u}}^r(\mu_{\mathrm{FF}}) + 2 f_{d\bar{u}}^r(\mu_{\mathrm{FF}})) \cr
    &+ \frac{8}{3}e^2(K_{1}^r(\mu_\mathrm{ChPT}) + K_2) + \frac{20}{9}e^2(K_5^r(\mu_\mathrm{ChPT})+K_6) \cr & + \frac{3\alpha}{8\pi} \log\frac{\mu_\mathrm{ChPT}^2}{M_V^2} 
    + \frac{3 \alpha}{4\pi} \log \frac{\mu_\mathrm{FF}^2}{\mu_\mathrm{ChPT}^2}
    + \frac{\alpha}{4\pi} \log
    \left(\frac{\mathit{\Delta}m^2 M_V^4}{m_\pi^6}\right)
    + \frac{2\alpha}{\pi} \log 2
    \ ,
\end{align}
where
\begin{align}
    g_{{\chi}}(\zeta,\eta) =& 
-\frac{3 \left(1+6 \zeta ^2-7 \zeta ^4+4 \left(\zeta ^2+3 \zeta ^4\right) \log (\zeta
   )\right)}{\left(1-\zeta^2\right)^2}-16 \log \left(\frac{4 \eta ^2}{\zeta ^2}\right)\ .
\end{align}
The combination of $f^r$'s in the second line is given by
\begin{align}
\label{eq:counterterms from EW}
     e^2(2f_{\chi\chi}^r(\mu_{\mathrm{FF}}) +& f_{\chi d}^r(\mu_{\mathrm{FF}})-f_{\chi \bar{u}}^r(\mu_{\mathrm{FF}})+ 2f_{d\bar{u}}^r(\mu_{\mathrm{FF}}))\cr
     =& \frac{\alpha}{2\pi}
  \left[-\frac{3}{2} \log\frac{\mu_{\mathrm{FF}}^2}{M_Z^2}
     +\left(
      \frac{3}{s_W^2} 
     - \frac{3}{s_W^4}
     \right)\log c_W
     -\frac{3}{s_W^2}
     -\frac{7}{4}
   \right] \cr
   &-\frac{\alpha(4 + 6c_W + c_W^2)}{2(1+c_W)s_W^2}\frac{M_W}{m_{{\chi}}} + \order{M_W^2/m_\chi^2}\ ,
\end{align}
which is determined by the matching condition in Eq.\,\eqref{eq:FFEWmatching}.
The full $M_W/m_\chi$ dependence of this combination of $f^r$'s
is given in Appendix\,\ref{sec:FullMW}.

The result Eq.\,\eqref{eq:dGamchi} is free from the UV and IR singularities.
The dependences on $\mu_{\mathrm{ChPT}}$ and $\mu_{\mathrm{FF}}$
are also cancelled by the running of the $f^r$'s and $K^r$'s. 
Besides, Eq.\,\eqref{eq:dGamchi} does not have $\order{\alpha\log m_\chi}$ enhanced contributions and depends on $m_\chi$ only through $\order{\alpha m_{\chi}^{s}}$ ($s \le 0$).
As a result, Eq.\,\eqref{eq:dGamchi} becomes constant in the limit of $m_\chi \to \infty$.
This shows that 
the decoupling theorem similar to the Appelquist-Carazzone theorem
holds in the Wino decay.

\subsection{Ratio Between Wino and Pion Decay Rates}
The radiative correction to the pion decay rate including the real photon emissions
is given by Ref.\,\cite{Marciano:1993sh, Decker:1994ea}\footnote{Note also that we need to correct $19\to 17$ and $13\to 11$ in 
Eq.\,(7b)
of Ref.\,\cite{Marciano:1993sh}.
The relation between the structure dependent constant $C_1$ in Ref.\,\cite{Marciano:1993sh}
and the constants $K$'s and $X$'s is given in Ref.\,\cite{Knecht:1999ag}.}
\begin{align}
\label{eq:GpiNLO}
  \frac{\delta\Gamma_\pi}
   {\Gamma_{\pi}} =& \frac{\alpha}{2\pi}
    \left[
    -\frac{3}{2}\left(\frac{1}{\bar{\epsilon}} + \log \frac{\bar{\mu}^2}{m_\pi^2}\right)
    +6\log \frac{m_\mu}{m_\pi} + \frac{11}{4}-\frac{2}{3}\pi^2 + f_\pi\left(\frac{m_\mu}{m_\pi}\right)
    \right]  \cr
    &
    + 
    e^2 
    \Bigg[\frac{8}{3} (K_1+K_2) + \frac{20}{9} (K_5+K_6)+4 K_{12}\cr &
    \hspace{1cm}-\hat{X}_6 -\frac{4}{3} (X_1+\hat{X}_1)
    -4 (X_2+ \hat{X}_2)
    +4  X_3
    \Bigg]\ , 
    \end{align}
where
    \begin{align}
        f_\pi(r)=&4 \left(\frac{1+r^2}{1-r^2}\log r - 1\right)\log(1-r^2) + 4 \frac{1+r^2}{1-r^2}
    \mathrm{Li}_2(r^2) \cr
& -\frac{r^2(8-5r^2)}{(1-r^2)^2}\log r - \frac{r^2}{1-r^2} \left(\frac{3}{2}+\frac{4}{3}\pi^2\right)\ .
\end{align}

In D\&M analysis, three 
arbitrary mass scales, $\mu$, $\mu_0$, $\mu_1$
are introduced, although final result does not depend on $\mu$'s.
In the following, we set 
$\mu = \mu_0=\mu_1=\bar{\mu}$.
The constants, $X$'s, are given by,
\begin{align}
    e^2(X_1+\hat{X}_1) =& \frac{3\alpha}{16\pi}
    \left(
    \log\frac{M_V^2}{M_Z^2}
    + 1-\frac{c_V}{2M_V^2}\right)\ , \\
    e^2(X_2 + \hat{X}_2)=&
    \frac{\alpha}{16\pi}
    \Bigg[\frac{5}{\bar{\epsilon}}+
    \frac{3M_V^2M_A^2}{(M_A^2-M_V^2)^2}
    \log\frac{M_A^2}{M_V^2}
    -\frac{3M_A^2}{M_A^2-M_V^2}
    +\frac{5}{2}
    \Bigg]\cr
    &-\frac{1}{4}e^2(g_{02}^{r(\mathrm{PV})}(\bar{\mu})-g_{03}^{r(\mathrm{PV})}(\bar{\mu}))\ ,
    \\
    e^2 X_3 =&
        \frac{3\alpha}{8\pi}
    \left(\frac{1}{\bar{\epsilon}}+\log\frac{\bar{\mu}^2}{M_V^2} 
    + \frac{M_V^2}{M_A^2-M_V^2}\log\frac{M_A^2}{M_V^2} -
    \frac{1}{6}\right)\ ,\\
    e^2 \hat{X}_6 =& - \frac{\alpha}{4\pi}
        \frac{1}{\bar{\epsilon}} -2 e^2g_{00}^{r(\mathrm{PV})}(\bar{\mu})+ \frac{\alpha}{8\pi}\ .
\end{align}
Here, $g^{r(\mathrm{PV})}$'s are determined by the subtraction scheme based on the Pauli-Villars regularization in D\&M analysis.
The constant $K_{12}$ in Eq.\,\eqref{eq:GpiNLO} is given by
\begin{align}
\label{eq:K12PV}
 e^2K_{12} =& -\frac{1}{8}
 \frac{\alpha}{4\pi}\frac{1}{\bar{\epsilon}}   +\frac{1}{2}g_{23}^{r(\mathrm{PV})}(\bar{\mu})\cr
 &+ \frac{1}{8}
 \frac{\alpha}{4\pi}\left(
 -3 \log \frac{\bar{\mu}^2}{M_V^2}
 + \frac{3(M_A^2 + M_V^2)M_V^2}{(M_A^2-M_V^2)^2}
 \log\frac{M_A^2}{M_V^2} - \frac{6 M_A^2}{M_A^2 - M_V^2}
 + \frac{1}{2}\right)
\ ,
\end{align}
which reproduces Eq.\,\eqref{eq:K12}
by substituting $e^2g_{23}^{r(\mathrm{PV})}(\bar{\mu})=e^2f_{d\bar{u}}^r(\bar{\mu}) +\alpha/16\pi$.
As a result, we obtain,
\begin{align}
      \frac{\delta \Gamma_{\pi}}{\Gamma_{\pi}}
    =& \frac{\alpha}{16\pi} g_{\pi}\!\left(\frac{M_V}{M_A}\right)+
    \frac{\alpha}{2\pi} 
    \left(6\log \frac{m_\mu}{m_\pi} + \frac{11}{4}-\frac{2}{3}\pi^2 +f_{\pi}\left( \frac{m_\mu}{m_\pi}\right)
    \right)
    \cr
    &+ e^2(2 g_{00}^{r(\mathrm{PV})}(\bar{\mu}) + g_{02}^{r(\mathrm{PV})}(\bar{\mu})-g_{03}^{r(\mathrm{PV})}(\bar{\mu}) + 2 g_{23}^{r(\mathrm{PV})}(\bar{\mu})) \cr
    &+ \frac{8}{3}e^2(K_{1}^r(\bar{\mu}) + K_2) + \frac{20}{9}e^2(K_5^r(\bar{\mu})+K_6) \cr & + \frac{3\alpha}{8\pi} \log\frac{\bar{\mu}^2}{M_V^2} 
    -\frac{3\alpha}{4\pi} \log \frac{M_V^2}{m_\pi^2}
    + \frac{\alpha}{4\pi} \log
    \frac{M_Z^2}{M_V^2}
    \ ,
\end{align}
where
\begin{align}
    g_{\pi}(\zeta) = 
-19-\frac{36 \zeta ^2 \log \zeta }{1-\zeta ^2}+2
\hat{c}_V\ ,
\end{align}
with $\hat{c}_V= c_V/M_V^2$.
The combination of the counterterms is given by D\&M,
\begin{align}
\label{eq:pion counterterm}
    e^2(2 g_{00}^{r(\mathrm{PV})}(\bar{\mu}) + g_{02}^{r(\mathrm{PV})}(\bar{\mu})- g_{03}^{r(\mathrm{PV})}(\bar{\mu}) + 2
    g_{23}^{r(\mathrm{PV})}(\bar{\mu})) = \frac{3\alpha}{4\pi}
    \log\frac{M_Z^2}{\bar{\mu}^2}\ .
\end{align}
Indeed, with the $\bar{\mu}$ dependence of the constants $g^r$'s and $K$'s, we confirm that $\delta \Gamma_\pi/\Gamma_\pi$ does not depend on $\bar{\mu}$.
The total $M_Z$ dependence of $\delta \Gamma_\pi/\Gamma_\pi$ reproduces the logarithmically enhanced $\log M_Z$ term in Ref.\,\cite{Marciano:1993sh}.

Finally, let us take the ratio between the pion decay rate and the Wino decay rate at the NLO,
\begin{align}
\label{eq:ratio_summary}
    \frac{\Gamma(\chi^-\to \chi^0 + \pi^- (+ \gamma))}{\Gamma(\pi^-\to \mu^- + \nu_\mu (+ \gamma)) }
    =   
    \frac{\Gamma_{\chi}}{\Gamma_{\pi}}
    \times\Bigg(
    1 + \frac{\delta\Gamma_{{\chi}}}{\Gamma_{{\chi}}}- \frac{\delta\Gamma_{\pi}}{\Gamma_{\pi}}
    \Bigg)\ .
\end{align}
The ratio at the tree-level decay rates $\Gamma_\chi$ and $\Gamma_\pi$ is given in Eq.\,\eqref{eq:treeRatio}.
The difference of the radiative correction is, on the other hand,
given by,
\begin{align}
\label{eq:analyticNLO}
     \frac{\delta\Gamma_{{\chi}}}{\Gamma_{{\chi}}}- \frac{\delta\Gamma_{\pi}}{\Gamma_{\pi}}=& -\frac{\alpha M_AM_V}{\mathit{\Delta}m (M_A+M_V)}
   + \frac{\alpha}{16\pi} g_{{\chi}}\left(\frac{M_V}{M_A},\frac{\mathit{\Delta}m}{M_A}\right)
    -\frac{\alpha}{16\pi} g_\pi\left(\frac{M_V}{M_A}\right)\cr
    &+
    \frac{\alpha}{\pi}
    f_{{\chi}}\left(\frac{m_\pi}{\mathit{\Delta}m}\right) 
    - \frac{\alpha}{2\pi} 
    \left(6\log \frac{m_\mu}{m_\pi} + \frac{11}{4}-4\log2-\frac{2}{3}\pi^2 +f_{\pi}\left(\frac{m_\mu}{m_\pi}\right)
    \right)
    \cr
    & 
    +\frac{\alpha}{2\pi}
  \left(
     \left(
      \frac{3}{s_W^2} 
     - \frac{3}{s_W^4}
     \right)\log c_W
     -\frac{3}{s_W^2}
     -\frac{7}{4}
   \right)-\frac{\alpha(4 + 6 c_W+ c_W^2)}{2(1+c_W)s_W^2}\frac{M_W}{m_{{\chi}}}\cr
     &-\frac{\alpha}{4\pi} \log\frac{M_Z^2}{M_V^2}+\frac{\alpha}{4\pi}
     \log
     \left(\frac{\mathit{\Delta}m^2 M_V^{10}}{m_\pi^{12}}\right) 
     \ ,
\end{align}
where we have used Eqs.\,\eqref{eq:counterterms from EW} and \eqref{eq:pion counterterm}
with $\mu_{\mathrm{ChPT}}=\bar{\mu}$.
We have neglected the terms higher than  $\order{M_W^2/m_\chi^2}$.
The full $M_W^2/m_\chi^2$ dependence
appearing through $f^r$'s can be 
found in Appendix\,\ref{sec:FullMW}.

\subsection{Estimation of Error from Hadron Model}
The largest contribution to the NLO decay rate turns out to be%
\footnote{Note that 
$e^2 Y_L$ is always much smaller than $\order{1}$ since we assume that $\mathit{\Delta m}$ is larger than $m_\pi$ for kinematical reasons while 
$m_\pi^2-m_{\pi^0}^2=\order{\alpha M_V^2}$  \cite{Das:1967ek} with $m_{\pi^0}$ being the neutral pion mass.}
\begin{align}
e^2Y_L \equiv
\left.e^2\left(Y_2+\hat{Y}_2-
        \frac{m_\chi}{\mathit{\Delta}m} Y_3\right)\right|_{\mathrm{lead}} &=  \frac{\alpha}{4} \frac{M_A M_V}{\mathit{\Delta}m(M_A+M_V)}\ .
\end{align}
This expression is obtained by the phenomenological hadron model, i.e., the MRM.
In this subsection, we discuss uncertainties originate from the hadron model.

For this purpose, let us first note that the leading contribution
is obtained from the current correlator in the limit of $r\to 0$ and $\mathit{\Delta}m \to 0$,
\begin{align}
e^2 Y_L\mathit{\Delta}m\times\bar{u}_{0}(p)u_{-}(p)
&= \frac{ie^2}{2 F_0}  
\int \frac{d^4\ell}{(2\pi)^4}
 \Gamma_{VA}^{\mu\nu}(\ell,0)
 \times \frac{-i}{\ell^2 - m_\gamma^2} 
  \bar{u}_{0}(p)\gamma_\nu 
      \frac{i(\slashed{p} - \slashed{\ell} + m_{{\chi}})}{(p-\ell)^2-m_{{\chi}}^2}
      \gamma_\mu
      u_{-}(p) \ .
      \end{align}
The correlator $\Gamma_{VA}^{\mu\nu}(\ell,r)$ at $r=0$ is,
on the other hand, related to another current correlator, $\Pi_{VV-AA}^{\mu\nu}(\ell)$, via~\cite{Das:1967ek},%
\begin{align}
     \Gamma_{VA}^{\mu\nu}(\ell,0) =\frac{1}{F_0}\Pi_{VV-AA}^{\mu\nu}(\ell)\ .
\end{align}
Here, the correlator $\Pi_{VV-AA}^{\mu\nu}(k)$ is defined by,
\begin{align}
   \Pi_{VV-AA}^{\mu\nu}(k)=&
    i \int d^4x e^{-ik x} 
      \langle 0|T
     J_{V\mu}^1(x)
     J_{V\nu}^1(0)-
     J_{A\mu}^1(x)
     J_{A\nu}^1(0)
     |0\rangle \\
     =&F_0^2 (k_\mu k_\nu-k^2 g_{\mu\nu}) \Pi_{VV-AA}(k)\ .
\end{align}
In terms of $\Pi_{VV-AA}(k)$, the leading contribution is rewritten as,
  \begin{align}
  e^2 Y_L\mathit{\Delta}m\times\bar{u}_{0}(p)u_{-}(p) &= \frac{ie^2}{2 F_0^2}  
\int \frac{d^4\ell}{(2\pi)^4}
 \Pi_{VV-AA}^{\mu\nu}(\ell)
 \times \frac{-i}{\ell^2 - m_\gamma^2} 
  \bar{u}_{0}(p)\gamma_\nu 
      \frac{i(\slashed{p} - \slashed{\ell} + m_{{\chi}})}{(p-\ell)^2-m_{{\chi}}^2}
      \gamma_\mu
      u_{-}(p)  \ .
\end{align}

In the MRM, 
$\Pi_{VV-AA}$ is given by
\begin{align}
    \Pi_{VV-AA}(\ell)\simeq \frac{M_A^2 M_V^2}{\ell^2(\ell^2 - M_V^2)(\ell^2 - M_A^2)}\ .
\end{align}
By comparing with the lattice simulation in Ref.\,\cite{Boyle:2009xi},
we find that the MRM well fits the lattice estimation of $\Pi_{VV-AA}(\ell)$ for $|\ell^2| \lesssim \Lambda_{\mathrm{cut}}^2 \simeq 5$\,GeV$^2$ by taking,
\begin{align}
    M_V = 0.6\,\mathrm{GeV}\mbox{--}0.8\,\mathrm{GeV}\ ,
\end{align}
with the assumption $M_A^2/M_V^2 = 2$.%
\footnote{The parameters $M_{V,A}$ should be taken as the fit parameter for $\Pi_{VV-AA}$ instead of the physical masses of the corresponding (pseudo-)vector mesons.
The function $\Pi^{(1)}$ in Ref.\,\cite{Boyle:2009xi} is normalized as $\Pi^{(1)}=\Pi_{VV-AA}\times (2F_\pi^2)$.
}

The contribution to 
$e^2 Y_L$ from
the larger loop momentum,
$|\ell^2| > \Lambda_{\mathrm{cut}}^2 \simeq 5$\,GeV$^2$ is, on the other hand, estimated to be,
\begin{align}
   e^2 Y_L|_{|\ell^2|>\Lambda_{\mathrm{cut}}^2} &\simeq  \frac{\alpha}{4} \frac{M_A^2 M_V^2}{2\mathit{\Delta}m\Lambda_{\mathrm{cut}}^3} \cr
    &\simeq \frac{M_AM_V(M_A+M_V)}
    {2\Lambda_{\mathrm{cut}}^3}\times e^2 Y_L\ \cr
    &\simeq 6.5\times 10^{-2} \times e^2 Y_L\ . 
\end{align}
Thus, the errors caused by the contributions from $|\ell^2| > \Lambda_{\mathrm{cut}}^2$ is minor.

From these arguments, we estimate the uncertainty of the leading hadronic contribution 
by varying $M_V =0.6$--$0.8$\,GeV.
For the other contributions obtained by the MRM, $g_\chi$ and $g_\pi$ in Eq.\,\eqref{eq:analyticNLO}, 
we put $\pm 50$\% following D\&M, although their contributions are subdominant.

Several comments are in order.
In our analysis, we have taken the chiral limit to derive the matching conditions between the Four-Fermi theory and the ChPT.
The effects of the pion mass 
to the matching conditions are expected to be of $\order{m_\pi^2/M_V^2}$, which is minor than the uncertainties of our estimate discussed above.
We also note that there are mixed QED and QCD corrections of $\order{\alpha\alpha_s}$
at the two-loop level,
where $\alpha_s$ is the QCD coupling.
Those corrections are, however, negligibly small~\cite{Marciano:1993sh}, and hence, we do not take into account.

\subsection{Numerical Estimate}
Let us move on to the numerical estimate of the NLO decay rate.
The numerical values of the terms in Eq.\,\eqref{eq:analyticNLO} are given by,
\begin{align}
\label{eq:fchiexpand}
    &\frac{\alpha}{\pi}f_{{\chi}}
    \left(\frac{m_\pi}{\mathit{\Delta}m}\right)  \simeq -0.64\times 10^{-2} 
    -0.84\times10^{-2}\times \log \left(\frac{\mathit{\Delta}m}{164\,\mathrm{MeV}}\right)+0.78\times 10^{-2}\times\log^2 \left(\frac{\mathit{\Delta}m}{164\,\mathrm{MeV}}\right)\ ,    \\&- \frac{\alpha}{2\pi} 
    \left(6\log \frac{m_\mu}{m_\pi} + \frac{11}{4} -4\log 2-\frac{2}{3}\pi^2 +f_{\pi}\left(\frac{m_\mu}{m_\pi}\right)
    \right) \simeq 0.75\times 10^{-2}\ , \\
&\frac{\alpha}{2\pi}
  \left[
    \left(
      \frac{3}{s_W^2} 
     - \frac{3}{s_W^4}
     \right)\log c_W
     -\frac{3}{s_W^2}
     -\frac{7}{4}
   \right] -\frac{\alpha(4 + 6 c_W+ c_W^2)}{2(1+c_W)s_W^2}\frac{M_W}{m_{{\chi}}} \cr
&\hspace{6.5cm}\simeq -1.07\times 10^{-2} - 0.70\times 10^{-2}\times \left(\frac{1\,\mathrm{TeV}}{m_{{\chi}}}\right)\ , \\
\label{eq:MZdependence}
   &-\frac{\alpha}{4\pi} \log\frac{M_Z^2}{M_V^2}+\frac{\alpha}{4\pi}
     \log
     \left(\frac{\mathit{\Delta}m^2 M_V^{10}}{m_\pi^{12}}\right)
 \simeq (0.39\pm 0.1)\times 10^{-2} + 0.12\times 10^{-2}\times \log
     \left(\frac{\mathit{\Delta}m}{164\,\mathrm{MeV}}\right)\ ,\\
&
\label{eq:leading}
-\frac{\alpha M_AM_V}{\mathit{\Delta}m (M_A+M_V)} \simeq -(1.8\pm 0.3) \times10^{-2}\times \left(\frac{164\,\mathrm{MeV}}{\mathit{\Delta}m}\right)\ , \\
&\frac{\alpha}{16\pi} g_{{\chi}}\left(\frac{M_V}{M_A},\frac{\mathit{\Delta}m}{M_A}\right)
    - \frac{\alpha}{16\pi}g_\pi\left(\frac{M_V}{M_A}\right)
    \simeq 
    (0.23\pm 0.11)\times 10^{-2}  - (0.46\pm 0.23)\times 10^{-2}
    \log
    \left(
    \frac
    {\mathit{\Delta}m}{164\,{\mathrm{MeV}}}\right)\ .\cr
\end{align}
Note that we expand the $\mathit{\Delta}m$ dependences 
around $\log(\mathit{\Delta}m/164\,\mathrm{MeV}) = 0$. 
The errors in Eqs.\,\eqref{eq:MZdependence} and 
\eqref{eq:leading} are caused by
 the choice of $M_V=0.6\mbox{--}0.8$\,GeV.
We also put $\pm 50$\% to the hadron model contributions, $g_\chi$ and $g_\pi$ 
as mentioned in the previous section. 
For the estimate of $\hat{c}_V$, we have used $F_\pi$ instead of $F_0$ in Eq.\,\eqref{eq:cV}.
Note that the errors of $\alpha$, $m_\pi$, $m_\mu$, $M_Z$, and $M_W$ are negligible at the accuracy of the current analysis.

Combining all the contributions, we obtain the radiative correction to the ratio of the decay rates as,
\begin{align}
\label{eq:numericalNLO}
     \frac{\delta\Gamma_{{\chi}}}{\Gamma_{{\chi}}}- \frac{\delta\Gamma_{\pi}}{\Gamma_{\pi}}=&(-0.59\pm 0.1)\times 10^{-2}
     - 0.70\times10^{-2}\times
     \left(\frac{1\,\mathrm{TeV}}{m_{{\chi}}}\right)\cr
     & -0.72\times10^{-2}\times \log \left(\frac{\mathit{\Delta}m}{164\,\mathrm{MeV}}\right)\cr
     &+0.78\times 10^{-2}\times\log^2 \left(\frac{\mathit{\Delta}m}{164\,\mathrm{MeV}}\right)\cr
     & -(1.8 \pm 0.3)\times 10^{-2}
     \times \left(\frac{164\,\mathrm{MeV}}{\mathit{\Delta}m}\right)\cr
     & + (0.23\pm 0.11) \times 10^{-2}-(0.5\pm 0.2)\times 10^{-2}
     \times
     \log \left(\frac{\mathit{\Delta}m}{164\,\mathrm{MeV}}\right)
     \ .
\end{align}

\begin{figure}[t]
    \centering
    \includegraphics[width=0.6\linewidth]{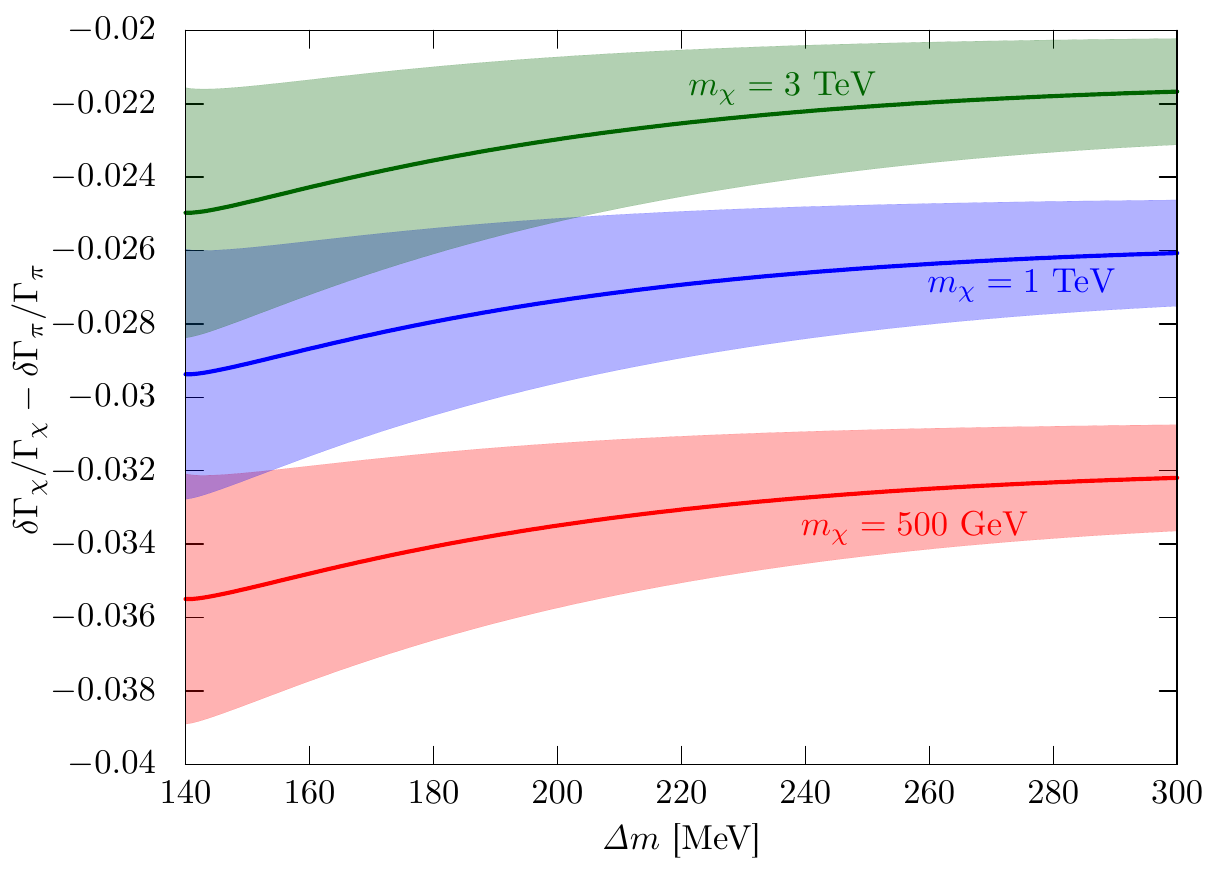}
    \caption{ The radiative correction to the ratio of the decay rate, ${\delta\Gamma_{{\chi}}}/{\Gamma_{{\chi}}}- {\delta\Gamma_{\pi}}/{\Gamma_{\pi}}$,
    as a function of $\mathit{\Delta}m$.
Each band corresponds to the error of 
NLO analysis which is dominated by 
the uncertainties of the MRM.
For the pure Wino case, the mass differences ranges in $\mathit{\Delta}m \simeq 163$--$165$\,MeV for $m_\chi >600$\,GeV~\cite{Ibe:2012sx}.
}
    \label{fig:dGamma}
\end{figure}

In Fig.\,\ref{fig:dGamma}, we show the radiative correction to the ratio of the decay rate as a function of $\mathit{\Delta}m$ for given values of $m_\chi$.
In the figure, we use the full expression of $f_\chi(z)$ in Eq.\,\eqref{eq:fchi}.
We also take into account the full $M_W/m_\chi$
dependence of $f^r$'s.
The figure shows that the one-loop radiative correction 
to the ratio of the decay rates is 
about 
$-(2\mbox{--}4)$\% for $\mathit{\Delta}m \simeq 164$\,MeV.
The error bands are dominated by the uncertainty of the MRM.

By using the charged pion lifetime, the branching fraction of $\pi^\pm \to \mu^\pm + \nu  (+\gamma)$,
and the two-loop estimation of $\mathit{\Delta}m$ as a function 
of $m_\chi$ (see Ref.\,\cite{Ibe:2012sx}), 
we can now make a prediction of the Wino lifetime at the $\order{1}$\% precision.
In Fig.\,\ref{fig:lifetime}, we show the 
Wino decay length as a function of the Wino mass.
In the figure, we show the central value of our estimation in a black solid line.
We also show the tree-level decay length in a black dashed line.
The blue bands show the uncertainty  from the hadronic model.
The red bands show the uncertainty of the prediction of the Wino mass difference, $\mathit{\Delta}m$, from the three-loop correction,
$\delta\mathit{\Delta}m = \pm 0.3\, \mathrm{MeV}$.
In the figure, we have included the three body decay modes 
$\chi^- \to \chi^0 + \ell + \bar{\nu}_\ell$,
\begin{align}
    \tau_{\chi}^{-1} \equiv& \Gamma(\chi^-\to \chi^0 + 
    \pi^- + (\gamma)) + \Gamma_e + \Gamma_\mu  \cr
    =&\frac{\Gamma(\chi^-\to \chi^0 + \pi^- + (\gamma))}{\Gamma(\pi^-\to \mu^-+\bar{\nu}_\mu + (\gamma))} \times B(\pi^\pm \to \mu^\pm + \nu(+\gamma))\times\tau_\pi^{-1} + \Gamma_e + \Gamma_\mu \ ,
\end{align}
where $\Gamma_\ell$ denotes the Wino decay rate into $\chi^0+\ell+\bar{\nu}_\ell$ in Eq.\,\eqref{eq:3body}.

\section{Conclusions and Discussions}
\label{sec:conclusions}

\begin{figure}[t]
    \centering
    \includegraphics[width=0.6\linewidth]{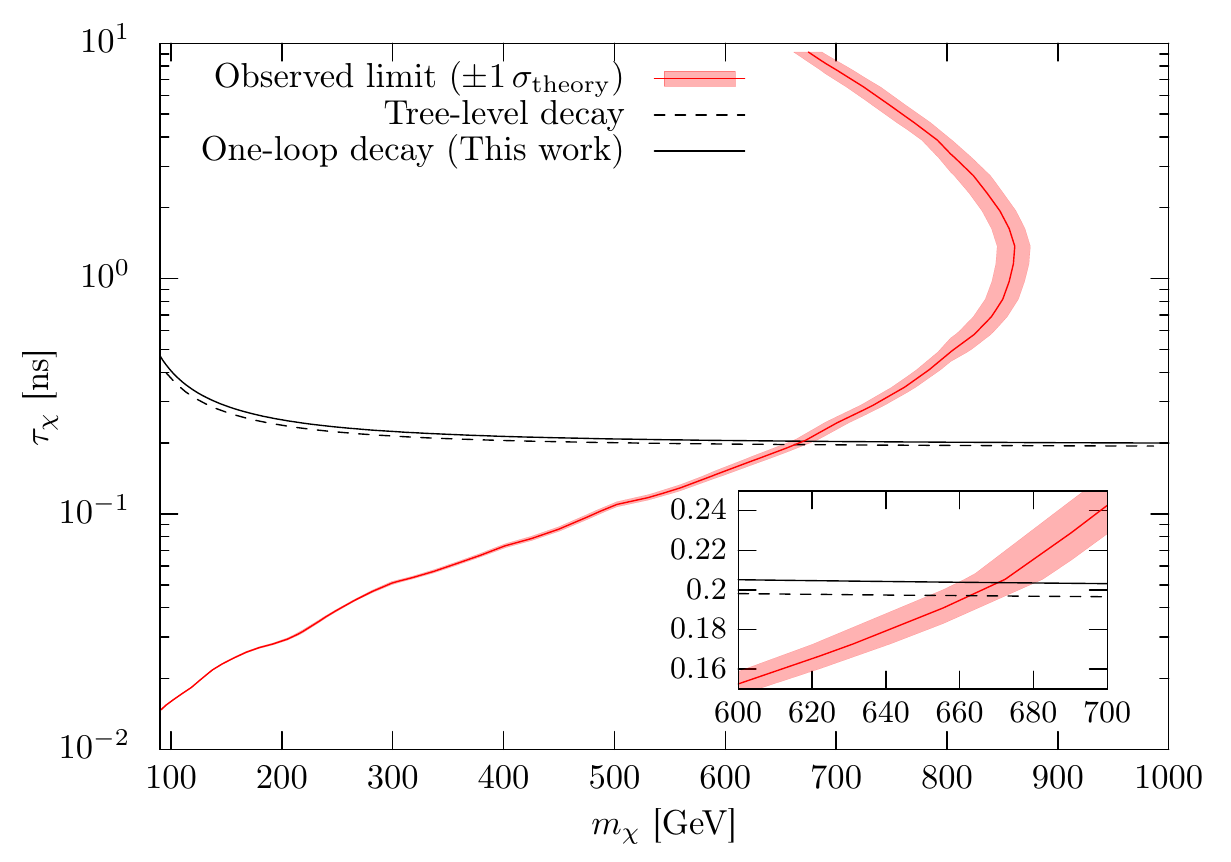}
    \caption{
    The LHC constraint on the charged Wino mass and lifetime based on the disappearing charged track search.
    The red solid line shows the 95\% CL limit by the ATLAS with data of $\int  \mathcal{L} dt =  136\,\mathrm{fb}^{-1}$ and $\sqrt{s} = 13$\,TeV\,\cite{ATLAS:2022rme}.
    The red band shows $\pm 1\,\sigma$ uncertainty of the production cross section of the Wino.
    The black solid line is the prediction with the two-loop mass difference and the one-loop decay rate, whereas the dashed line shows the result of the tree-level calculation. 
}
    \label{fig:ATLAS}
\end{figure}

In this paper, 
we have computed the NLO correction of the charged Wino decay and made the most precise estimate of the lifetime of the charged Wino.
In our analysis, we have constructed the ChPT which includes the Winos and QED.
By matching the ChPT to the electroweak theory, we have derived the NLO decay rate free from the UV divergence.
We have also taken into account the real photon emission, so that the NLO rate is free from the IR divergence.
As a result, we found the NLO correction gives a minor impact on the lifetime of 2--4\% increase.
The effect on search of the Wino at the LHC is also minor, with only a 5--10 GeV increase in the pure Wino mass limit as shown in Fig.\,\ref{fig:ATLAS}.

We have also confirmed that a decoupling theorem similar to the Appelquist-Carazzone theorem holds for the Wino decay at the one-loop level.
The radiative corrections depend on the Wino mass only through $\order{\alpha m_{\chi}^{s}}$  with $s \le 0$ and there is no logarithmically enhanced dependences on $m_\chi$. 
The decay rate becomes constant in the limit of $m_\chi \to \infty$.
This result is non-trivial 
since the Appelquist--Carazzone decoupling theorem is not applicable to the decay of the Wino, where the external lines of the diagrams include heavy particles.
Besides, we have explicitly found
that
the $\mathit{\Delta}m/m_{\chi}$ enhanced NLO contributions appear in Eq.\,\eqref{eq:WinoVirtual}.
Such enhanced effects can only be cancelled by preparing counterterms matched to the electroweak theory.
It is not clear whether this cancellation takes place or not at the higher-loop order.
We will study this aspect in more general setup in future.

\section*{Acknowledgements}
This work is supported by Grant-in-Aid for Scientific Research from the Ministry of Education, Culture, Sports, Science, and Technology (MEXT), Japan, 18H05542, 21H04471, 22K03615 (M.I.), 18K13535, 20H01895, 20H05860 and 21H00067 (S.S.) and by World Premier International Research Center Initiative (WPI), MEXT, Japan. This work is also supported by the JSPS Research Fellowships for Young Scientists (Y.N.) and International Graduate Program for Excellence in Earth-Space Science (Y.N.).
\appendix 

\section{Gordon Equations}
\label{sec:GordonEquations}
Let us consider the spinor wave functions $u_1(p_1)$ and $u_2(p_2)$ with real valued masses $m_1$ and $m_2$, respectively.
They satisfy
\begin{align}
    &(\slashed{p}_i-m_i)u_i(p_i) = 0 \ , \\
    &\bar{u}_i(p_i)(\slashed{p}_i-m_i) = 0 \ ,
\end{align}
where $\bar{u}(p)= u^\dagger(p)\gamma^0$ and we have used $\gamma^{\mu\,\dagger} = \gamma^0 \gamma^\mu \gamma^0$ and  $(\gamma^{0})^2=1$.
We find
\begin{align}
    &\bar{u}_2(p_2)\gamma^\mu (m_1 u_1(p_1)) = \bar{u}_2(p_2) \gamma_\mu\slashed{p}_1 u_1(p_1) \ ,\\
   &(\bar{u}_2(p_2)m_2)\gamma^\mu u_1(p_1) = \bar{u}_2(p_2) \slashed{p}_2\gamma_\mu u_{1}(p_1) \ ,
\end{align}
and hence,
\begin{align}
    (m_1 + m_2)\bar{u}_2(p_2)\gamma^\mu u_1(p_1) &= \bar{u}_2(p_2) (\gamma^\mu\slashed{p}_1+\slashed{p}_2\gamma^\mu
    ) u_1(p_1) \notag \\
    & = \bar{u}_2(p_2) \left((p_1+p_2)^\mu
    + \frac{1}{2}(p_1-p_2)_\rho[\gamma^\mu,\gamma^\rho]\right) u_1(p_1) \notag \\
    &= \bar{u}_2(p_2) \left((p_1+p_2)^\mu
- i \sigma^{\mu\rho}(p_1-p_2)_\rho
\right) u_1(p_1)\ .
\end{align}
Here, $\sigma^{\mu\nu}=i[\gamma^\mu,\gamma^\nu]/2$.
As a result, we obtain the Gordon identity,
\begin{align}
    \bar{u}_2(p_2)\gamma^\mu u_1(p_1)
    = \bar{u}_2(p_2) \left(\frac{(p_2+p_1)^\mu}{m_1+m_2}
+ i \sigma^{\mu\rho}\frac{(p_2-p_1)_\rho}{m_1+m_2}
\right) u_1(p_1)\ .
\end{align}

Similarly, we find 
\begin{align}
  &\bar{u}_2(p_2)\gamma^\mu\gamma_5 (m_1  u_1(p_1)) = \bar{u}_2(p_2) \gamma_\mu\gamma_5\slashed{p}_1 u_1(p_1) = - \bar{u}_2(p_2) \gamma_\mu\slashed{p}_1 \gamma_5 u_1(p_1) \ ,\\
   &(\bar{u}_2(p_2)m_2)\gamma^\mu \gamma_5 u_1(p_1) = \bar{u}_2(p_2) \slashed{p}_2\gamma_\mu\gamma_5 u_{1}(p_1) \ .
\end{align}
Thus, we obtain,
\begin{align}
    \bar{u}_2(p_2)\gamma^\mu
    \gamma_5 u_1(p_1)
    &= \bar{u}_2(p_2) \left(\frac{(p_2-p_1)^\mu}{m_1+m_2}\gamma_5
+ i \sigma^{\mu\rho}\gamma_5\frac{(p_2+p_1)_\rho}{m_1+m_2}
\right) u_1(p_1)\ \notag \\
&= \bar{u}_2(p_2) \left(\frac{(p_2+p_1)^\mu}{m_2-m_1}\gamma_5
- i \sigma^{\mu\rho}\gamma_5\frac{(p_2-p_1)_\rho}{m_2-m_1}
\right) u_1(p_1)\ .
\end{align}
In the limit of $p_1=p_2$
and $m_1 = m_2$,
the Gordon equations are reduced to
\begin{align}
    &p^\mu\bar{u}_2(p)u_{1}(p) = m  \bar{u}_{2}(p)\gamma^\mu u_{1}(p)\ , \\
     &p^\mu\bar{u}_2(p)\gamma_5 u_{1}(p) = 0\ . 
\end{align}

\section{Full 
Expression of Eq.\texorpdfstring{\,\eqref{eq:counterterms from EW}}{}}\label{sec:FullMW}
In Eq.\,\eqref{eq:counterterms from EW},
we only showed the matching condition to the $\order{M_W/m_\chi}$ terms.
In Fig.\,\ref{fig:lifetime} and Fig.\,\ref{fig:dGamma}, 
we keep the full $M_W/m_\chi$ dependence at one-loop level;
\begin{align}
\label{eq:FullMW_mod}
      e^2&(2f_{\chi\chi}^r(\mu_{\mathrm{FF}}) + f_{\chi d}^r(\mu_{\mathrm{FF}})-f_{\chi \bar{u}}^r(\mu_{\mathrm{FF}})+ 2f_{d\bar{u}}^r(\mu_{\mathrm{FF}}))= -\frac{3\alpha}{4\pi}\log\frac{\mu_{\mathrm{FF}}^2}{M_Z^2} \cr
      &+\frac{\alpha}{8\pi}
      \Bigg\{
     \frac{-7 s_W^2+18 z_W^2-12}{s_W^2}+\frac{2 z_W \left[8+2 z_W^2-z_W^4+8 s_W^2 \left(-5-2
   z_W^2+z_W^4\right)\right]\cot
   ^{-1}\frac{z_W}{\sqrt{4-z_W^2}}}{s_W^4 \sqrt{4-z_W^2}}
      \cr
      &\hspace{1cm}+\frac{2}{s_W^4 c_W^2}\times
   \Big[-\left(s_W^2+1\right) z_W
   \sqrt{4 c_W^2-z_W^2} \left(2
   c_W^2+z_W^2\right) \cot
   ^{-1}\frac{z_W}{\sqrt{4
   c_W^2-z_W^2}}  \cr  &\hspace{1cm} +2
   s_W^2 \left(4 s_W^2-5\right) z_W^4 \log
   z_W +\left(\left(s_W^2+1\right)
   z_W^4-6 c_W^4\right)  \log c_W  \Big]
   \Bigg\} \, ,
\end{align}
where $z_W = M_W/m_\chi < 1$.

\section{Quark Wave Function Renormalization}
\label{sec:quark counterterm}
To see the equivalence between
the quark counterterms in Eqs.\,\eqref{eq:quark counterterms} and \eqref{eq:kinetic counterterm}, let us rewrite the kinetic term of the quark,
\begin{align}
    \mathcal{L} = \frac{1}{2}
\left(\bar{\psi}_L i \overleftrightarrow{\slashed{D}}\psi_L
+(L\leftrightarrow R)
\right)\ ,
\end{align}
and consider field redefinition,
\begin{align}
&\psi_{L} =  \psi_L' + e^2 f_{d\bar{u}} \mathbf{q}_L^2 \psi_L'  \ , \\
&\bar{\psi}_{L} =  \bar{\psi}_L' + e^2 f_{d\bar{u}}\bar{\psi}_L' \mathbf{q}_L^2 \ , 
\end{align}
where we have assumed $\mathbf{q}_L^\dagger = \mathbf{q}_L$.
With the field redefinition, the kinetic term leads to
\begin{align}
\label{eq:spurion counterterm}
    \mathcal{L}|_{L,\order{e^2}} = 
    &\frac{i}{2}\bigg[e^2 f_{d\bar{u}} \bar{\psi}_L' \mathbf{q}_L^2 \gamma^\mu D_\mu \psi_L' +   e^2 f_{d\bar{u}} \bar{\psi}_L'
    \mathbf{q}_L \gamma^\mu D_{\mu}(\mathbf{q}_L \psi_L')
    + e^2 f_{d\bar{u}} \bar{\psi}_L'\gamma^\mu (D_{\mu}\mathbf{q}_L) \mathbf{q}_L\psi_L' \cr
    &-e^2 f_{d\bar{u}} (D_\mu\bar{\psi}_L') \mathbf{q}_L^2 \gamma^\mu \psi_L' -   e^2 f_{d\bar{u}} D_\mu(\bar{\psi}_L'
    \mathbf{q}_L) \gamma^\mu \mathbf{q}_L \psi_L'
    - e^2 f_{d\bar{u}}\bar{\psi}_L'\mathbf{q}_L(D_\mu\mathbf{q}_L) \gamma^\mu \psi_L' \bigg] \cr
    =& ie^2 f_{d\bar{u}} \bar{\psi}_L'
    \mathbf{q}_L \gamma^\mu D_{\mu}(\mathbf{q}_L \psi_L')
    +i e^2 f_{d\bar{u}}\bar{\psi}_L' \gamma^\mu (D_{\mu}\mathbf{q}_L) \mathbf{q}_L \psi_L' \cr
    & -  ie^2 f_{d\bar{u}} D_\mu(\bar{\psi}_L'
    \mathbf{q}_L) \gamma^\mu \mathbf{q}_L \psi_L'
    -ie^2 f_{d\bar{u}}\bar{\psi}_L' \mathbf{q}_L(D_\mu\mathbf{q}_L) \gamma^\mu \psi_L'\ \cr
    = & ie^2 f_{d\bar{u}} \bar{\psi}_L'\mathbf{q}_L \overleftrightarrow{\slashed{D}} \mathbf{q}_L \psi_L'
    - i e^2 f_{d\bar{u}}\bar{\psi}_L' [\mathbf{q}_L,D_\mu \mathbf{q}_L]\gamma^\mu \psi'_L
    \ .
\end{align}
In the second equality, we have used the partial integration to remove the terms proportional to $\mathbf{q}_L^2$.
As a result, the counterterm in Eqs.\,\eqref{eq:quark counterterms} 
can be transformed into the one in Eq.\,\eqref{eq:kinetic counterterm}
by renaming $\psi_L'$ to $\psi_L$
to the order of $\order{e^2}$.

\section{Wino-Pion Interaction}
\label{sec:Wino-Pion Interaction}
The pion weak interactions with leptons can be 
rewritten 
as Yukawa interactions~\cite{Kinoshita:1959ha}.
Here, we show that the Wino--pion interaction in Eq.\,\eqref{eq:TreeLevel} is also equivalent to the Yukawa interaction 
\begin{align}
\label{eq:WinoPionYukawa}
    \mathcal{L} &=2 \sqrt{2}V_{ud}F_0 G_F (\hat{m}_\chi
    -m_0) (i\pi^- \bar{\psi}_- \psi_0 - i\pi^+\bar{\psi}_0 \psi_-) \\
    &=2 \sqrt{2}V_{ud}F_0 G_F (\mathit{\Delta}{m} + \delta m_\chi) (i\pi^- \bar{\psi}_- \psi_0 - i\pi^+\bar{\psi}_0 \psi_-) \ ,
\end{align}
where we defined $\hat{m}_\chi \equiv m_\chi + \delta m_\chi$.
The physical charged and the neutral Wino masses are given by $m_{\chi}$ and $m_{0}$, respectively, while
$\delta m_\chi$ is the Wino mass counterterm.
By noting that the QED correction to the charged Wino self-energy at $\slashed{p}=m_{\chi}$ is given by,
\begin{align}
    \Sigma_{-}^{\mathrm{(ChPT)}}(m_{\chi}) =  \frac{3\alpha}{4\pi}m_\chi
    \left(\frac{1}{\bar{\epsilon}_{\mathrm{ChPT}}} + \log \frac{\mu_{\mathrm{ChPT}}^2}{m_{\chi}^2} +
    \frac{4}{3} \right)\ , 
\end{align}
we find 
\begin{align}
\label{eq:delta mchi}
    \delta m_\chi =  -\frac{3\alpha}{4\pi}m_\chi
    \left(\frac{1}{\bar{\epsilon}_{\mathrm{ChPT}}} + \log \frac{\mu_{\mathrm{ChPT}}^2}{m_{\chi}^2} +
    \frac{4}{3} \right)\ .
\end{align}

To prove the equivalence between Eq.\,\eqref{eq:TreeLevel} and Eq.\,\eqref{eq:WinoPionYukawa}, let us consider the field redefinitions,
\begin{align}
    &\psi_0 = \psi_0' + i \varepsilon \pi^+ \psi_-' -
    i \varepsilon \pi^- \psi_-^{\prime c}\ , \\
     &\bar{\psi}_0 = \bar{\psi}_0' - i \varepsilon \pi^{-} \bar{\psi}_-' + i \varepsilon \pi^+ \bar{\psi}_-^{\prime c}\ , \\
    &\psi_- = \psi_-' + i \varepsilon \pi^- \psi_0'\ ,\\
    &\bar{\psi}_- = \bar{\psi}_-' - i \varepsilon \pi^+ \bar{\psi}_0'\ .
\end{align}
Here, $\varepsilon$ will be a parameter which will be determined below.
The $S$-matrix is not affected by this change. 

\begin{figure}[t]
    \centering
    \includegraphics[]{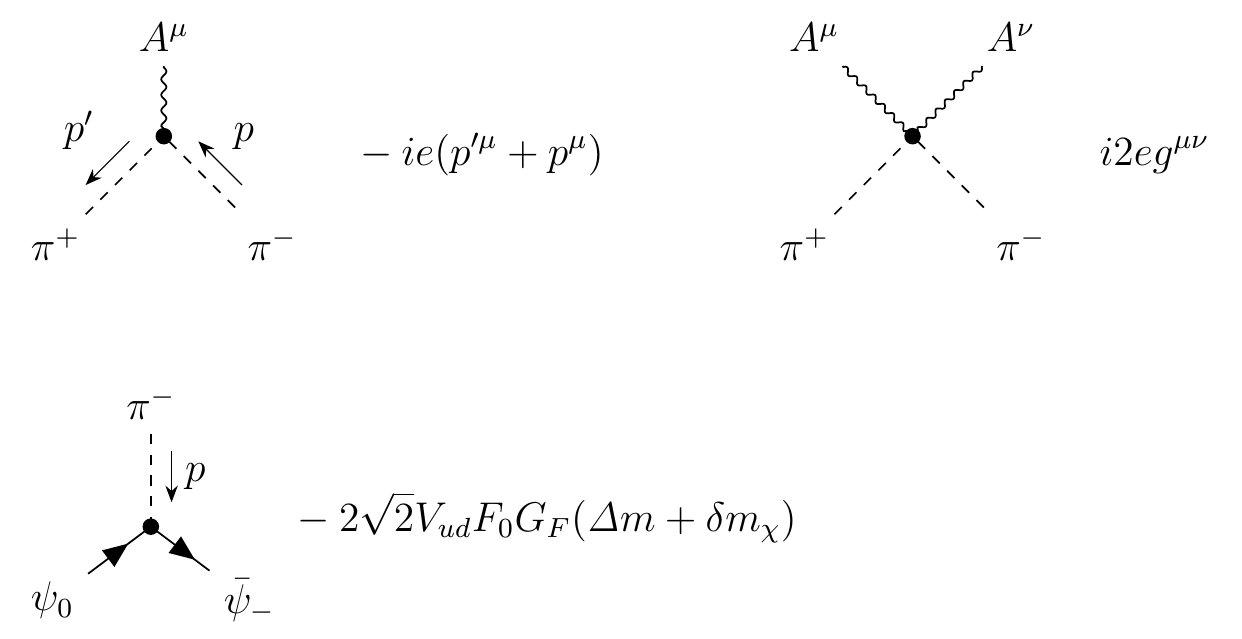}
    \caption{The Feynman rules of the Wino--pion interactions in Eq.\,\eqref{eq:WinoPionYukawa}. }
    \label{fig:Wino-Pion-QED2}
\end{figure}

From the kinetic term of $\psi_0$, we obtain
\begin{align}
    \delta\mathcal{L}_0&=-\frac{1}{2}(i\varepsilon \pi^- \bar{\psi}_-' - i\varepsilon \pi^+ \bar{\psi}_-^{\prime c})(i\slashed{\partial}-m_0)\psi_0^{\prime } + \frac{1}{2} \bar{\psi}_0' (i\slashed{\partial}-m_0) (i \varepsilon \pi^+ \psi_-' -
    i \varepsilon \pi^- \psi_-^{\prime c}) \cr
     &= \varepsilon \pi^-\bar{\psi}'_-\slashed{\partial}\psi'_0  
     -  \varepsilon \bar{\psi}'_0\slashed{\partial}(\pi^+\psi'_-) +  i\varepsilon m_0 (\pi^-\bar{\psi}'_- \psi'_0  -  \pi^+\bar{\psi}'_0 \psi'_-) \cr
    &= \varepsilon \pi^-\bar{\psi}'_-\slashed{\partial}\psi'_0  -  \varepsilon \bar{\psi}'_0(\slashed{D}\pi^+)\psi'_- -  \varepsilon \bar{\psi}'_0\pi^+(\slashed{D}\psi'_-) +  i \varepsilon m_0 (\pi^-\bar{\psi}'_- \psi'_0  -  \pi^+\bar{\psi}'_0 \psi'_-)\ .
\end{align}
Similarly, we obtain 
\begin{align}
    \delta \mathcal{L}_- &=  - i\varepsilon \pi^+ \bar{\psi}'_0(i\slashed{D}-\hat{m}_{{\chi}}) \psi'_- + i\varepsilon \bar{\psi}'_-(i \slashed{D}-\hat{m}_{{\chi}})(\pi^-\psi'_0) \cr
    &= \varepsilon \pi^+ \bar{\psi}'_0 \slashed{D}\psi'_- - \varepsilon\bar{\psi}'_- (\slashed{D}\pi^-) \psi'_0 - \varepsilon \pi^-\bar{\psi}'_- i\slashed{\partial}\psi'_0
    - i  \varepsilon \hat{m}_{{\chi}}
    (\pi^-\bar{\psi}'_- \psi'_0-\pi^+ \bar{\psi}'_0 \psi'_-)\ .
\end{align}
Here, we have used
\begin{align}
    \bar{\psi}^{\prime c}_i \psi_j^{\prime c} &= 
    (C\bar{\psi}^{\prime T}_i)^\dagger \gamma^0 (C \bar{\psi}^{\prime T}_j) =  (\psi_i^{\prime T} \gamma^0 C^\dagger) \gamma^0( C \bar{\psi}^{\prime T}_j)
    = - (\psi_i^{\prime T} \bar{\psi}^{\prime T}_j) = \psi'_j \bar{\psi}'_i
   \ ,\\
\bar{\psi}_i^{\prime c} \gamma^\mu\psi_j^{\prime c} &= 
    (C\bar{\psi}^{\prime T}_i)^\dagger \gamma^0 \gamma^\mu(C \bar{\psi}^{\prime T}_j) =  (\psi^{\prime T}_i \gamma^0 C^\dagger) \gamma^0\gamma^\mu( C \bar{\psi}^{\prime T}_j)
    = (\psi^{\prime T}_i \gamma^{\mu T}\bar{\psi}^{\prime T}_j) = -\psi'_j \gamma^\mu\bar{\psi}'_i\ .
\end{align}

Altogether, we obtain, 
\begin{align}
\delta\mathcal{L}&= - \varepsilon\bar{\psi}'_-(\slashed{D}\pi^-)\psi'_0 - \varepsilon\bar{\psi}'_0 \slashed{D}\pi^+ \psi'_- -i  \varepsilon (\hat{m}_\chi - m_0)
    (\pi^-\bar{\psi}'_- \psi'_0-\pi^+ \bar{\psi}'_0 \psi'_-)\cr
    &= - \varepsilon\bar{\psi}'_-(\slashed{D}\pi^-)\psi'_0 - \varepsilon\bar{\psi}'_0 \slashed{D}\pi^+ \psi'_- -i  \varepsilon (\mathit{\Delta}{m} + \delta m_\chi)
    (\pi^-\bar{\psi}'_- \psi'_0-\pi^+ \bar{\psi}'_0 \psi'_-)\ ,
\end{align}
which does not affect the $S$-matrix.
Thus, by choosing
\begin{align}
    \varepsilon = -2\sqrt{2}V_{ud}F_0{G_F}\ ,
\end{align}
we can replace the interaction in Eq.\,\eqref{eq:TreeLevel}
with the one in Eq.\,\eqref{eq:WinoPionYukawa}.
The Feynman rules for Eq.\,\eqref{eq:WinoPionYukawa} are
given in Fig.\,\ref{fig:Wino-Pion-QED2}.

\section{Real Photon Emission in Wino Decay}
\label{sec:Details of Real Emission}
In this appendix, we present the details of the computation of the real photon emission.
It is convenient to 
use the Wino-pion interaction 
in Eq.\,\eqref{eq:WinoPionYukawa},
which is equivalent to the Wino-pion interaction in Eq.\,\eqref{eq:TreeLevel}.
In this picture, 
the relevant diagrams are,
\begin{align}
      i\mathcal{M}_1^\mathrm{emit} &= \begin{tikzpicture}[baseline=-0.1cm] 
\begin{feynhand}    
    \vertex [dot](w1) at (0,0){};
     \vertex [dot](w3) at (-1,0){};
 \vertex [particle] (a) at (1.95,1.3) {$\qquad\pi^-(r)$};
     \vertex [particle] (b) at (1.5,-1.) {$\bar{\psi}_0(q)$};
     \vertex [particle] (e) at (1.5,1.66667) {$\gamma(k)$};
  \vertex [particle] (c) at (-1,0) ;
    \vertex [particle] (d) at (-2.5,0) {$\psi_-(p)$};
     \propag [sca] (w1) to [] (a);
     \propag [photon] (w3) to 
     [](e);
       \propag [fer] (c) to [] (w1);
       \propag [fer] (d) to [] (c) ;
\propag [fer] (w1) to (b);
\end{feynhand}
\end{tikzpicture} \cr
 & =-(-ie)(2\sqrt{2}\mathit{\Delta} m V_{ud} F_0 G_F) \bar{u}_{0}(q) \frac{i(\slashed{p}-\slashed{k}+m_{{\chi}})}{(p-k)^2 - m_{{\chi}}^2} \slashed{\epsilon}^*  u_{-}(p)\ ,
 \label{eq:Winoemit1}\\[10pt]
 i\mathcal{M}^\mathrm{emit}_2 & = \begin{tikzpicture}[baseline=-0.1cm] 
\begin{feynhand}    
    \vertex [dot](w1) at (0,0){};
     \vertex [dot](w2) at (0.707,0.471333){};
     \vertex [particle] (a) at (1.95,1.3) {$\qquad\pi_-(r)$};
     \vertex [particle] (b) at (1.5,-1.) {$\bar{\psi}_0(q)$};
  \vertex [particle] (c) at (-1,0) {};
    \vertex [particle] (d) at (-2.5,0) {$\psi_-(p)$};
     \vertex [particle] (e) at (1.5,1.66667) {$\gamma(k)$};
     \propag [sca] (w1) to [] (w2);
     \propag [photon] (w2) to 
     [](e);
       \propag [fer] (d) to [] (w1) ;
     \propag [fer] (w1) to (b);
     \propag [sca] (w2) to (a);
\end{feynhand}
\end{tikzpicture} \cr
&=-(-ie)(2\sqrt{2}\mathit{\Delta} m V_{ud}F_0 G_F) \epsilon^{*}_\mu(2r+k)^\mu \frac{i}{(r+k)^2-m_\pi^2} \bar{u}_{0}(q)u_{-}(p)\ ,
\label{eq:Winoemit2}
\end{align}
where $\epsilon^\mu$ denotes the polarization vector of the photon.
We have neglected $\delta m_\chi$ term as it gives higher order contribution for $\order{e\alpha}$. 
When we use the Wino-pion interaction in the form of Eq.\,\eqref{eq:WinoPionYukawa},
the third diagram in Fig.\,\ref{fig:WinoQEDSoft} does not appear.

The denominators of Eqs.\,\eqref{eq:Winoemit1} and \eqref{eq:Winoemit2} are reduced to
\begin{align}
(p - k)^2 - m_{{\chi}} = -2 p\cdot k \ , \quad 
    (r + k)^2 - m_{\pi}^2 = 2 r\cdot k \ , 
\end{align}
where we have used $k^2=0$.
Besides,
\begin{align}
    (\slashed{p} - \slashed{k} + m_{{\chi}})
    \slashed{\epsilon}^* u_{-}(p)
    =( 2 \epsilon^*\cdot p - \slashed{k}\slashed{\epsilon}^*) u_{-}(p)  \ , 
    \quad {\epsilon}^*\cdot (2r + k) =  2 {\epsilon}^*\cdot r\ ,
\end{align}
where we have used $\epsilon^*_\mu k^\mu = 0$.
Thus, the total real emission amplitude is
\begin{align}
\label{eq:Winoemit}
    i \mathcal{M}^\mathrm{emit} =2\sqrt{2}e\mathit{\Delta} m V_{ud} F_0 G_F
    \bar{u}_{0}(q)u_{-}(p) 
    \left(\frac{p\cdot \epsilon^* }{p\cdot k}
    -\frac{r\cdot\epsilon^*}{r\cdot k}\right) -\sqrt{2}e\mathit{\Delta} m V_{ud} F_0 G_F
    \bar{u}_{0}(q)\frac{\slashed{k}\slashed{\epsilon}^*}{p\cdot k}u_{-}(p)\ .
\end{align}
The spin summed squared matrix averaged by the charged Wino spin is given by,
\begin{align}
\overline{|{\mathcal{M}}^\mathrm{emit}|^2}
=& \frac{1}{2}\sum_{\mathrm{spin}}|{\mathcal{M}}^\mathrm{emit}|^2  \cr
=& 16 e^2 \mathit{\Delta} m^2 V_{ud}^2 F_0^2 G_F^2
\Bigg[(p\cdot q +m_0m_{{\chi}})
\left(
-\frac{m_\chi^2 }{(p\cdot k)^2}
    -\frac{m_\pi^2}{(r\cdot k)^2}
    + \frac{2 p\cdot r}{(k\cdot p)(k\cdot r)}
    \right) \cr
    &+\left(
    \frac{r\cdot q}{k\cdot r}
    -\frac{p\cdot q}{k\cdot p}
    + \frac{k\cdot q}{k\cdot p} 
     - \frac{  (k\cdot q)(p\cdot r)}{(k\cdot p)(k\cdot r)}
    +\frac{m_{{\chi}}^2 k\cdot q}{(k\cdot p)^2}
    \right)\Bigg]\ ,
    \label{eq:WinoMSQemit}
\end{align}
where we have used $\sum_{\mathrm{polarization}}
\epsilon_\mu \epsilon_{\nu}^*
=-g_{\mu\nu}$.
The Wino decay rate with a real emission is given by,
\begin{align}
    \Gamma^{\mathrm{emit}} = \frac{1}{(2\pi)^3} \frac{1}{m_{{\chi}}^3} \int dm_{0\pi}^2 dm_{0\gamma}^2 |\bar{\mathcal{M}}_{{\chi}}^\mathrm{emit}|^2\ ,
\end{align}
where $m_{0\pi}^2 = (q+r)^2 = (p-k)^2$ and $m_{0\gamma}^2 = (q+k)^2 = (p-r)^2$.

For a given value of $m^2_{0\pi}$, 
the range of $m^2_{0\gamma}$ is given by,
\begin{align}
    (m^2_{0\gamma})^\mathrm{max} &= (E_0^*+E_\gamma^*)^2 - \left(\sqrt{E_0^{*2}-m_0^2}-\sqrt{E_\gamma^{*2}-m_\gamma^2}\right)^2\ , \\
    (m^2_{0\gamma})^\mathrm{min} &= (E_0^*+E_\gamma^*)^2 - \left(\sqrt{E_0^{*2}-m_0^2}+\sqrt{E_\gamma^{*2}-m_\gamma^2}\right)^2\ , \\
     E_0^* &= \frac{m_{0\pi}^2 - m_{{\chi}}^2 + m_0^2}{2 m_{0\pi}}\ , \\
     E_\gamma^* &= \frac{m_{{\chi}}^2 - m_{0\pi}^2 - m_\gamma^2}{2 m_{0\pi}}\ .
\end{align}
The kinematical range of $m_{0\pi}^2$ is given by,
\begin{align}
    (m_{\pi} + m_0)^2< m_{0\pi}^2 < (m_{{\chi}} - m_\gamma)^2 \ .
\end{align}
We also use
\begin{align}
    p\cdot k &= (m_{{\chi}}^2 + m_\gamma^2 - m_{0\pi}^2)/2 \ , \\
    r \cdot k &= (m_{{\chi}}^2 + m_\nu^2 -m_{0\pi}^2 - m_{0\gamma}^2)/2 \ , \\
    q\cdot k&= (m_{0\gamma}^2 - m_\gamma^2-m_0^2)/2\ , \\
    p\cdot r &= (m_{{\chi}}^2 + m_\pi^2 - m_{0\gamma}^2)/2\ ,\\
    p\cdot q &= (m_{\pi0}^2 + m_{0\gamma}^2 - m_\pi^2 - m_\gamma^2)/2\ , \\
    r\cdot q &= (m_{0\pi}^2 - m_\pi^2-m_0^2)/2 \ .
\end{align}
The integration over $m_{0\gamma}^2$ can be performed simply.

Let us consider the first term in 
Eq.\,\eqref{eq:WinoMSQemit} which contributes to the IR divergence.
We divide the integration region of $m_{0\pi}^2$ into the soft region
\begin{align}
    m_{{\chi}}^2 + m_\gamma^2 - 2 m_{{\chi}} E_\mathrm{soft} < m_{0\pi}^2 < (m_{{\chi}}-m_\gamma)^2
\end{align}
and the hard region
\begin{align}
(m_\pi+m_0)^2 <m_{0\pi}^2< m_{{\chi}}^2 + m_\gamma^2 - 2 m_{{\chi}} E_\mathrm{soft}\ ,
\end{align}
in the charged Wino rest frame.
We take $E_\mathrm{soft}$ in
\begin{align}
    m_\gamma \ll E_\mathrm{soft} \ll \frac{m_{{\chi}}^2 + m_\gamma^2-(m_0+m_\pi)^2}{2m_{{\chi}}} \simeq \mathit{\Delta} m - m_\pi \ .
\end{align}
In the soft region, integration over $m_{0\pi}^2$ can be performed by choosing 
the photon velocity $v=\sqrt{1-m_\gamma^2/E_\gamma^2}$, as a new integration  variable~\cite{Kinoshita:1958ru}. 
The integration region of $v$ is,
\begin{align}
    0<v < \frac{\sqrt{x_\mathrm{soft}^2-4 \lambda^2 }}{x_\mathrm{soft}} \ , 
\end{align}
where $\lambda = m_{\gamma}/m_{{\chi}}$ and $x_\mathrm{soft} = 2 E_\mathrm{soft}/m_{{\chi}}$.

As a result, we find the contribution from the first term in Eq.\,\eqref{eq:WinoMSQemit} is given by,
\begin{align}
\label{eq:Wino Emit soft}
    \frac{\Gamma_{{\chi}}^{\mathrm{emit,soft,1}}}{\Gamma_{{\chi}}} = \frac{\alpha}{\pi}
    \Bigg\{&
    1+2 
    \left( 1+\frac{1}{2\sqrt{1-z^2}}
    \log
    \frac{1-\sqrt{1-z^2}}{1+\sqrt{1-z^2}}\right)\log\frac{m_{\gamma}}{2 E_\mathrm{soft}}\cr
    &+ \frac{1}{2\sqrt{1-z^2}}\bigg[-
    \frac{\pi^2}{3}-
   \left(
   1-\log
    \frac{4(1-z^2)}{z^2}
   \right)
   \log
    \frac{1-\sqrt{1-z^2}}{1+\sqrt{1-z^2}} \cr
    & + \frac{1}{2}\log^2\frac{1-\sqrt{1-z^2}}{1+\sqrt{1-z^2}}+ 2\,\mathrm{Li_2}\left(
    \frac{1-\sqrt{1-z^2}}{1+\sqrt{1-z^2}}
    \right)\bigg]\Bigg\}\ ,
\end{align}
in the limit of $m_{{\chi}}\gg \mathit{\Delta} m> m_\pi $.
By comparing Eqs.\,\eqref{eq:WinoVirtual} and \eqref{eq:Wino Emit soft}, we find that 
the IR divergence is cancelled between the virtual correction and the real emission.
The hard part contribution from the first term in Eq.\,\eqref{eq:WinoMSQemit} is given by
\begin{align}
\label{eq:Wino Emit Hard1}
    \frac{\Gamma_{{\chi}}^{\mathrm{emit,hard,1}}}{\Gamma_{{\chi}}} =\frac{\alpha}{\pi}
    \Bigg\{4&
    -2\log
    \frac{4(1-z^2)}{z}
    +2 
    \left( 1+\frac{1}{2\sqrt{1-z^2}}\log
    \frac{1-\sqrt{1-z^2}}{1+\sqrt{1-z^2}}
    \right)\log \frac{2E_\mathrm{soft}}{\mathit{\Delta}m} \cr
&+\frac{1}{\sqrt{1-z^2}}
    \bigg[
    -\frac{\pi^2}{3} - \log z \log 
    \frac{1-\sqrt{1-z^2}}{1+\sqrt{1-z^2}}
    \cr
    &
    + \frac{1}{2}\log^2\frac{1-\sqrt{1-z^2}}{1+\sqrt{1-z^2}} +2\,\mathrm{Li}_2\left(\frac{1-\sqrt{1-z^2}}{1+\sqrt{1-z^2}}    \right)\bigg]\Bigg\}\ ,
\end{align}
 in the limit of the large Wino mass.
 The contribution from the second line in Eq.\,\eqref{eq:WinoMSQemit} is given by,
 \begin{align}
 \label{eq:Wino Emit Hard2}
     \frac{\Gamma_{{\chi}}^{\mathrm{emit,2}}}{\Gamma_{{\chi}}}
     = \frac{\alpha}{4\pi}\frac{\mathit{\Delta} m^2}{m_\chi^2}
     \left(
     1+2z^2 - \frac{3z^2\log z}{\sqrt{1-z^2}}
     + \frac{3z^2\log\left(1-\sqrt{1-z^2}\right)}{\sqrt{1-z^2}}
     \right)\ ,
 \end{align}
 which is free from the IR divergence, and we can directly integrate over $m_{0\gamma}^2$ and $m_{0\pi}^2$.
 As this contribution is further suppressed by $(\mathit{\Delta}m/m_\chi)^2\ll 1$,
 this contribution is negligible compared with the above contributions.

\subsection*{Recalculation of QED Virtual Correction}
Incidentally, 
we recalculate the virtual correction using the Wino-pion interaction 
in Eq.\,\eqref{eq:WinoPionYukawa}.
In this picture, the QED correction is given by a single diagram,
\begin{align}
\label{eq:deltaMWinoPicture2}
      i\delta\mathcal{M}_{{\chi}} &= 
      \begin{tikzpicture}[baseline=0cm] 
 \begin{feynhand}    
     \vertex [dot](w1) at (0,0){};
      \vertex [dot](w2) at (0.707,0.707){};
      \vertex [dot](w3) at (-1,0){};
      \vertex [particle] (a) at (1.5,1.5) {$\pi^-(r)$};
      \vertex [particle] (b) at (1.5,-1.5) {$\psi_0(q)$};
   \vertex [particle] (c) at (-1,0);
     \vertex [particle] (d) at (-2.5,0) {$\psi_-(p)$};
      \propag [photon,mom=$\ell$] (w3) to 
      [half left, looseness=1.2](w2);
       \propag [fer] (c) to  (w1);
       \propag [fer] (d) to  (c) ;
       \propag [sca] (w1) to [] (a);
      \propag [fer] (w1) to (b);
 \end{feynhand}
 \end{tikzpicture} \cr
 & = 
 -2\sqrt{2}e^2V_{ud}F_0 G_F \mathit{\Delta} m \mu^{2\epsilon}\int \frac{d^d\ell}{(2\pi)^d}
\frac{-i}{\ell^2 - m_\gamma^2}
\frac{i(2r -\ell)_\mu}{\ell_\pi^2 - m_\pi^2}
\left[\bar{u}(q)
\frac{i(\slashed{\ell}_{\chi}+m_{{\chi}})}{\ell_\chi^2 - m_{{\chi}}^2}\gamma^\mu
u(p)\right]\ ,
\end{align}
where $\ell_\chi = \ell - p$ and $\ell_\pi = \ell -r$.
Here, we have dropped $\delta m_\chi$ in the one-loop diagram since $\delta m_{\chi}= \order{\alpha}$.

The $\order{\alpha}$ contribution to 
the virtual correction including the
Wino mass counterterm is given by
\begin{align}
    \frac{ {\mathcal{M}}_{{\chi}}^{\mathrm{Virtual}}}{\hat{\mathcal{M}}_{\mathrm{tree}}}
    \Bigg|_{\mathrm{Eq.\,\eqref{eq:WinoPionYukawa}}}
    = \frac{ \delta{\mathcal{M}}_{{\chi}} + \mathcal{M}^{\mathrm{WF}}}
    {\hat{\mathcal{M}}_{\mathrm{tree}}} +  \frac{\delta m_{{\chi}}}{\mathit{\Delta} m} \ .
\end{align}
This expression reproduces the virtual correction in Eq.\,\eqref{eq:WinoVirtual}
obtained by the original Wino-pion interaction 
in Eq.\,\eqref{eq:TreeLevel}.
Here, we have use $\delta m_\chi$ in Eq.\,\eqref{eq:delta mchi}.
In this picture, we find that the $m_\chi/\mathit{\Delta}m$ enhanced term originates from the Wino mass counterterm.

\bibliographystyle{apsrev4-1}
\bibliography{oneloop}

\end{document}